\documentclass{emulateapj}

\usepackage{geometry}
\usepackage{graphicx}
\usepackage{epstopdf}
\usepackage{amsmath}
\usepackage{amsfonts}
\usepackage{amssymb}

\usepackage{CJKutf8}
\usepackage[T1]{fontenc}
\usepackage{times}

\renewcommand{\(}{\left(}
\renewcommand{\)}{\right)}
\newcommand{\vb}{{\bf v}_b}
\newcommand{\xvec}{{\bf x}}

\newcommand{\rhob}{\rho_b}

\newcommand{\mn}{{\tt n}}

\newcommand{\hi}{H {\footnotesize I}~}
\newcommand{\hii}{H {\footnotesize II}~}

\begin{document}

\title{Direct Numerical Simulation of Reionization II: Recombinations, Clumping Factors, and the Photon Budget for Reionization}

\author{Geoffrey C. So \begin{CJK}{UTF8}{bkai}(蘇治平)\end{CJK}\altaffilmark{1}}
\author{Michael L. Norman\altaffilmark{1,2}}
\author{Daniel R. Reynolds\altaffilmark{3}}
\author{Robert P. Harkness\altaffilmark{2,4}}

\affiliation{
\altaffilmark{1}CASS, University of California, San Diego, 9500 Gilman Drive La Jolla, CA 92093-0424\\
\altaffilmark{2}SDSC, University of California, San Diego, 9500 Gilman Drive La Jolla, CA 92093-0505\\ 
\altaffilmark{3}Southern Methodist University, 6425 Boaz Ln, Dallas, TX 75205\\
\altaffilmark{4}NICS, Oak Ridge National Laboratory, 1 Bethel Valley Rd, Oak Ridge, TN 37831\\
}

\begin{abstract}
We use a fully self-consistent cosmological simulation including dark
matter dynamics, multispecies hydrodynamics, chemical ionization,
flux limited diffusion radiation transport, and a parameterized model of star formation and feedback (thermal and radiative) to investigate the epoch of hydrogen
reionization in detail. 
Our numerical method is scalable with respect to the number of radiation sources, size of the mesh, and the number of computer processors employed, and is described in Paper I of this series.  
In this the first of several application papers, we investigate the mechanics of reionization from stellar sources forming in high-z galaxies, the utility of various formulations for the gas clumping factor on accurately estimating the effective recombination time in the IGM, and the photon budget required to achieve reionization. We also test the accuracy of the static and time-dependent models of Madau et al. as predictors of reionization completion/maintenance.  

We simulate a WMAP7 $\Lambda$CDM cosmological model in a 20 Mpc comoving cube, resolved with $800^3$ uniform fluid cells and dark matter particles. By tuning our star formation recipe to approximately match the observed high redshift star formation rate density and galaxy luminosity function, we have created a fully coupled radiation hydrodynamical realization of hydrogen reionization which begins to ionize at $z \approx 10$ and completes at $z \approx 5.8$ without further tuning. The complicated events during reionization that lead to this number can be generally described as inside-out, but in reality the narrative depends on the level of ionization of the gas one attributes to as ionized.  We find that roughly 2 ionizing photons per H atom are required to convert the neutral IGM to a highly ionized state, which supports the ``photon starved'' reionization scenario discussed by Bolton \& Haehnelt. We find that the formula for the ionizing photon production rate $\dot{\mathcal{N}}_{ion}(z)$ needed to maintain the IGM in an ionized state derived by Madau et al. should not be used to predict the epoch of reionization completion because it ignores history-dependent terms in the global ionization balance which are not ignorable. We find that the time-dependent model for the ionized volume fraction $Q_{\hii}$ is more predictive, but overestimates the redshift of reionization completion $z_{reion}$ by $\Delta z \approx 1$. We propose a revised formulation of the time-dependent model which agrees with our simulation to high accuracy. Finally, we use our simulation data to estimate a globally averaged ionizing escape fraction due to circumgalactic gas resolved on our mesh $\bar{f}_{esc}(CGM) \approx 0.7$. 

\end{abstract}

\maketitle

\section{Introduction}
\label{sec:introduction}

The Epoch of Reionization (EoR) is an active area of research observationally,
theoretically, and computationally. Observations constrain the tail end of hydrogen reionization
to the redshift range $z=6-8$ \citep{RobertsonEtAl2010}. These observations include the presence of Gunn-Peterson
troughs in the Ly $\alpha$ absorption spectra of high redshift quasars \citep{FanEtAl2006}, and 
the strong evolution of Lyman $\alpha$ emitter luminosity function (Robertson et al. 2010 and references
therein.) 
Observations from the WMAP and Planck satellites tell us that the universe was substantially 
ionized by $z \approx 10$ but can say little about the
reionization history or topology \citep{JarosikEtAl2011,Planck2013}. 
High redshift 21cm observations hold forth great promise of elucidating the details of this transition \citep{BarkanaLoeb2007, PritchardLoeb2012}, but these results are still in the future. 

It is believed that early star forming galaxies provided the bulk of the UV photons responsible for
reionization \citep{RobertsonEtAl2010,RobertsonEtAl2013}, but early QSOs may have also contributed \citep{MadauEtAl1999, BoltonHaehnelt2007, HaardtMadau2012}.  The ``galaxy reionizer" hypothesis has been greatly strengthened by the recent advances in the study of high redshift galaxies afforded by the IR-sensitive Wide Field Camera 3 (WFC3) aboard the Hubble Space Telescope \citep[e.g.][]{RobertsonEtAl2010, RobertsonEtAl2013, BouwensEtAl2011, BouwensEtAl2011b, OeschEtAl2013}.  Within uncertainties, the luminosity function of $z=6$ Lyman break galaxies (LBGs) appears to be sufficient to account for reionization at that redshift from a photon counting argument \citep{BoltonHaehnelt2007, RobertsonEtAl2010, BouwensEtAl2012}. Among the observational uncertainties are the faint-end slope of the galaxy luminosity function \citep{WiseCen2009,LabbeEtAl2010,BouwensEtAl2012}, the spectral energy distribution of the stellar population \citep{CowieEtAl2009,WillotEtAl2010,HaardtMadau2012}, and the escape fraction of ionizing photons \citep{WyitheEtAl2010, YajimaEtAl2011, MitraEtAl2013}. Among the theoretical uncertainties are the number of ionizing photons per H atom required to bring the neutral IGM to its highly ionized state by $z=6$, the clumping factor correction to the mean IGM recombination time \citep{PawlikEtAl2009, RaicevicTheuns2011, FinlatorEtAl2012, ShullEtAl2012, RobertsonEtAl2013}, 
and the contribution of Pop III stars and accreting black holes to the early and late stages of reionization \citep{BoltonHaehnelt2007,TracGnedin2011,AhnEtAl2012}. 

When assessing whether an observed population of high-z galaxies is capable of 
reionizing the universe (e.g., Robertson et al. 2013), observers often use the criterion derived
by \cite{MadauEtAl1999} for the ionzing photon volume density $\dot{\mathcal{N}}_{ion}$ necessary
to maintain the clumpy IGM in an ionized state:

\begin{align}
\label{eq:ndot}
\dot{\mathcal{N}}_{ion}(z) &= \frac{\bar{n}_\mathrm{H}(0)}{\bar{t}_{rec}(z)} = (10^{51.2}s^{-1}Mpc^{-3})\left(\frac{C}{30}\right) \notag\\
&\times \left(\frac{1+z}{6}\right)^3 \left(\frac{\Omega_b h_{50}^2}{0.08}\right)^2,
\end{align}

where $\bar{n}_\mathrm{H}(0)$ is the mean comoving number density of H atoms, $C \equiv \langle n^2_\mathrm{H\,II} \rangle /\langle n_\mathrm{H\,II} \rangle ^2$ is the H {\footnotesize II} clumping factor (angle brackets denote volume average over a suitably large volume that the average is globally meaningful),
and the rest of the symbols have their usual meaning. 
The origin of this formula is a simple photon counting argument, which says 
that in order to maintain ionization at a given redshift $z$, 
the number of ionizing photons emitted in
a large volume of the universe multiplied by a characteristic recombination
time, denoted $\bar{t}_{rec}$,  must equal the number of hydrogen atoms: $\dot{\mathcal{N}}_{ion} \times \bar{t}_{rec} = \bar{n}_\mathrm{H}(0)$.
The clumping factor enters as a correction factor to account for the density inhomogeneties in the IGM
induced by structure formation. We note that $\bar{t}_{rec}$ is not the volume average of the local recombination time of the
ionized plasma, as this would heavily weight regions with the {\em longest} recombination
times; i.e. voids. A proper derivation of  Equation \eqref{eq:ndot} shows that $\bar{t}_{rec} \propto \langle t_{rec}^{-1} \rangle ^{-1}$, which weights regions
with the {\em shortest} recombination times; i.e. regions at the mean density and above. 

Equation \eqref{eq:ndot} is based on a number of simplifying assumptions discussed by
\cite{MadauEtAl1999}, including the assumption $\bar{t}_{rec} \ll t$. It is this assumption that allows history-dependent effects to be ignored, and a quasi-instantaneous analysis of the 
photon budget for reionization to be done. The validity of this assumption is naturally
redshift dependent, but it is also dependent upon the adopted definition of $\bar{t}_{rec}$.  A second comment about Equation \eqref{eq:ndot} is that it does not ask how many ionizing photons per H atom are required to convert a neutral IGM to a fully ionized one, only how many are required to {\em maintain} the IGM in an ionized state. Because the recombination time is short at high redshifts, it is expected that this number is greater than one. 

In this paper we examine these and related topics within the context of a direct numerical simulation of cosmic reionization based on a new flux-limited diffusion radiation transport solver installed in the {\em Enzo} code \citep{NormanEtAl2013} (hereafter Paper I).
Our approach self-consistently couples all the relevant physical processes 
(gas dynamics, dark matter dynamics, self-gravity, star formation/feedback, 
radiative transfer, nonequilibrium ionization/recombination, heating and cooling) and evolves the
system of coupled equations on the same high resolution mesh. We refer to this
approach as {\em direct numerical simulation} or {\em resolution matched}, in contrast to previous approaches 
which decouple and coarse-grain the radiative transfer and ionization balance 
calculations relative to the underlying dynamical calculation.
Our method is
scalable with respect to the number of radiation sources, size of the mesh, and the
number of computer processors employed.
This scalability permits us to simulate cosmological reionization in large cosmological
volumes (L $\sim100$ Mpc) while directly modeling the sources and sinks of ionizing 
radiation,  
including radiative feedback effects such as photoevaporation of gas from halos, 
Jeans smoothing of the IGM, and enhanced recombination due to small scale clumping. 
In this the first of several application papers, we investigate in a volume of modest size (L=$20$ Mpc) the mechanics of reionization from stellar sources forming in high-$z$ galaxies, the role of gas clumping, recombinations, and the photon budget required to complete reionization.

By analyzing this simulation we are able to critically examine the validity of  Equation \eqref{eq:ndot} as a predictor of when the end of EoR will occur, and we can calculate the integrated number of ionizing photons 
per H atom needed to ionize the simulated volume $\gamma_{ion}/H=\int dt \dot{\mathcal{N}}_{ion} / \bar{n}_\mathrm{H}(0)$. Ignoring recombinations within the virial radii of collapsed halos, we find $\gamma_{ion}/H \approx 2$. This result
supports the ``photon starved'' reionization scenario discussed by \cite{BoltonHaehnelt2007}.
We also examine whether modern revisions to Equation \eqref{eq:ndot} using alternatively defined 
clumping factors \citep{PawlikEtAl2009, RaicevicTheuns2011, FinlatorEtAl2012, ShullEtAl2012} are 
improvements over the original. We find they systematically overestimate the redshift of reionization completion $z_{reion}$ because the condition $\bar{t}_{rec}/t \ll 1$ is never obeyed.  We study the accuracy and validity of the time-dependent analytic model of \cite{MadauEtAl1999}, and find that while it is in better agreement with the simulation, it also overestimates $z_{reion}$ because it ignores important corrections to the ionization term at early and late times.

This paper is organized as follows: in \S\ref{Method} we discuss the design criteria 
for the simulation and briefly outline the basic equations and implementation of the FLD radiation transport model, referring the
reader to Paper I for a more complete description of the numerical algorithms and tests.  
In \S\ref{GeneralResults},  we present some general features of the simulation and demonstrate its broad consistency with observed star formation rate density and high redshift 
galaxy luminosity function. 
%Because we track the ionization fraction of the gas at every point in the volume as a function of time, a more quantitative language is required to say when the simulated volume is X\% ionized; this is provided in \S\ref{QuantitativeLanguage}. 
%In \S\ref{IOOI} we examine the mechanics of reionization in our simulation, and in particular question whether ``inside-out'' or ``outside-in'' is a better description of certain phases of evolution. 
In \S\ref{sec:ClumpingFactors} we examine the accuracy of different clumping factor approaches to estimating the redshift of complete reionization.
In \S\ref{escape} we derive a global estimate for the circumgalactic absorption of ionizing radiation from our simulation. 
In \S\ref{Qdot} we test a simple analytic model for the evolution of the ionized volume fraction $Q_\mathrm{H\,II}$ and present an improvement to the model which better agrees with our simulation. 
In \S\ref{Discussion} we
discuss implications of our results on the current understanding of
reionization.  And finally, in \S\ref{Conclusions} we end with a
summary of our main results and conclusions.

\section{Method}
\label{Method}
\subsection{Simulation Goals and Parameters}
We use the Enzo code \citep{TheEnzoCollaboration}, augmented with a flux-limited diffusion radiative transfer solver and a parameterized model of star formation and feedback \citep{NormanEtAl2013} to simulate inhomogeneous hydrogen reionization in a 20 Mpc comoving box in a WMAP7 $\Lambda$CDM cosmological model. Details of the numerical methods and tests are provided in Paper I.  Here we briefly describe the simulation's scientific goals and design considerations to put it into perspective with other reionization simulations. For completeness, the physical equations we solve and the treatment of the ionizing sources and radiation field are included below.

Our principle
goal is to simulate the physical processes occuring in the IGM outside the virial radii of high redshift galaxies in a {\em representative} realization of inhomogenous reionization. We wish to simulate the early, intermediate, and late phases of reionization  in a radiation hydrodynamic cosmological  framework so that we may study the nonequilibrium ionization/recombination processes in the IGM at reasonably high resolution self-consistently coupled to the dynamics. In this way we can study such effects as optically thick heating behind the I-fronts \citep{AbelHaehnelt1999}, Jeans smoothing \citep{ShapiroEtAl1994,Gnedin2000b}, photoevaporation of dense gas in halos \citep{ShapiroEtAl2004}, and nonequilibrium effects in the low density voids. Because we carry out our simulation on a fixed Eulerian grid, we do not resolve the internal processes of protogalaxies very well. In this sense, our simulation is not converged on all scales. Nonetheless Equations \eqref{eq:gravity} to  \eqref{eq:cons_radiation} are solved everywhere on the mesh self-consistently, including ionization/recombination and radiative transfer inside protogalaxies. The escape of ionizing radiation from galaxies to the IGM is thus simulated directly, and not introduced as a parameter. We use a star formation recipe that can be tuned to closely reproduce the observed high-$z$ galaxy luminosity function (LF), star formation rate density (SFRD), and redshift of reionization completion. This gives us confidence that we are simulating IGM processes in a realistic scenario of reionization. 

We simulate a WMAP7 \citep{JarosikEtAl2011} $\Lambda$CDM cosmological model with the following parameters: 
$\Omega_{\Lambda} = 0.73$, $\Omega_m = 0.27$, $\Omega_b = 0.047$, 
$h = 0.7$, $\sigma_8 = 0.82$, $n_s = 0.95$, where the symbols have their usual meanings.  
A Gaussian random
field is initialized at $z=99$ using the {\em Enzo} initial conditions generator {\em inits} 
using the \cite{EisensteinHu1999} fits to the transfer functions.The simulation is performed in a comoving volume of (20 Mpc)$^3$ with a grid
resolution of $800^3$ and the same number of dark matter particles. This yields a comoving spatial resolution of 25 kpc and
dark matter particle mass of $4.8 \times 10^5 M_{\odot}$. This resolution yields a dark matter halo mass function that
is complete down to $M_h = 10^8 M_{\odot}$, which is by design, since this is the mass scale below 
which gas cooling becomes inefficient. However, due to our limited boxsize, our halo mass function is incomplete above 
$M_h \approx 10^{11} M_{\odot}$ (see Figure \ref{HMF}). In a forthcoming paper we will report on a simulation of identical design and resolution as this one, but in a volume 64 times as large, which contains the rarer, more massive halos. With regard to resolving the diffuse IGM, our $25$ kpc resolution equals the value recommended by \cite{BryanEtAl1999} to converge on the properties of the Ly $\alpha$ forest at lower redshifts, is $3\times$ better than the optically thin high resolution IGM simulation described in \cite{ShullEtAl2012}, and nearly $4\times$ better than the inhomogeneus reionization simulation described in \cite{TracEtAl2008}. 

As described below in \S\ref{starformationandfeedback}, we use a parameterized model of star formation calibrated to observations of high redshift galaxies. The star formation efficiency parameter $f_*$ is adjusted to match the observed star formation rate density in the interval $6 \leq z \leq 10$ from \cite{BouwensEtAl2011}.  
The simulation consumed 255,000 core-hrs running on 512 cores of the Cray XT5 system {\em Kraken} operated by the 
National Institute for Computational Science at ORNL. 

\subsection{Governing Equations}
\label{GoverningEquations}

The equations of cosmological radiation hydrodynamics implemented in the Enzo code used for this research are given by the following system of partial differential equations (Paper I):

\begin{align}
  \nabla^2 \phi &= \frac{4\pi g}{a}(\rhob + \rho_\mathrm{dm} - \langle \rho \rangle),
  \label{eq:gravity}\\
  \partial_\mathrm{t} \rhob + \frac1a \vb \cdot \nabla
    \rhob &= -\frac1a \rhob \nabla\cdot\vb -\dot{\rho}_{SF},
  \label{eq:cons_mass}\\
  \partial_\mathrm{t} \vb + \frac1a\(\vb\cdot\nabla\)\vb &=
    -\frac{\dot{a}}{a}\vb - \frac{1}{a\rhob}\nabla p - \frac1a
    \nabla\phi,
  \label{eq:cons_momentum}\\
  \partial_\mathrm{t} e + \frac1a\vb\cdot\nabla e &=
    - \frac{2\dot{a}}{a}e
    - \frac{1}{a\rhob}\nabla\cdot\left(p\vb\right) \nonumber\\
    &- \frac1a\vb\cdot\nabla\phi + G - \Lambda + \dot{e}_{SF}
  \label{eq:cons_energy}\\
  \partial_\mathrm{t} \mn_\mathrm{i} + \frac{1}{a}\nabla\cdot\(\mn_\mathrm{i}\vb\) &=
    \alpha_\mathrm{i,j} \mn_\mathrm{e} \mn_\mathrm{j} - \mn_\mathrm{i} \Gamma_\mathrm{i}^{ph}, \qquad \nonumber\\
    &i=1,\ldots,N_\mathrm{s}
  \label{eq:chemical_ionization}\\
  \partial_\mathrm{t} E + \frac1a \nabla\cdot\(E \vb\) &= 
    \nabla\cdot\(D\nabla E\) - \frac{\dot{a}}{a}E \nonumber\\
    &- c \kappa E + \eta
  \label{eq:cons_radiation}
\end{align}
Equation \eqref{eq:gravity} describes the modified gravitational
potential $\phi$ due to baryon density $\rho_\mathrm{b}$ and dark matter
density $\rho_\mathrm{dm}$, with $a$ being the cosmological scale factor, $g$
being the gravitational constant, and $\langle \rho \rangle$ being the
cosmic mean density.  The collisionless dark matter density
$\rho_\mathrm{dm}$ is evolved using the Particle Mesh method (equation not
shown above), as described in 
\citealt{HockneyEastwood1988, TheEnzoCollaboration}. 
Equations \eqref{eq:cons_mass}, \eqref{eq:cons_momentum} and
\eqref{eq:cons_energy} are conservation of mass, momentum and energy,
respectively, in a comoving coordinate system \citep{BryanEtAl1995,TheEnzoCollaboration}.
In the above equations, $\vb\equiv a(t)\dot{\xvec}$ is the proper
peculiar baryonic velocity, $p$ is the proper pressure, $e$ is the
total energy per unit mass, and $G$ and $\Lambda$ are the heating and
cooling coefficients.  Equation \eqref{eq:chemical_ionization}
describes the chemical balance between the different ionization
species (in this paper we used H {\footnotesize I}, 
H {\footnotesize II}, He {\footnotesize I}, He {\footnotesize II}, 
He {\footnotesize III} densities) and electron density. Here, $\mn_\mathrm{i}$ is the
comoving number density of the $i^{th}$ chemical species, $\mn_\mathrm{e}$ is
the electron number density, $\mn_\mathrm{j}$ is the ion that reacts with
species $i$, and $\alpha_\mathrm{i,j}$ are the reaction rate coefficient
between species $i$ and $j$ \citep{AbelEtAl1997, HuiGnedin1997}, and
finally $\Gamma^{ph}_\mathrm{i}$ is the photoionization rate for species $i$. 

\subsection{Radiation Transport}
\label{RadiationTransport}

Equation \eqref{eq:cons_radiation} describes radiation transport in the Flux Limited
Diffusion (FLD) approximation in an expanding
cosmological volume \citep{ReynoldsEtAl2009,NormanEtAl2013}.  $E$ is the
comoving grey radiation energy density.  The {\em flux limiter} $D$ is
a function of $E$, $\nabla E$, and the opacity $\kappa$
\citep{Morel2000}, and has the form:
\begin{align}
  D &= \mbox{diag}\left(D_1, D_2, D_3\right), \quad\mbox{where} \\
  D_\mathrm{i} &= c \(9\kappa^2 + R_\mathrm{i}^2\)^{-1/2},\quad\mbox{and} \\
  R_\mathrm{i} &= \max\left\{\frac{|\partial_\mathrm{x_i} E|}{E},10^{-20}\right\}
\end{align}
In the calculation of the grey energy density $E$, we assume
$E_\nu(\mathbf{x},t,\nu)=\tilde{E}(\mathbf{x},t)\,\chi_E(\nu)$, therefore:
\begin{align}
\label{eq:grey_definition}
  E(\mathbf{x},t) &= \int_{\nu_1}^{\infty} E_\nu(\mathbf{x},t,\nu)\,\mathrm d\nu \nonumber \\
  &=\tilde{E}(\mathbf{x},t) \int_{\nu_1}^{\infty} \chi_E(\nu)\,\mathrm d\nu,
\end{align}
Which separates the dependence of $E$ on coordinate $\mathbf{x}$ and
time $t$ from frequency $\nu$. Here $\chi_E$ is the spectral energy
distribution (SED) taken to be that of a Pop II stellar population
similiar to one from \citep{RicottiEtAl2002}.

\subsection{Star Formation and Feedback}
\label{starformationandfeedback}

Because star formation occurs on scales not resolved by our uniform mesh simulation, 
we rely on a subgrid model which we calibrate to observations of star formation in high
redshift galaxies. The subgrid model is a variant of the \cite{CenOstriker1992}
prescription with two important modifications as described in \cite{SmithEtAl2011}. In the original \cite{CenOstriker1992} recipe, a computational cell forms a collisionless ``star particle" if a number of criteria are met: the baryon density exceeds a certain numerical threshold; the gas velocity divergence is negative, indicating collapse; the local cooling time is less than the dynamical time; and the cell mass exceeds the Jeans mass. In our implementation, the last criterion is removed because it is always met in large scale, fixed-grid simulations, and the overdensity threshold is taken to be $\rho_b/(\rho_{c,0}(1+z)^3) > 100$, where $\rho_{c,0}$ is the critical density at $z=0$. If the three remaining criteria are met, then a star particle representing a large collection of stars is formed in that timestep and grid cell with a total mass

\begin{equation}
m_* = f_* m_{cell} \frac{\Delta t}{t_{dyn}},
\end{equation}
where $f_*$ is an efficiency parameter we adjust to match observations of the cosmic star formation rate density (SFRD) \citep{BouwensEtAl2011}, $m_{cell}$ is the cell baryon mass, $t_{dyn}$ is the dynamical time of the combined baryon and dark matter fluid, and $\Delta t$ is the hydrodynamical timestep. An equivalent amount of mass is removed from the grid cell to maintain mass conservation. 

Although the star particle is formed instantaneously (i.e., within one timestep), the conversion of removed gas into stars is assumed to proceed over a longer timescale, namely $t_{dyn}$, which more accurately reflects the gradual process of star formation. In time $\Delta t$, the amount 
of mass from a star particle converted into newly formed stars is given by

\begin{equation}
\Delta m_{SF} = m_* \frac{\Delta t}{t_{dyn}} \frac{t-t_*}{t_{dyn}} e^{-(t-t_*)/t_{dyn}},
\end{equation}
where $t$ is the current time and $t_*$ is the formation time of the star particle. To make the 
connection with Equation \eqref{eq:cons_momentum}, we have $\dot{\rho}_{SF} =\Delta m_{SF}/(V_{cell}\Delta t)$, 
where $V_{cell}$ is the volume of the grid cell. 

Stellar feedback consists of the injection of thermal energy, gas, and radiation
to the grid, all in proportion to $\Delta m_{SF}$. The thermal energy $\Delta e_{SF}$ and gas
mass $\Delta m_g$ returned to the grid are given by

\begin{equation}
  \Delta e_{SF} = \Delta m_{SF} c^2 \epsilon_{SN}, \qquad
  \Delta m_g = \Delta m_{SF} f_{m*}, 
\end{equation}

where $c$ is the speed of light, $\epsilon_{SN}$ is the supernova energy efficiency parameter, and $f_{m*}=0.25$ is the fraction of the stellar mass returned to the grid as gas. Rather than add
the energy and gas to the cell containing the star particle, as was done in
the original \cite{CenOstriker1992} paper, we distribute it evenly among the cell and its
26 nearest neighbors to prevent overcooling. As shown by \cite{SmithEtAl2011}, this 
results in a star formation recipe which can be tuned to reproduce the observed SFRD. This is critical for us, as we use the observed high redshift SFRD to calibrate our reionization simulations. 

To calculate the radiation feedback, we define an emissivity field $\eta(x)$ on the grid which accumulates
the instantaneous emissivities $\eta_i(t)$ of all the star particles within each cell. To calculate the contribution of each star particle $i$ at time $t$ we assume an equation of the same form for supernova energy feedback, but with a different energy conversion efficiency factor $\epsilon_{UV}$. Therefore

\begin{equation}
\label{eq:emissivity}
  \eta= \sum_\mathrm{i}\epsilon_\mathrm{uv}\frac{\Delta m_\mathrm{SF} c^2}{V_\mathrm{cell}\Delta t}
\end{equation}

Emissivity $\eta$ is in units of erg s$^{-1}$cm$^{-3}$.   The UV efficiency factor $\epsilon_\mathrm{uv}$ is taken from \cite{RicottiEtAl2002} as 4$\pi\times 1.1 \times 10^{-5}$, where the factor $4\pi$ comes from the conversion from mean intensity to radiation energy density.

\subsection{Data Analysis}
\label{DataAnalysis}

Due to the enormous amount of data produced by the simulation (one output file is about 100 GB), 
we needed a scalable tool suited to the task of organizing and manipulating
the data into human readable form.  We use the analysis software tool \texttt{yt} \citep{TurkEtAl2011} specifically created for doing this type of vital task.  It is a python based software tool that does ``Detailed data analysis and visualizations, 
written by working astrophysicists and designed for pragmatic analysis needs."
\texttt{yt} is open source and publicly available at http://yt-project.org.

\section{General Results}
\label{GeneralResults}

%\subsection{General Aspects}
%\label{GeneralAspects}

% talk about the simulation in general terms, describing the parameters,
% features, how it matches observations, then lead into why we need
% a more quantitative language in describing the details of reionization

%In this paper, we analyze a simulation that serves as a test run for an upcoming large-scale %capability run.  As a result, the size of our computational domain is not representative of the %entire universe, but it is a good test case of the code's ability to simulate the desired physics in %a moderate amount of time on moderate computational resources.  We therefore take a physical %box size of 20 Mpc comoving,
%and choose initial conditions and cosmological parameters to match those of WMAP 7 data %\citep{JarosikEtAl2011}.  Although missing the higher power in the matter power spectrum, %which occurs on the scale of $\sim100$ Mpc comoving, we show that all the physics are correct %and behaving as expected.

%\begin{figure*}[ht]
%  \includegraphics[width=0.9\textwidth]{4_panel_HI_slice.png}
% \caption{H I density on slices through the 20 Mpc volume showing the growth, 
%percolation, and final overlap of HII regions. Panels show $z=9, 7.7,
%6.96, 6.3$. The box becomes fully ionized at z=6.13 as the last neutral islands
%are overrun by the I-fronts. Regions of extremely low HI density are shock-heated 
%bubbles due to supernova feedback.}
%  \label{HI_slices}
%\end{figure*}

\begin{figure*}[!tp]
    \begin{minipage}[h]{0.5\linewidth}
        \centering
        \includegraphics[trim = 15mm 5mm 0mm 15mm, clip, width=1.0\textwidth]{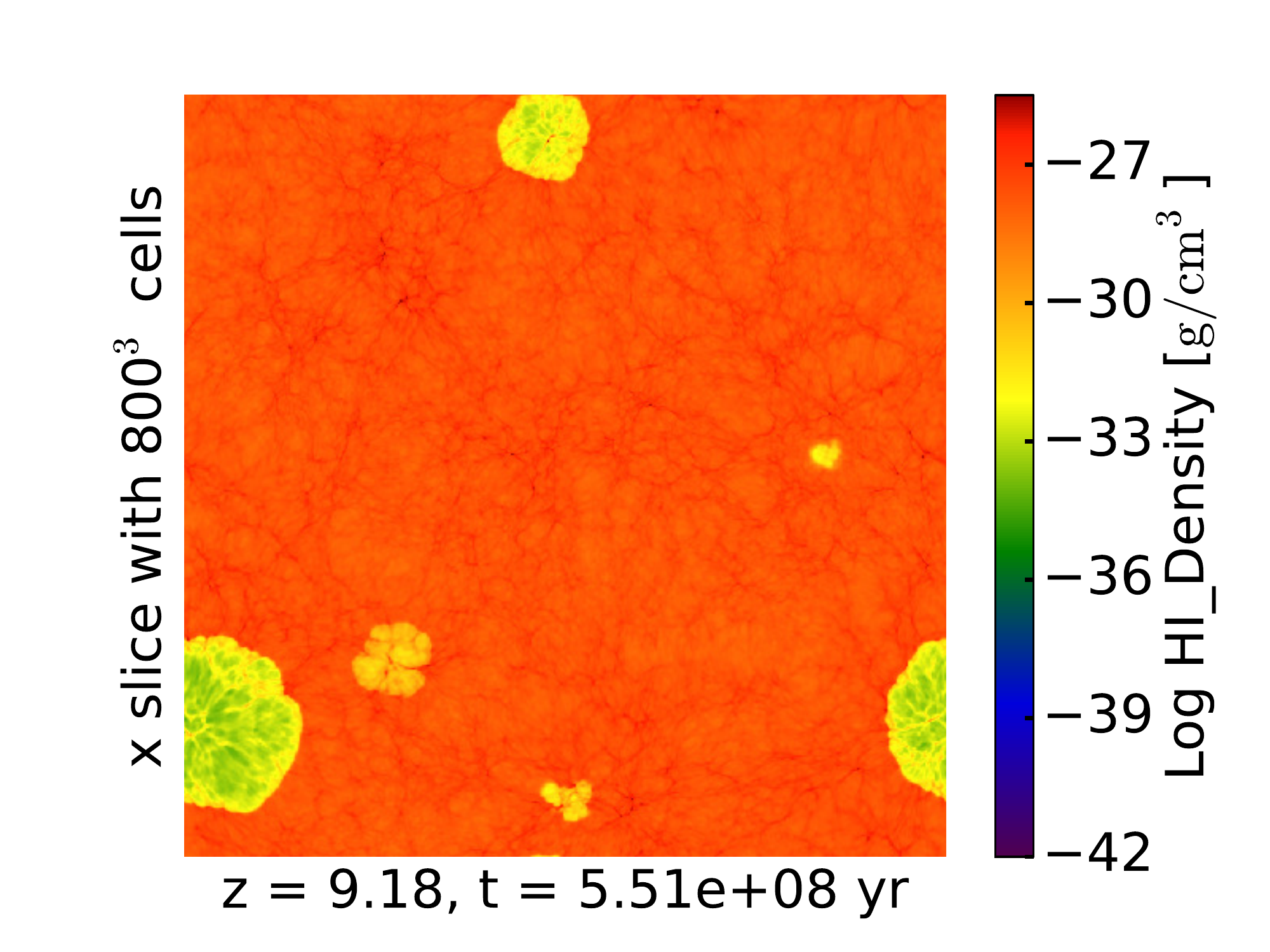}
    \end{minipage}
\hspace*{-4.00mm}
    \begin{minipage}[h]{0.5\linewidth}
        \centering
        \includegraphics[trim = 15mm 5mm 0mm 15mm, clip, width=1.0\textwidth]{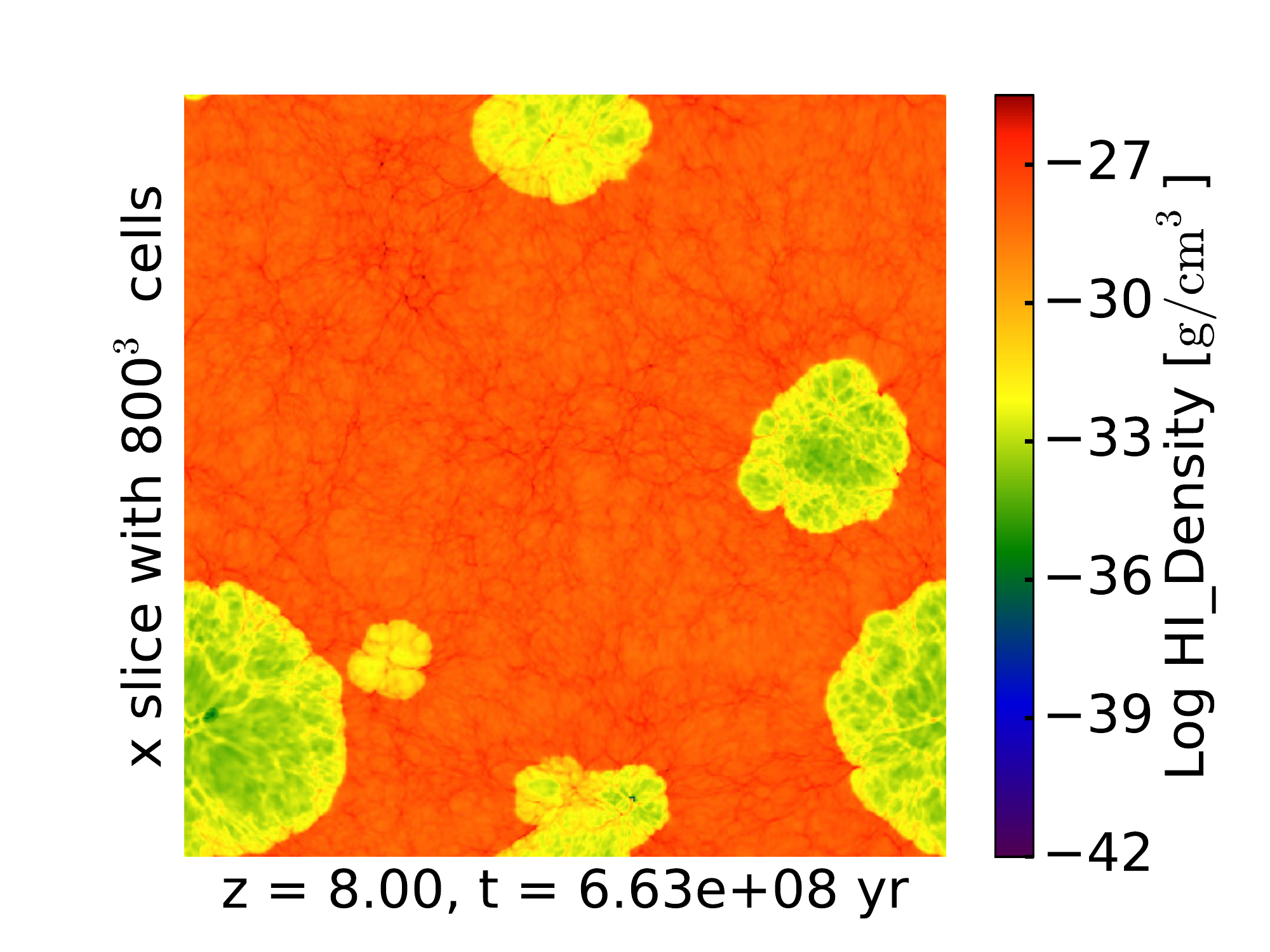}
    \end{minipage}
\\
    \begin{minipage}[h]{0.5\linewidth}
        \centering
        \includegraphics[trim = 15mm 5mm 0mm 15mm, clip, width=1.0\textwidth]{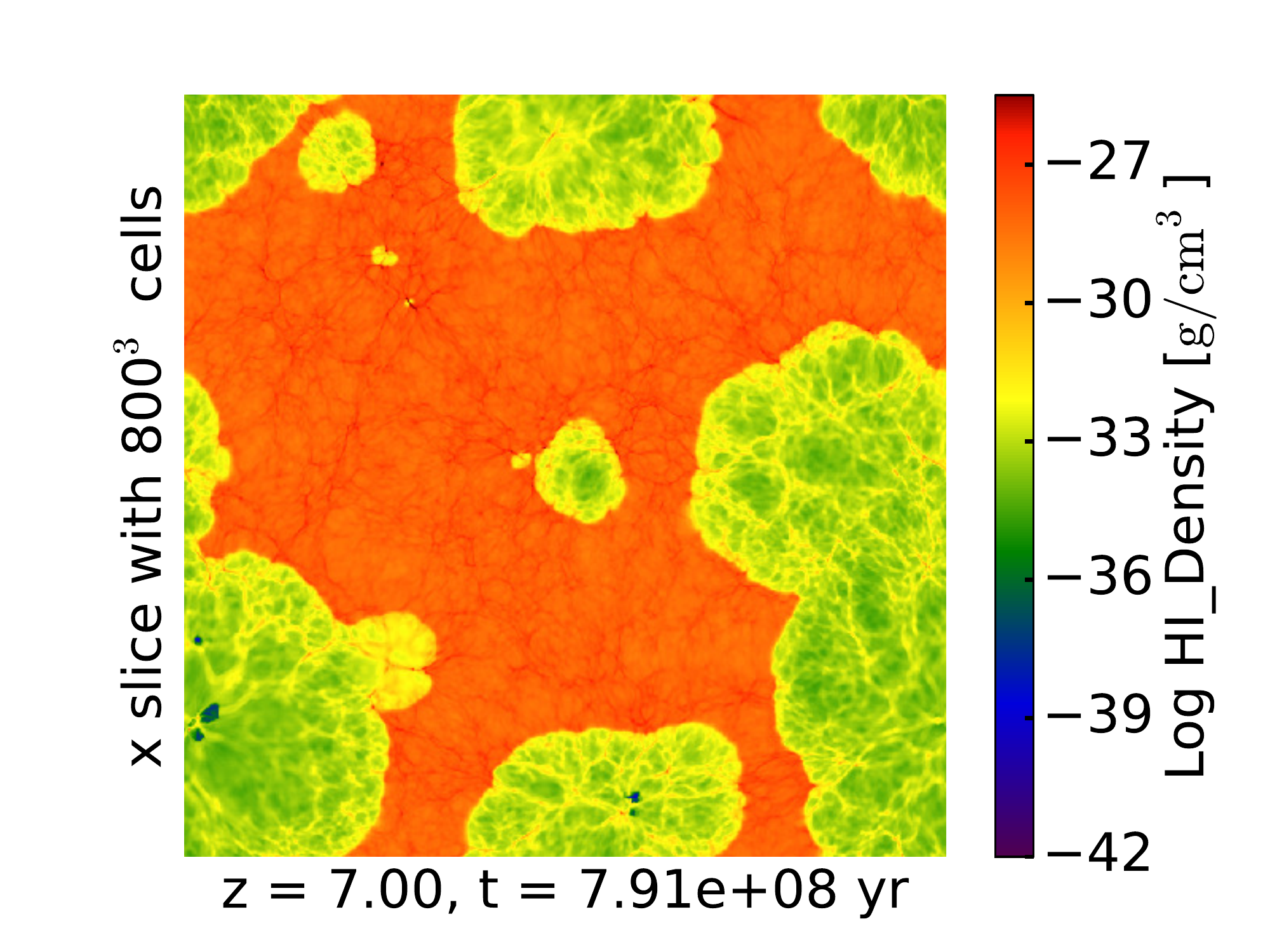}
    \end{minipage}
\hspace*{-4.00mm}
    \begin{minipage}[h]{0.5\linewidth}
        \centering
        \includegraphics[trim = 15mm 5mm 0mm 15mm, clip, width=1.0\textwidth]{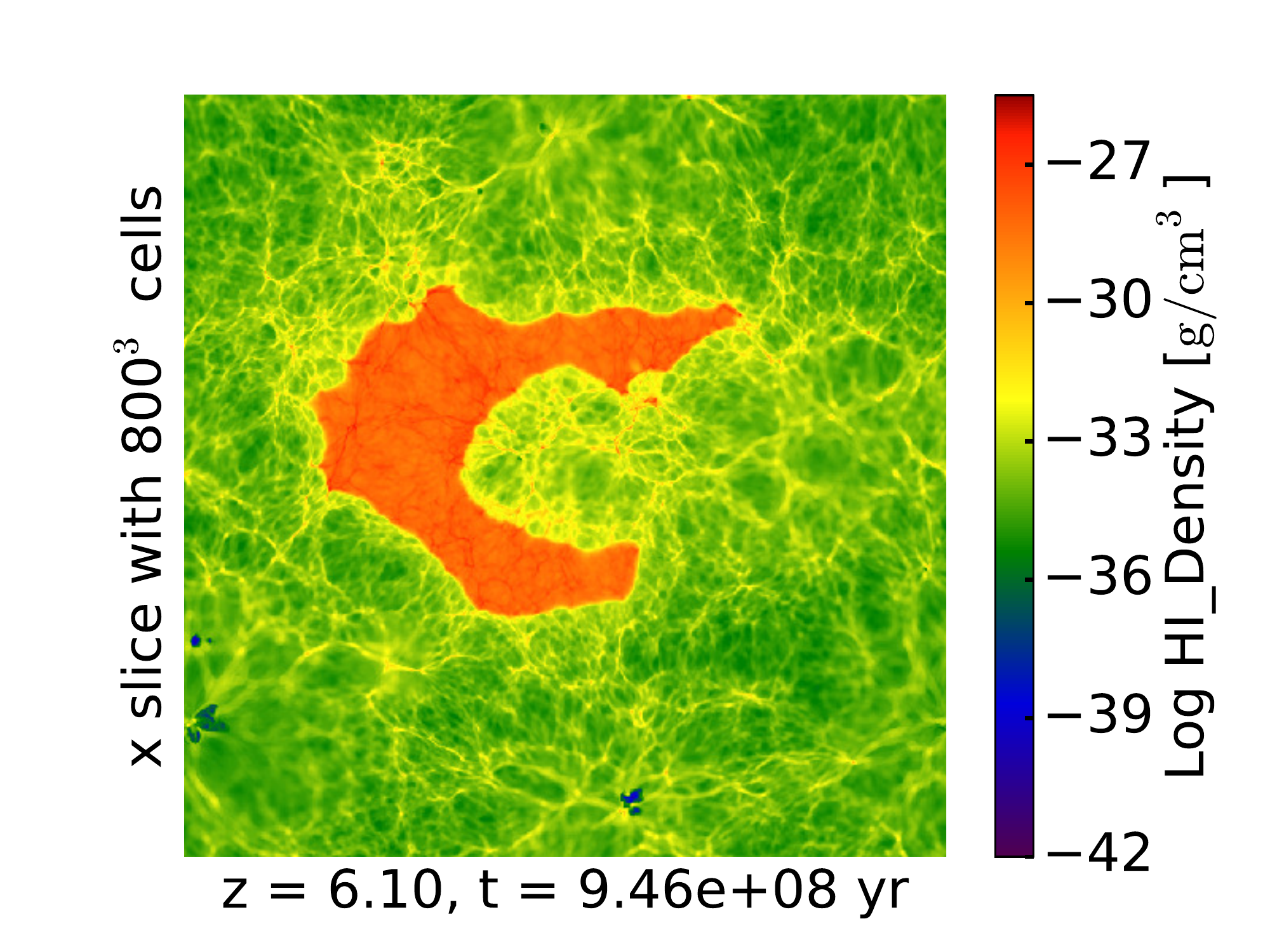}
    \end{minipage}
    \caption{H {\footnotesize I} density on slices through the 20 Mpc volume showing the growth, 
percolation, and final overlap of H II regions. Panels show $z=9.18, 8.0,
7.0, 6.1$. The box becomes fully ionized at $z=5.8$ as the last neutral islands
are overrun by the I-fronts. Regions of extremely low H {\footnotesize I} density are shock-heated 
bubbles due to supernova feedback.}
    \label{HI_slices}
\end{figure*}

Here we first present the basic properties of the simulation before delving into specific topics in subsequent sections. The star formation and feedback parameters for this simulation are $f_* =0.1, f_{m*}=0.25, \epsilon_{SN}=10^{-5}, \epsilon_{UV}=1.38 \times 10^{-4}$. Figure \ref{HI_slices} shows the reionization process as it proceeds through growth, percolation, and final overlap of ionized hydrogen (H {\footnotesize II}) regions driven by ionizing radiation from star forming galaxies. We plot the neutral hydrogen (H {\footnotesize I}) density on a slice through the densest cell in the volume at redshifts $z=9.18, 8.0, 7.0, 6.1$.  At $z=9.18$ several isolated quasi-spherical I-fronts are intersected by the slice plane.  These grow and have begun to merge by $z=8.0$. By $z=7.0$ the toplogy is beginning to invert, in that there are now isolated peninsula of H {\footnotesize I} gas embedded in an otherwise ionized IGM. By $z=6.1$ the remaining neutral island has almost disappeared as it is being irradiated from all sides. We can also see in the figure small patches of extremely low H {\footnotesize I} density; these correspond to bubbles of shock heated gas near galaxies heated to above $10^6$K in temperature by supernova feedback.

\begin{figure}
	\includegraphics[width=0.5\textwidth]{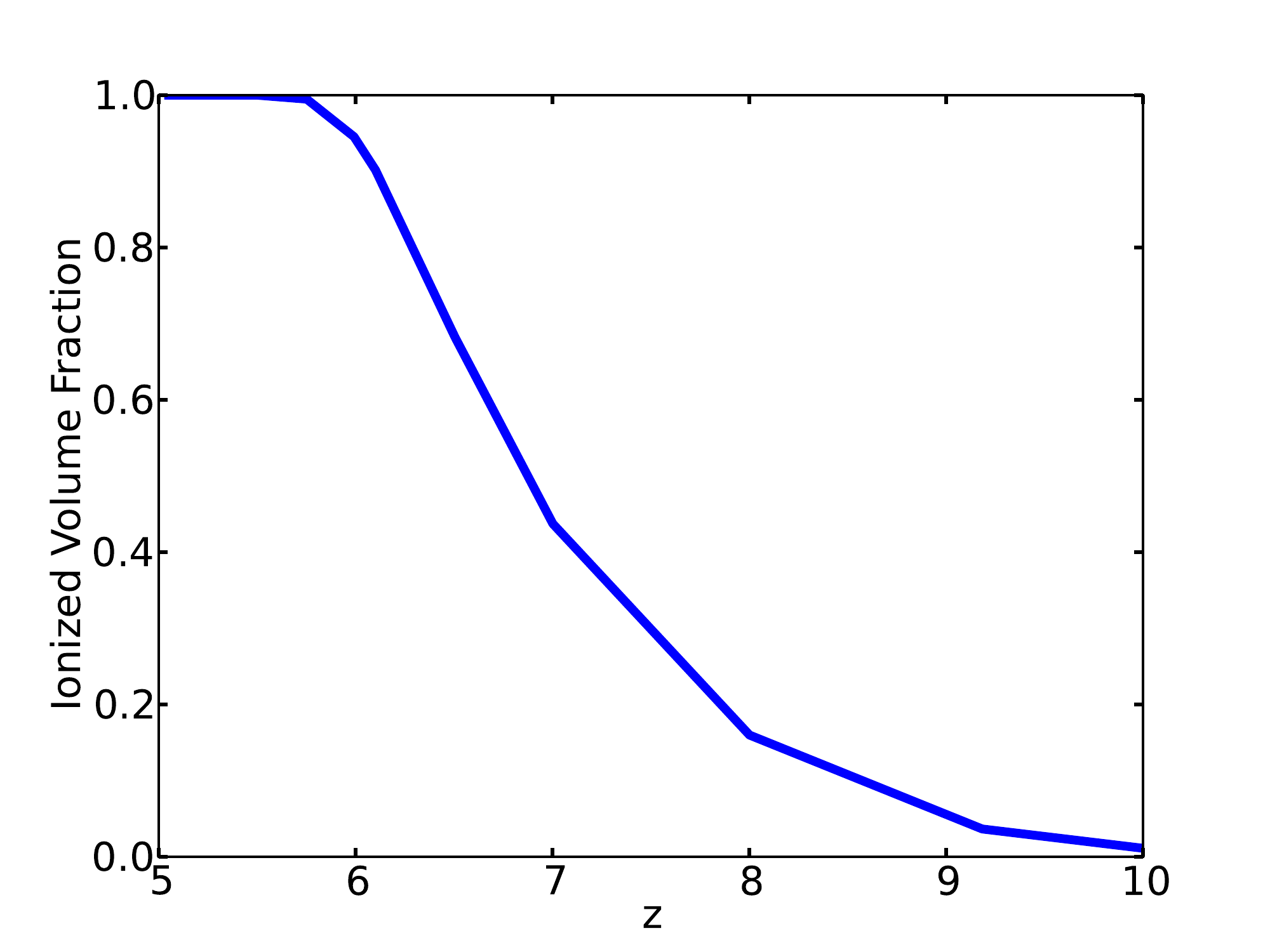}
	\caption{Evolution of the ionized volume fraction versus redshift for hydrogen ionized to less than 1 neutral in 10$^3$ atoms.  As redshift decreases, the volume filling fraction grows rapidly until around redshift of 6, at which time the rate of growth slows significantly as the last neutral island is ionized .  The sensitivity of this curve to ionization level is discussed in \S\ref{QuantitativeLanguage}.}
	\label{Ion1E3}
\end{figure}

Figure \ref{Ion1E3} plots the evolution of the ionized volume fraction $Q_\mathrm{H\,II}$ versus redshift. Here a cell is called ionized if $\rho_\mathrm{H\,II}/\rho_\mathrm{H} \geq 0.999$ (In \S\ref{QuantitativeLanguage} we discuss the sensitivity of this curve to level of ionization.) The first ionizing sources turn on at $z \sim 10$ in this simulation. The ionized volume fraction rises rapidly, reaching 0.5 at $z \approx 6.8$, 0.95 at $z \approx 6.0$, and near unity at $z \approx 5.8$. We compare this evolution with the predictions of the simple analytic model introduced by \cite{MadauEtAl1999} in \S\ref{Qdot}. For now we only draw attention to the flattening of the curve in the redshift interval $5.8 \leq z \leq 6$. This is the signature of neutral islands being ionized by I-fronts converging in 3D, as opposed to being ionized by internal sources. 

\begin{figure}
	\includegraphics[width=0.5\textwidth]{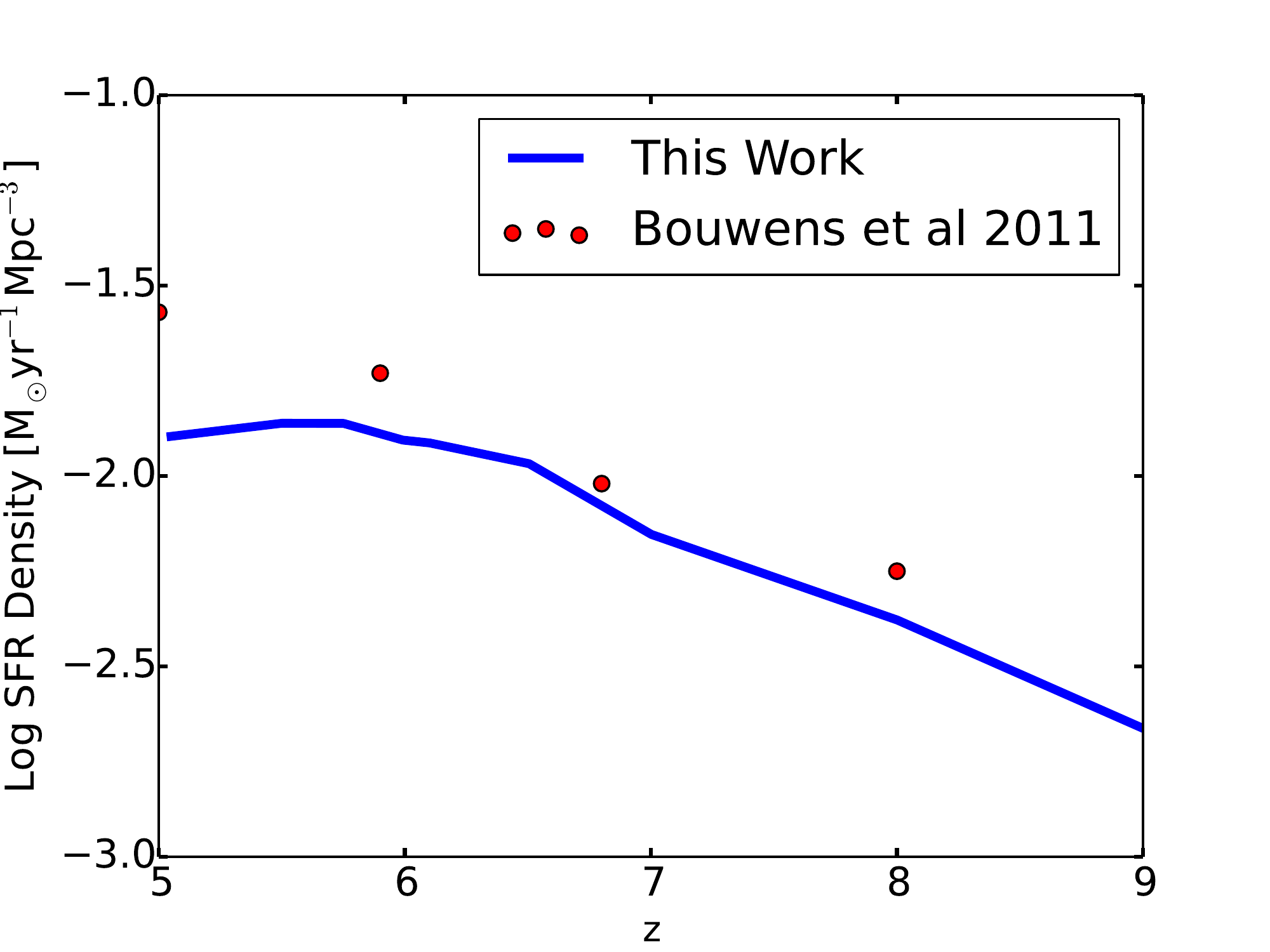}
	\caption{A comparison of simulated and observed star formation rate densities (SFRD) in units of M$_\odot$yr$^{-1}$Mpc$^{-3}$ comoving.  Blue curve labeled ``This Work'' is from our 20 Mpc / $800^3$ simulation, and ``Bouwens et al 2011'' are observationally derived data points from \cite{BouwensEtAl2011b} plotted without error bars. The leveling off of the simulated SFRD is an artifact of the small volume as a simulation carried out with identical physics, mass, and spatial resolution but in 64 times the volume does not show this effect.}
	\label{SFR}
\end{figure}

Our simulation was not designed to complete reionization by a certain fiducial redshift. Rather we adjusted our star formation efficiency parameter $f_*$ so that we can approximately match the star formation rate density (SFRD) in \citep{BouwensEtAl2011}.  Our SFRD is shown in Figure \ref{SFR}, along with the Bouwens data, plotted without error bars. For reference we also include the fitting function described in \citep{HaardtMadau2012}.  This shows that our simulated universe is one that produces approximately the same amount of stars in a given comoving volume, albeit a bit low relative to the data. We also note that the SFRD begins to flatten out at $z \approx 6.5$, and even turns over after overlap at $z \approx 5.8$, rather than continue to rise as indicated by the data points.  This is an artifact of the small box size as a simulation completed in a 80 Mpc comoving on a side box with identical physics, mass, and spatial resolution and star formation/feedback parameters does not show this slowing down of the SFRD. This will be reported on in a future paper. 

%Another general feature of the simulation is how fast the universe is reionized.  We look at a %specific ionization level (less than 1 neutral hydrogen in 10$^3$ hydrogen), and plot its volume %filling fraction versus redshift.  We shall discuss in depth what happens when we look at a range %of different ionization levels in \S\ref{QuantitativeLanguage}, but for now, consider the cell in %the simulation as ``completely'' ionized when it reaches this ionization state.  Figure %\ref{Ion1E3} shows in linear scale that the universe has slightly more than 10\% of its volume %completely ionized at a redshift of 8, and by a redshift of 6, more than 90\% of the volume is %completely ionized.

%One should note that the rate at which the volume filling fraction curve rises with respect to %decreasing redshift slows down significantly near redshift of 6.  This is an artifact of the way %radiation and ionization front is transported.  As the last region of space that is left mostly %neutral, it faces incoming ionization radiation waves from all sides.  Since the waves are %converging in 3 dimensions, the amount of H {\footnotesize I} volume that is swept through %with each unit of time gets smaller and smaller.  Although the propagation of the wave fronts %may not be exactly constant and the x axis of redshift is not exactly the same as time, the %analogy holds true.

\begin{figure}
	\includegraphics[width=0.5\textwidth]{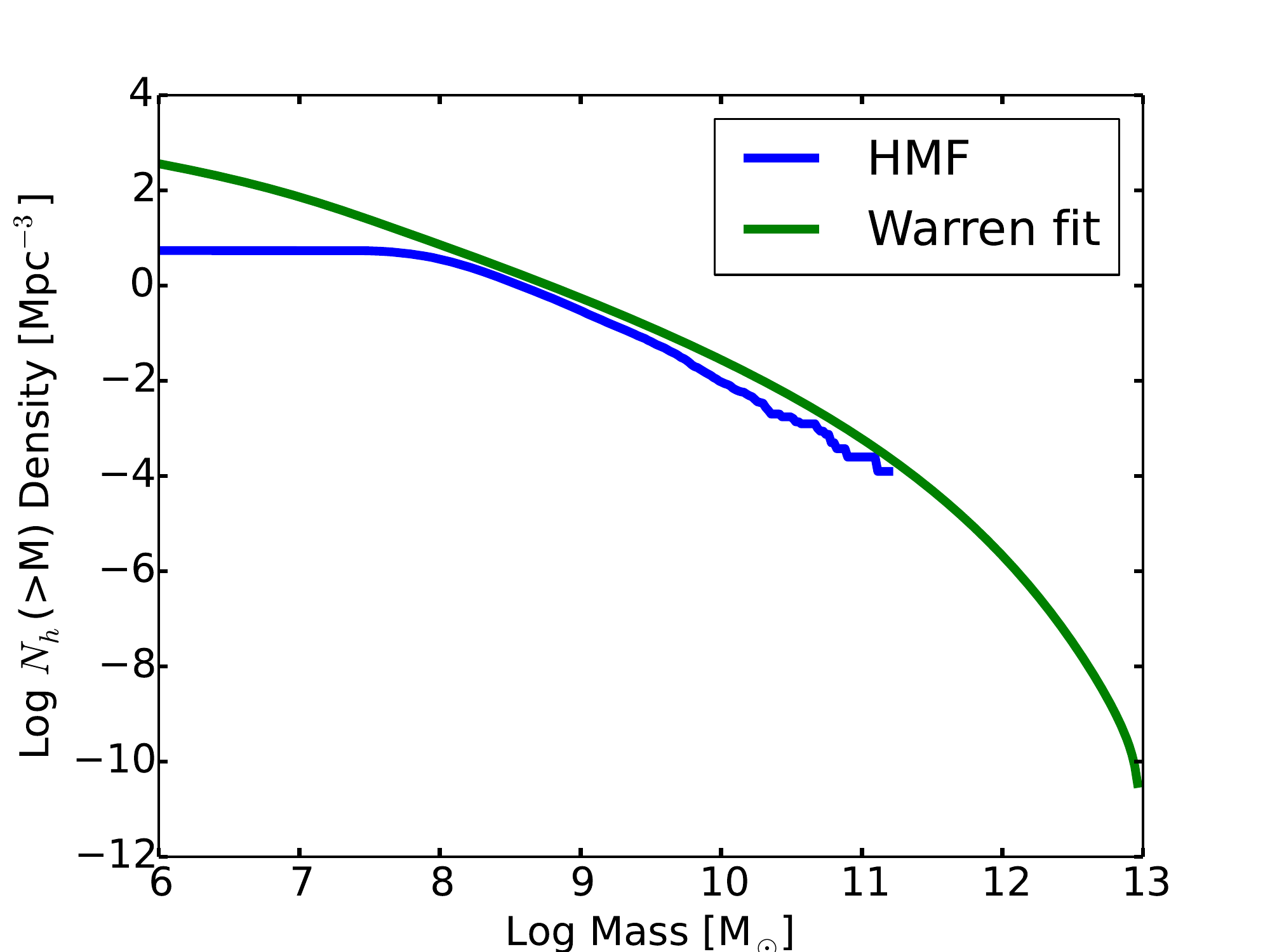}
	\caption{The dark matter halo mass function from our simulation (blue line).  Green line is the fit from \citep{WarrenEtAl2006}. Our low-mass HMF is reasonably complete down to $M_{halo} \approx 10^8 M_{\odot}$; i.e. halos believed to form stars efficiently due to atomic line cooling. Incompleteness at the high mass end is due to the limited volume sampled.}
	\label{HMF}
\end{figure}

To check and make sure that our simulation is giving us a fair representation of the universe, we plot several more quantities and look for any anomalies.  In Figure \ref{HMF}, we see that our halo mass function at redshift of $z \sim 6$ matches well with the Warren fit implemented in \texttt{yt} \citep{WarrenEtAl2006,TurkEtAl2011}.  The mass function captures haloes down to $\sim$10$^8$M$_\odot$, which as previously stated was a simulation design criterion.  The haloes are found by first running the parallelHOP halo finder installed in \texttt{yt} \citep{SkoryEtAl2010}, then taking the linked list of dark matter particles for each halo and wrapping the region around them in an ellipsoidal 3D container introduced in \texttt{yt} 2.4.  The 3D container enables the query of the fluid quantities of the haloes, such as baryonic, emissivity, radiation contents in addition to the particle information.  Since the dark matter particles used are $\sim 5 \times$ 10$^5$M$_\odot$, the 10$^8$M$_\odot$ dark matter haloes are considered to be resolved \citep{TrentiEtAl2010}.

\begin{figure}
	\includegraphics[width=0.5\textwidth]{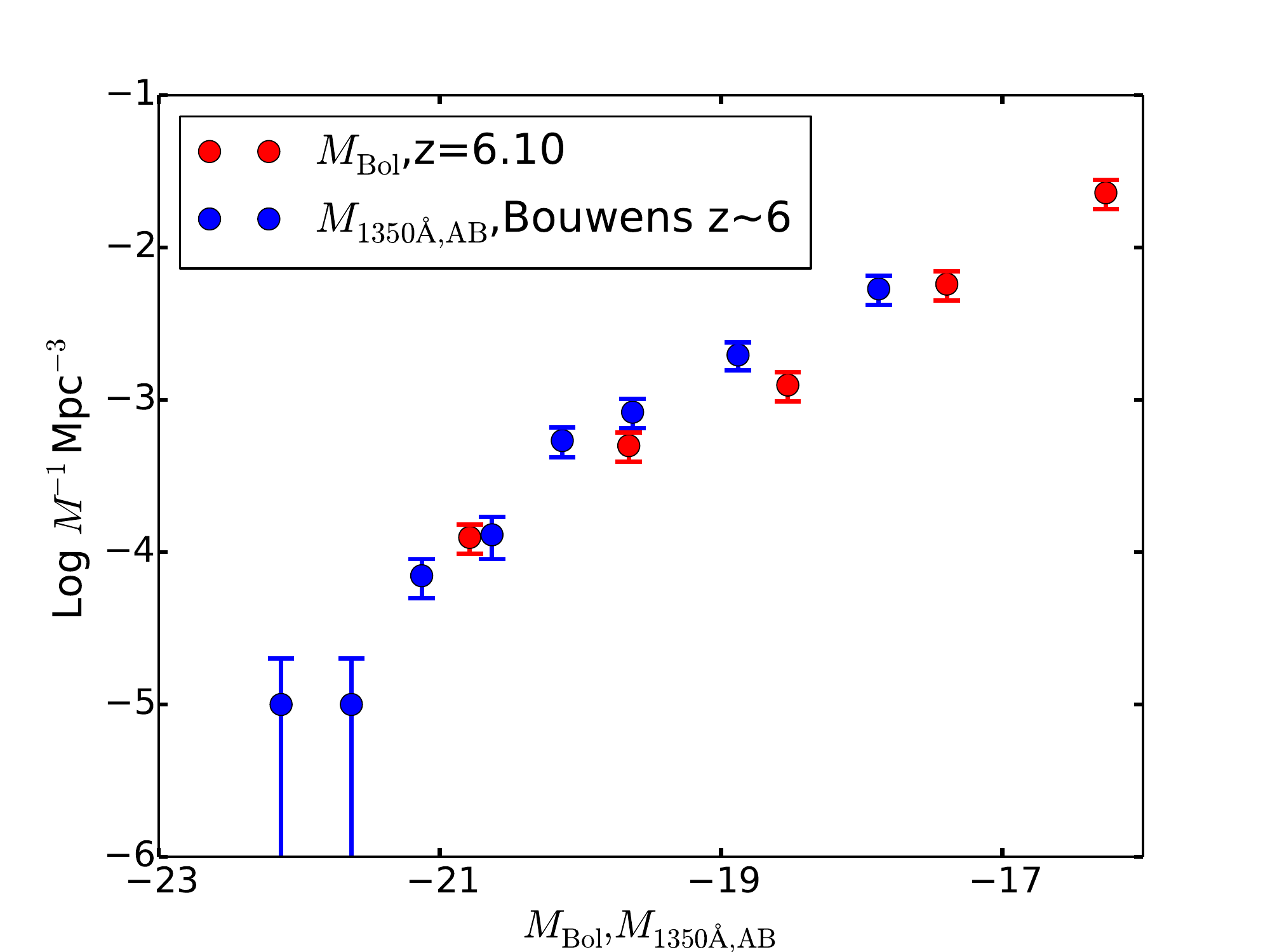}
	\caption{Bolometric luminosity function derived from our simulation data (red), compared with observational data points (blue) from \citep{BouwensEtAl2007}.}
	\label{scaledLF}
\end{figure}

As a final check that our ionizing source population is not wildly unrepresentative of the observed universe, in Figure \ref{scaledLF} we plot the luminosity function of our simulated galaxies at $z=6.1$ along side the observational data points from Table 5 of \citep{BouwensEtAl2007}. The points in red are the bolometric luminosities for our galaxy population calculated directly from the $z=6.1$ halo catalogue.  To calculate the luminosity of a given halo we sum the emissivity field within the 3D ellipsoidal containers defined by the halos' dark matter particles.  Our error bars are taken using one standard deviation of luminosity in the mass bins.  Although this is not proof that our simulation is matching observations exactly, it does lend support that our realization of reionization is being driven by sources not too dissimilar to those observed and is sufficient for the purposes of this study. 

\subsection{Quantitative Language}
\label{QuantitativeLanguage}

\begin{figure}
	\includegraphics[width=0.5\textwidth]{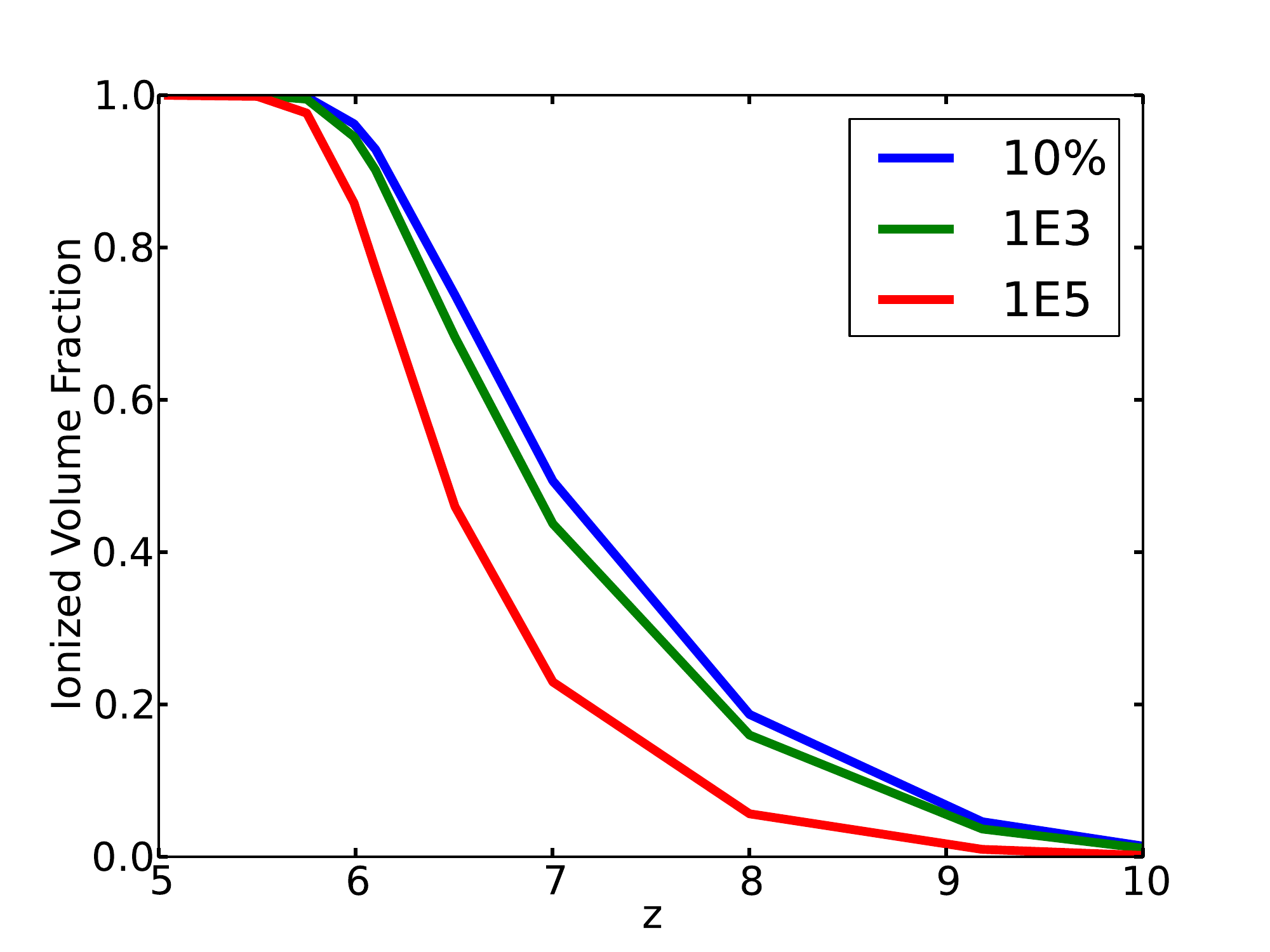}
	\includegraphics[width=0.5\textwidth]{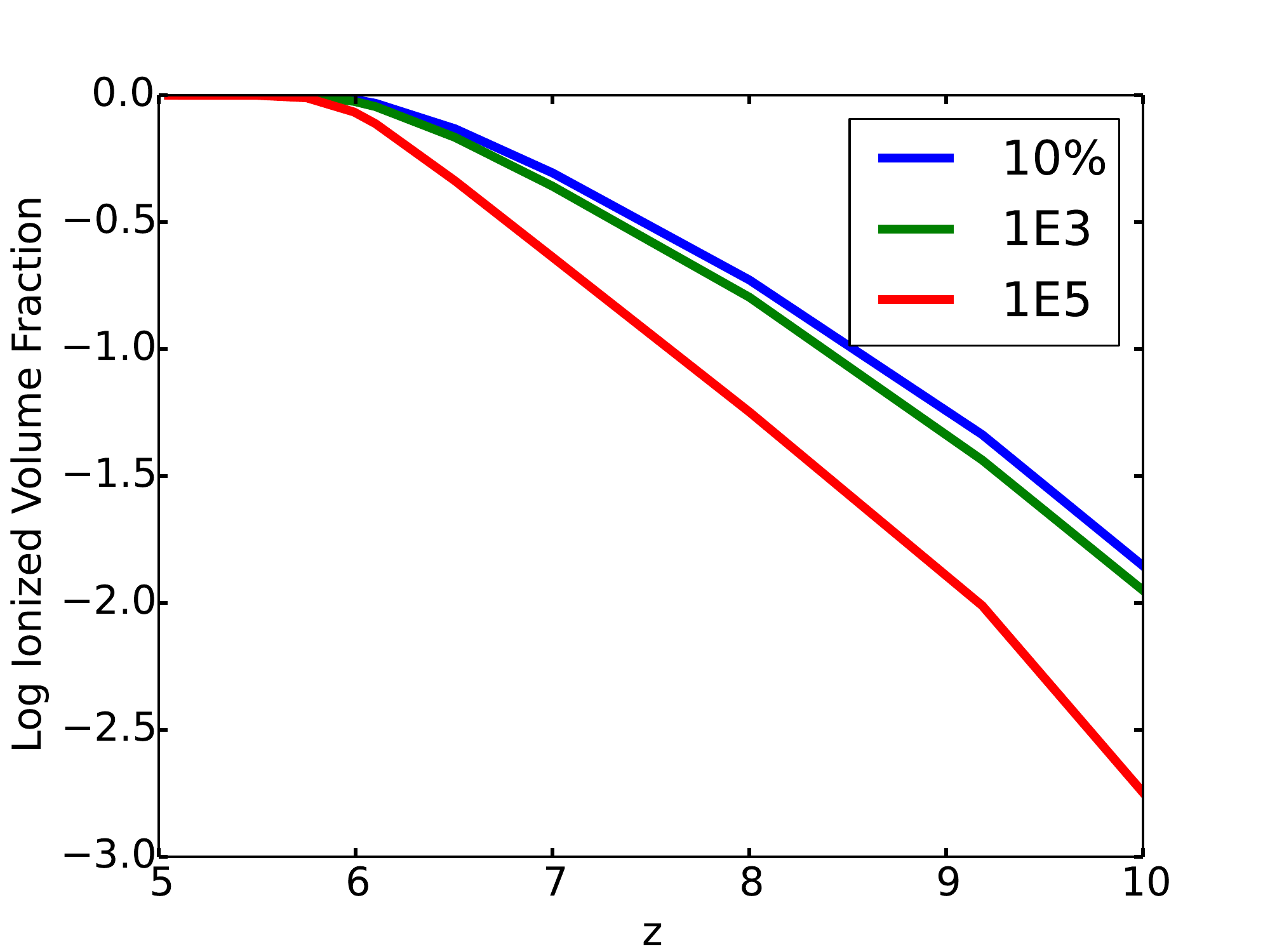}
	\caption{Volume filling fraction of ionized gas versus redshift for three ionized fraction thresholds. {\em Top} linscale; {\em Bottom} logscale. The three ionization levels are ``10\%'' in blue: fractional volume that have more than 1 ionized hydrogen atom per 10 hydrogen atoms.  ``1E3'' in green: fractional volume that have less than 1 neutral hydrogen atom per 10$^3$ hydrogen atoms.  ``1E5'' in red:  fractional volume that have less than 1neutral hydrogen atom per 10$^5$ hydrogen atoms.}
	\label{linearIonized}
\end{figure}

Earlier works on reionization such as \cite{ValageasSilk1999,Gnedin2000,MiraldaEscudeEtAl2000,IlievEtAl2006} speak of a two phase medium composed of completely neutral and completely ionized hydrogen gas, while more recent works \citep{CiardiEtAl2003,ZahnEtAl2007,ShinEtAl2008,PetkovaSpringel2011a,FinlatorEtAl2012} begin to consider the {\em degree of ionization} within ionized gas.  The simplification of considering a two phase medium helps reduce the simulation complexity and the language needed to describe the results.  However, as simulations become more sophisticated, the two phase paradigm becomes ill-suited to convey the wealth of information contained in the larger and more detailed simulations.  
As people begin to describe the new simulations, the old paradigm lingers and causes ambiguities. As a case in point, consider the ionized volume filling fraction versus redshift, one of the simplest quantitative metrics of any reionization simulation. Within the framework of a two-phase medium, this is uniquely defined at any redshift. For a simulation such as ours which tracks the ionization state in every cell, the volume filling fraction depends on the degree of ionization, as illustrated in Figure \ref{linearIonized}. 

%From the earliest N-body simulations to simulations with baryonic fluids to different ionization states of these quantities, simulations have made %big strides in terms of complexity.  If the language we use to describe the ionization state stagnates at distinguising only between two phases %using a single threshold, then we neglect the spectrum of information from the simulation.  This is why we propose that the community come up %with a more unified and quantitative way to describe the simulations.  For that end, we will try to raise two areas where this is apparent.

%As alluded to earlier in \S\ref{GeneralAspects}, we now investigate the ionization volume filling fraction more thoroughly.  When we plot other %ionization levels and their volume filling fraction versus redshift, the result is Figure \ref{linearIonized}.  
This figure shows the evolution of the volume filling fraction of ionized gas which exceeds a minimum local ionization fraction $f_i \equiv \rho_\mathrm{H\,II}/\rho_\mathrm{H}$. The three thresholds are $f_i=$ 0.1, 0.999, and 0.99999 and are labelled 10\%, 1E3, 1E5, respectively in Figure \ref{linearIonized}. 
We choose three specific levels not because we think they are more important than others, but because it suits our later narrative and gives a range values.  With the ionization state tracked by the simulation, we see that it is now ambiguous to ask at what redshift 50\% of the volume is ionized. In our simulation this occurs at $z \approx$ 7, 6.8 and 6.5 for $f_i$=0.1, 0.999, and 0.99999, respectively.

%We choose these numbers to demonstrate what can be done, but the values themselves are in no way set in stone.  The figure shows 10\% %ionization (having more than 1 ionized hydrogen per 10 hydrogen atoms), 1E3 (having less than 1 neutral hydrogen per 10$^3$ hydrogen atoms), %and 1E5 (having less than 1 neutral hydrogen per 10$^5$ hydrogen atoms), their volume filling fraction versus redshift.  In Figure %\ref{LogIonized}, the same information is plotted with the y-axis in Log$_{10}$ scale.

%\begin{figure}
% 	\includegraphics[width=0.5\textwidth]{Log_Ionized_vs_Redshift.eps}
%	\caption{Plot of the same information as Figure \ref{linearIonized}, however, this time theh y-axis is in Log$_{10}$ scale}
%	\label{LogIonized}
%\end{figure}

%One key element is that one should always talk about the level of ionization, and not throw around terms like ``ionized'' lightly, unless %predefined.  For example, we can tell when 10\% of the universe's volume is at a ionization level of 10\%.  This occurs around a redshift of 8.7 %(estimates are from simple linear interpolation of the available data output intervals).  Studying another curve, we see that 10\% of the universe's %volume is ionized to 1E5 around a redshift of 7.7.  

In the rest of this paper we will often report results as a function of these three ionization fraction thresholds. 
To make the text easier to read we will use the terms ``Ionized'' to designate $f_i$=0.1, ``Well Ionized'' to designate $f_i$=0.999, and ``Fully Ionized'' to designate $f_i$=0.99999 ionization levels.
%This graph ultimately shows that more of the universe becomes more highly ionized as redshift decreases.  Also, it is harder to make claims about %the ``entire'' universe having reached a specific ionization level, because there's often regions that may not have reached a specific level.

\subsection{Inside-out or Outside-in}
\label{IOOI}

Besides specifying the amount of ionized volume and levels of ionization, another area where quantitative language is useful is in the description of the reionization history.  Since the Outside-in model was proposed by \cite{MiraldaEscudeEtAl2000}, there is gathering support for the opposing view of the Inside-out model by \cite{SokasianEtAl2003,FurlanettoEtAl2004,IlievEtAl2006} to name a few.  In \cite{FinlatorEtAl2009}, the authors go even further and add to the lexicon ``Inside-outside-middle'', trying to describe the rich detail in a reionization scenario. The basic Inside-out picture is that galaxies form in the peaks of the dark matter density field and drive expanding H {\footnotesize II} regions into their surroundings ({\em expansion phase}). 
%by virtue of the UV radiation emitted from young, massive stars. 
These H {\footnotesize II} regions are initially isolated, but begin to merge into larger, Mpc-scale H {\footnotesize II} regions due to the clustering of the galaxy distribution ({\em percolation phase}). Driven by a steadily increasing global star formation rate and recombination time (due to cosmic expansion) this process goes on until H {\footnotesize II} regions completely fill the volume ({\em overlap phase}). In this picture, rare peaks in the density field ionize first while regions of lower density ionize later from local sources that themselves formed later.

\begin{figure*}[!tp]
	\begin{minipage}[h]{0.33\linewidth}
	\centering
	\includegraphics[trim = 7mm 9mm 1mm 7mm, clip, width=1.0\textwidth]{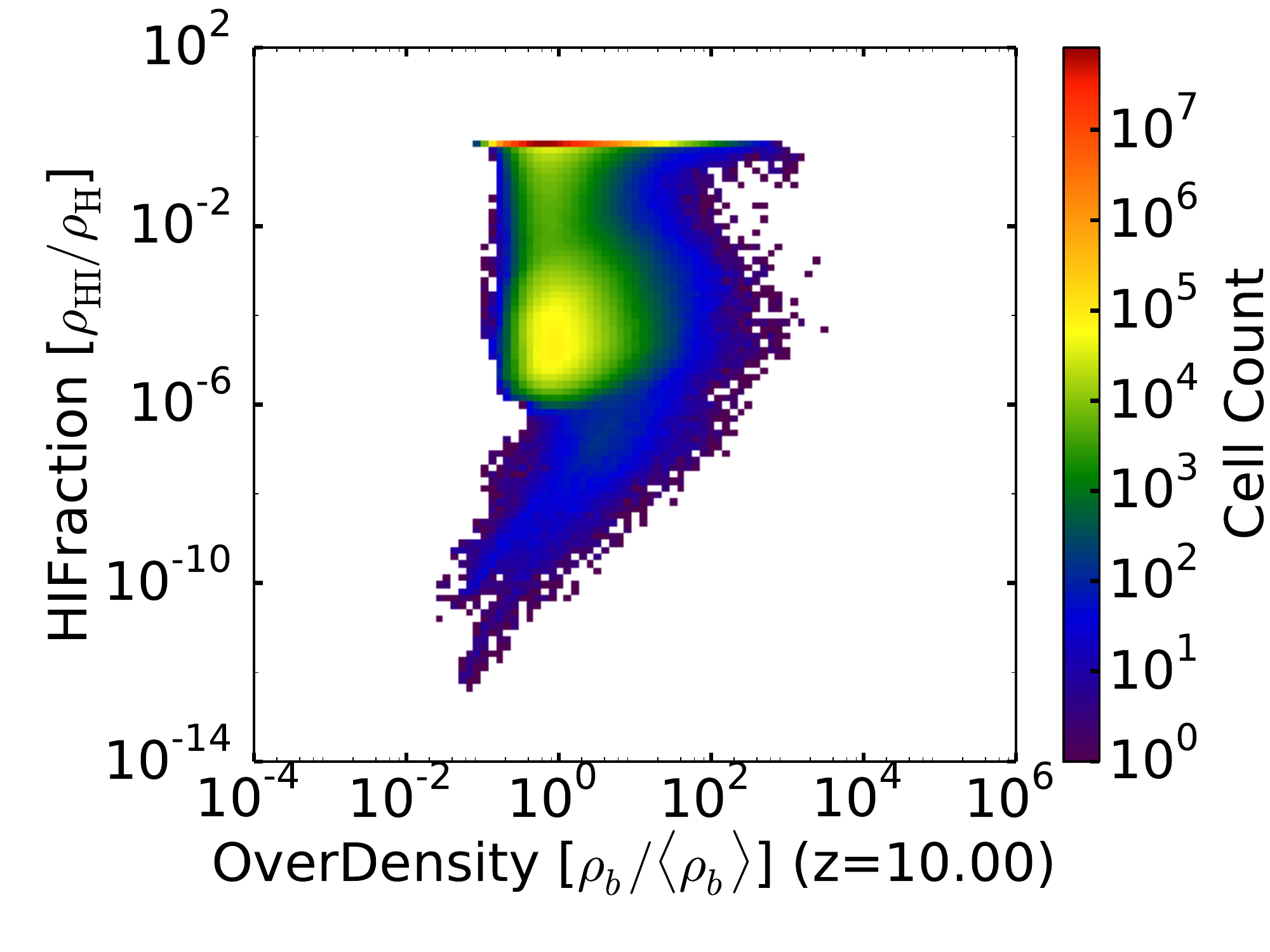}

	\end{minipage}
\hspace*{-2.00mm}
	\begin{minipage}[h]{0.33\linewidth}
	\centering
	\includegraphics[trim = 10mm 0mm 7mm 7mm, clip, width=1.0\textwidth]{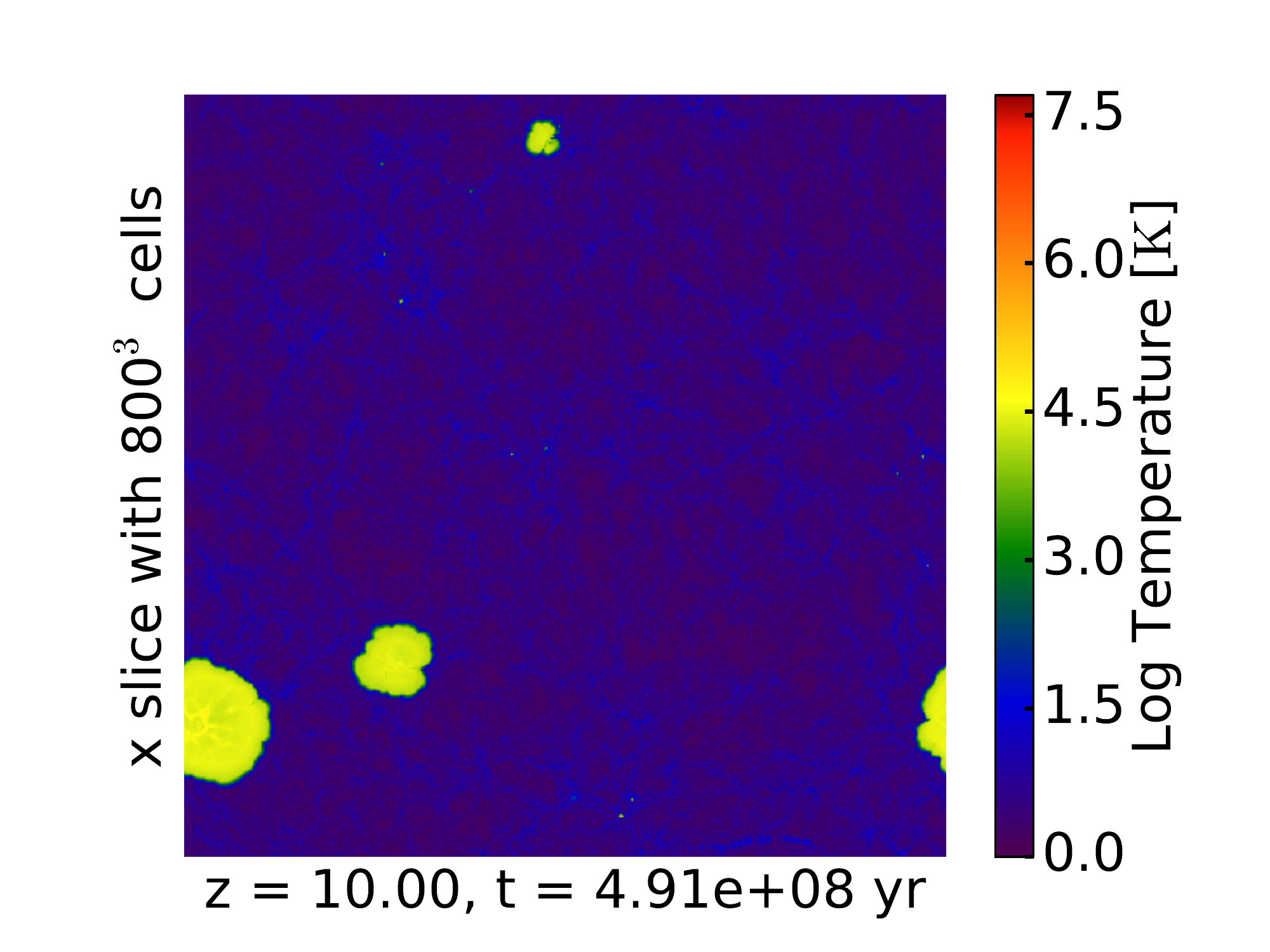}
	\end{minipage}
\hspace*{-2.00mm}
	\begin{minipage}[h]{0.33\linewidth}
	\centering
	\includegraphics[trim = 10mm 0mm 7mm 7mm, clip, width=1.0\textwidth]{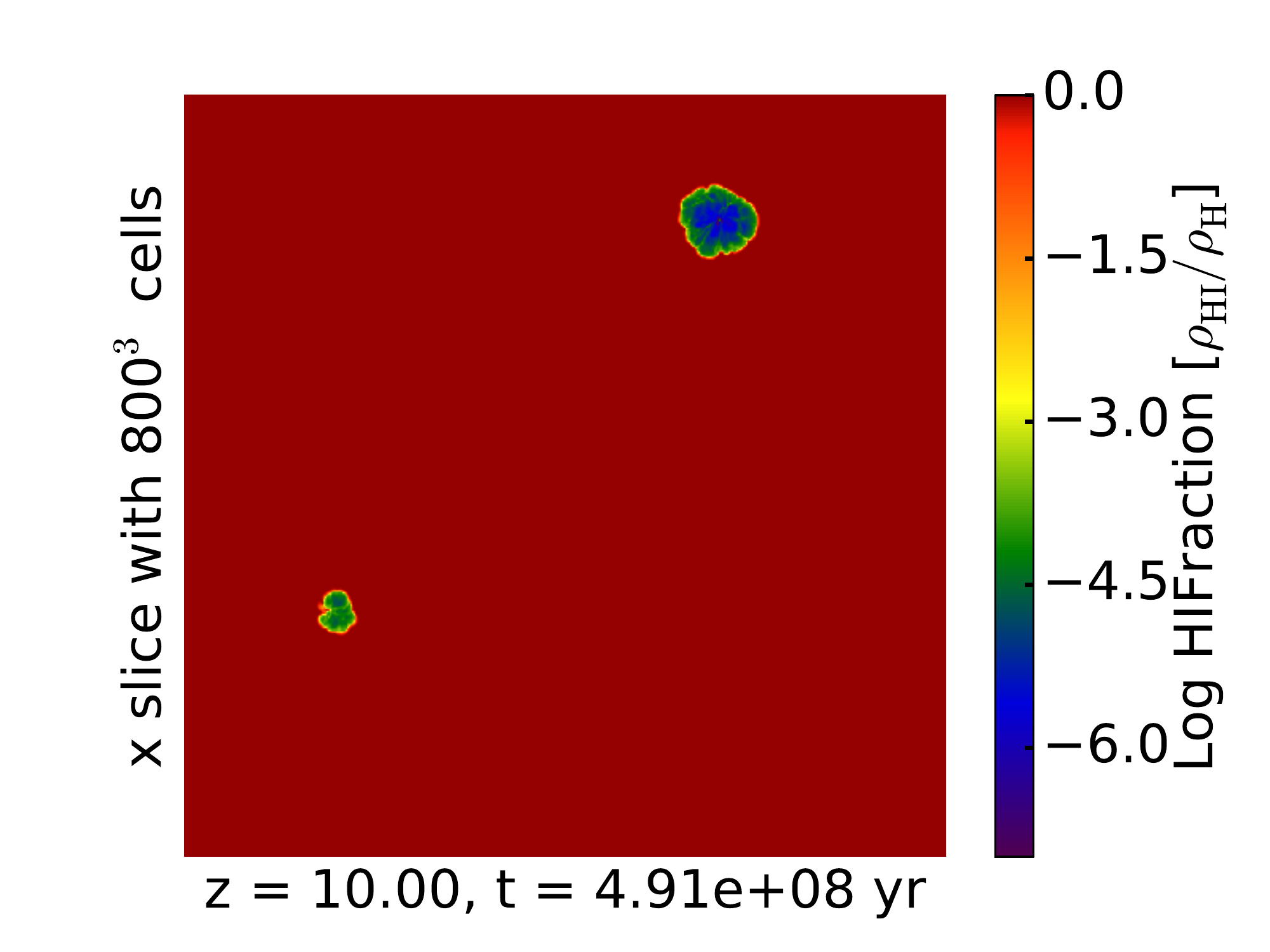}
	\end{minipage}fi
\vspace*{-2.00mm}\\
	\begin{minipage}[h]{0.33\linewidth}
	\centering
	\includegraphics[trim = 7mm 9mm 1mm 7mm, clip, width=1.0\textwidth]{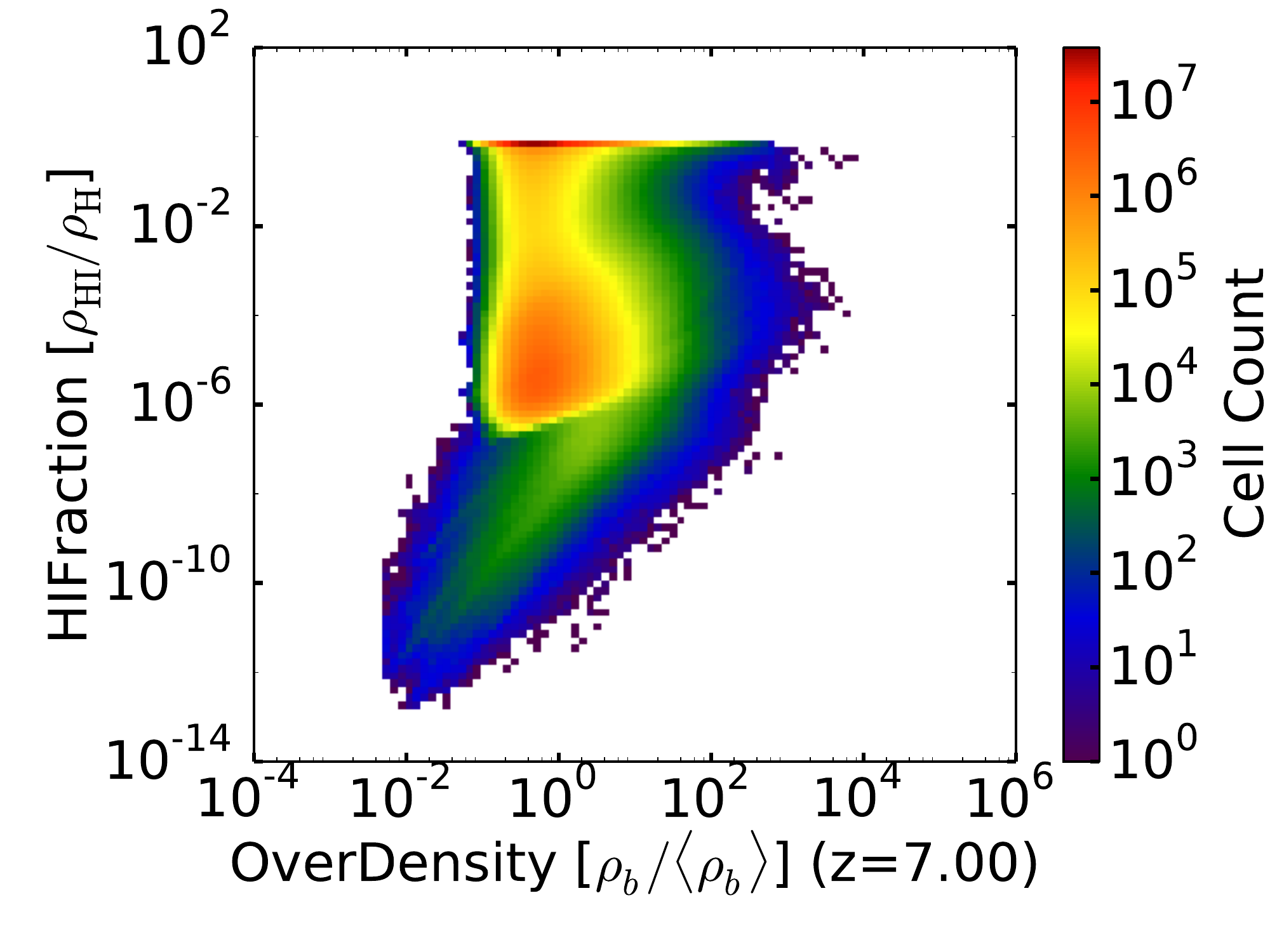}
	\end{minipage}
\hspace*{-2.00mm}
	\begin{minipage}[h]{0.33\linewidth}
	\centering
	\includegraphics[trim = 10mm 0mm 7mm 7mm, clip, width=1.0\textwidth]{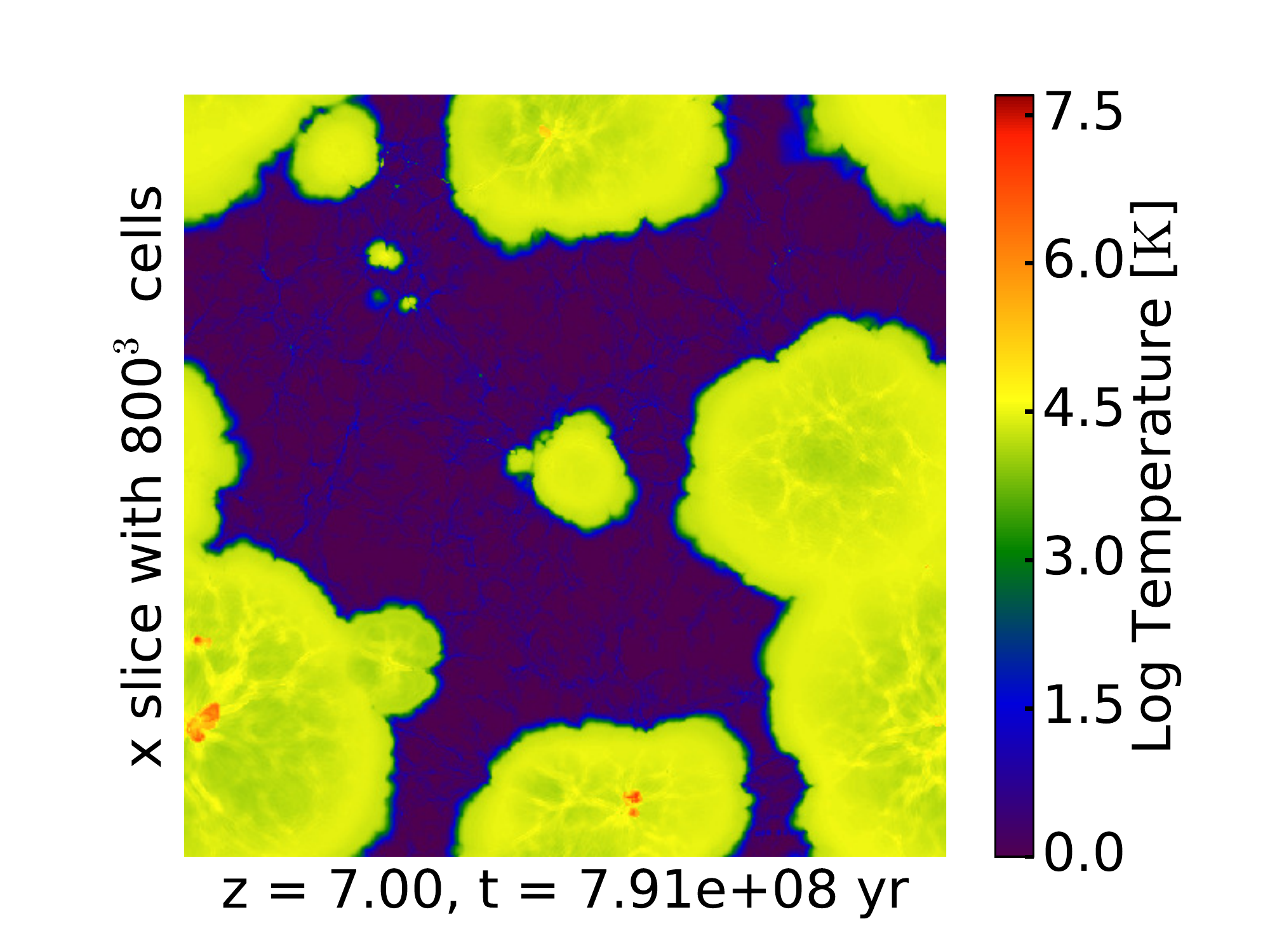}
	\end{minipage}
\hspace*{-2.00mm}
	\begin{minipage}[h]{0.33\linewidth}
	\centering
	\includegraphics[trim = 10mm 0mm 7mm 7mm, clip, width=1.0\textwidth]{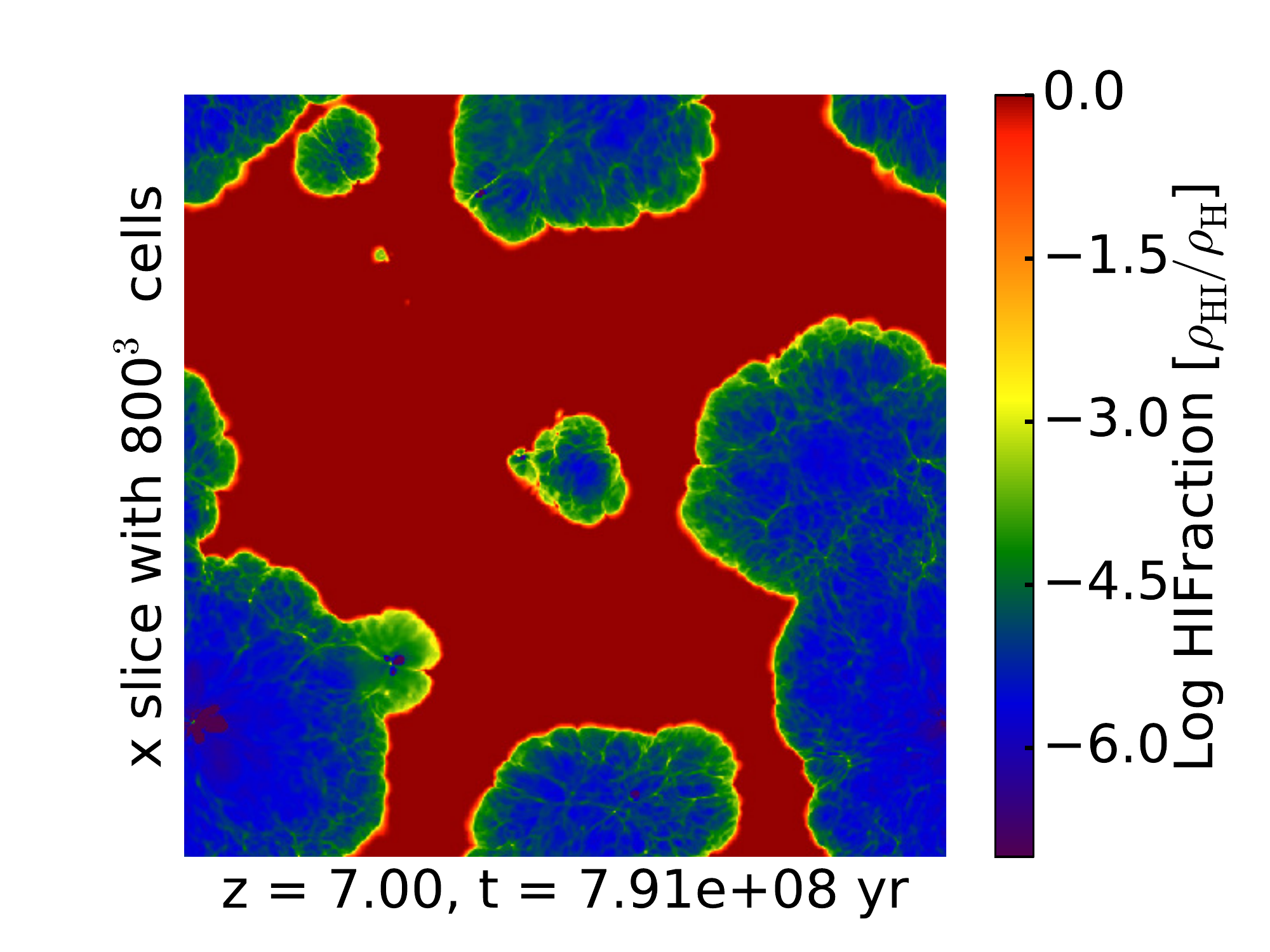}
	\end{minipage}
\vspace*{-2.00mm}\\
	\begin{minipage}[h]{0.33\linewidth}
	\centering
	\includegraphics[trim = 7mm 9mm 1mm 7mm, clip, width=1.0\textwidth]{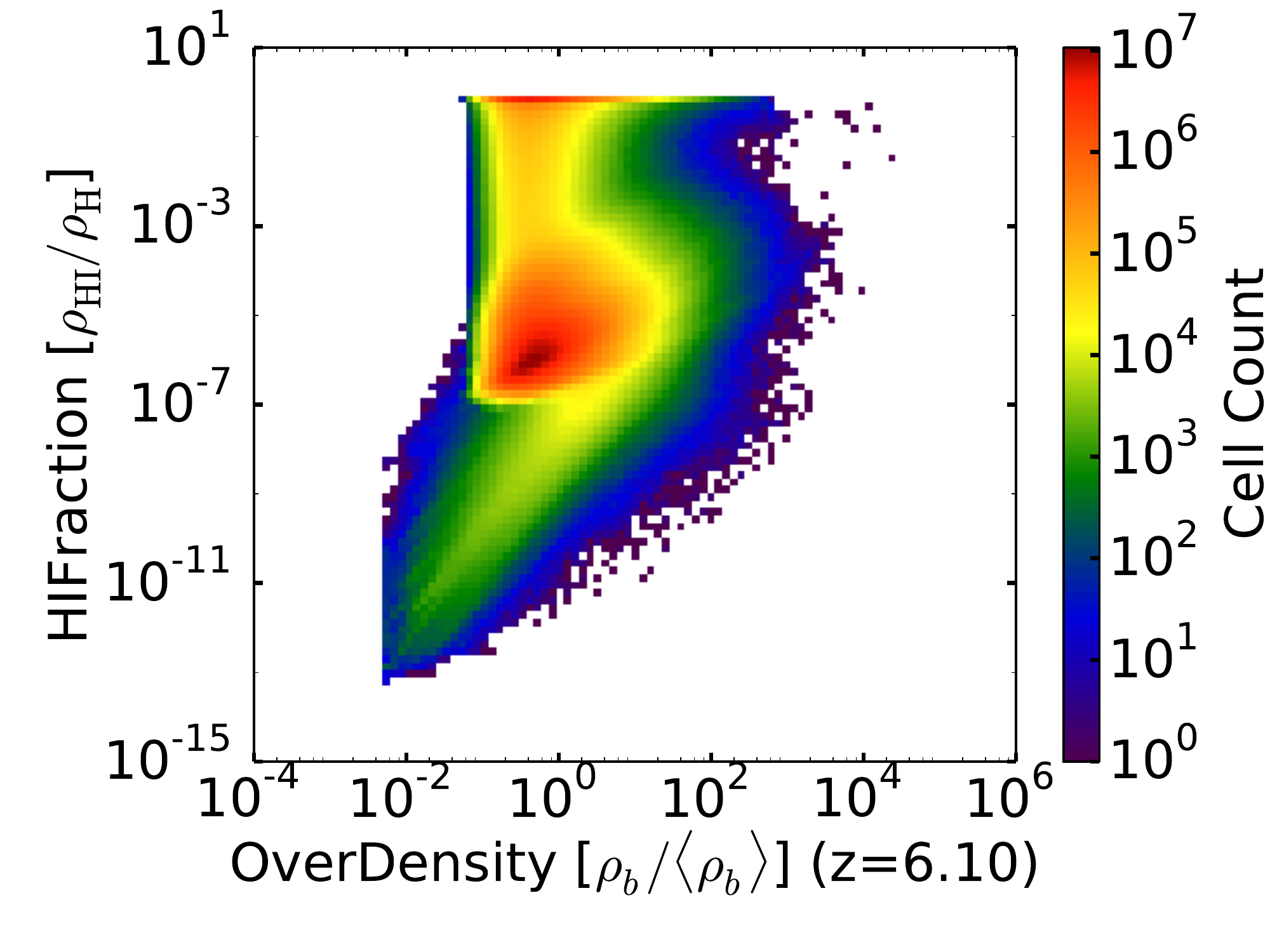}
	\end{minipage}
\hspace*{-2.00mm}
	\begin{minipage}[h]{0.33\linewidth}
	\centering
	\includegraphics[trim = 10mm 0mm 7mm 7mm, clip, width=1.0\textwidth]{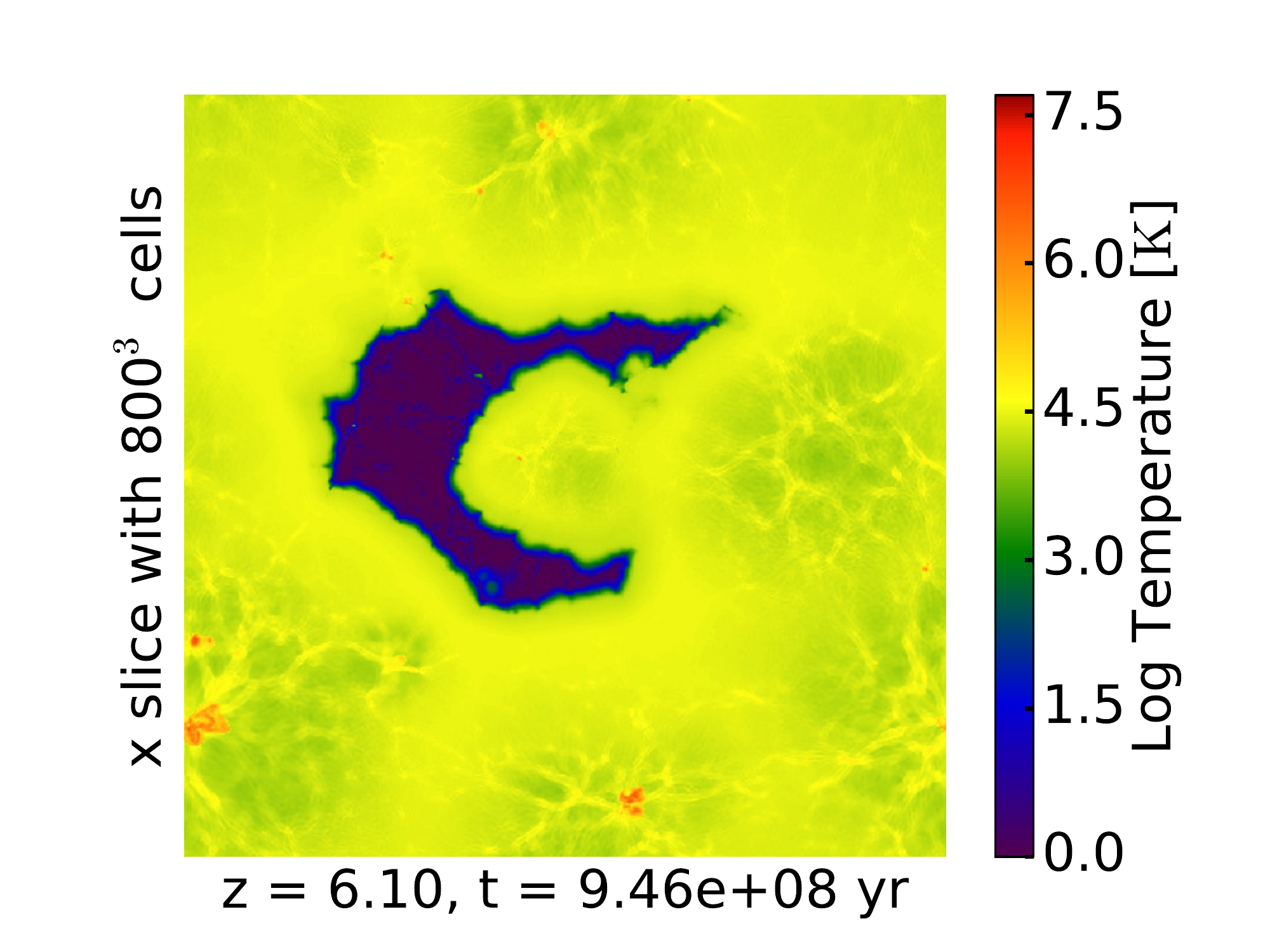}
	\end{minipage}
\hspace*{-2.00mm}
	\begin{minipage}[h]{0.33\linewidth}
	\centering
	\includegraphics[trim = 10mm 0mm 7mm 7mm, clip, width=1.0\textwidth]{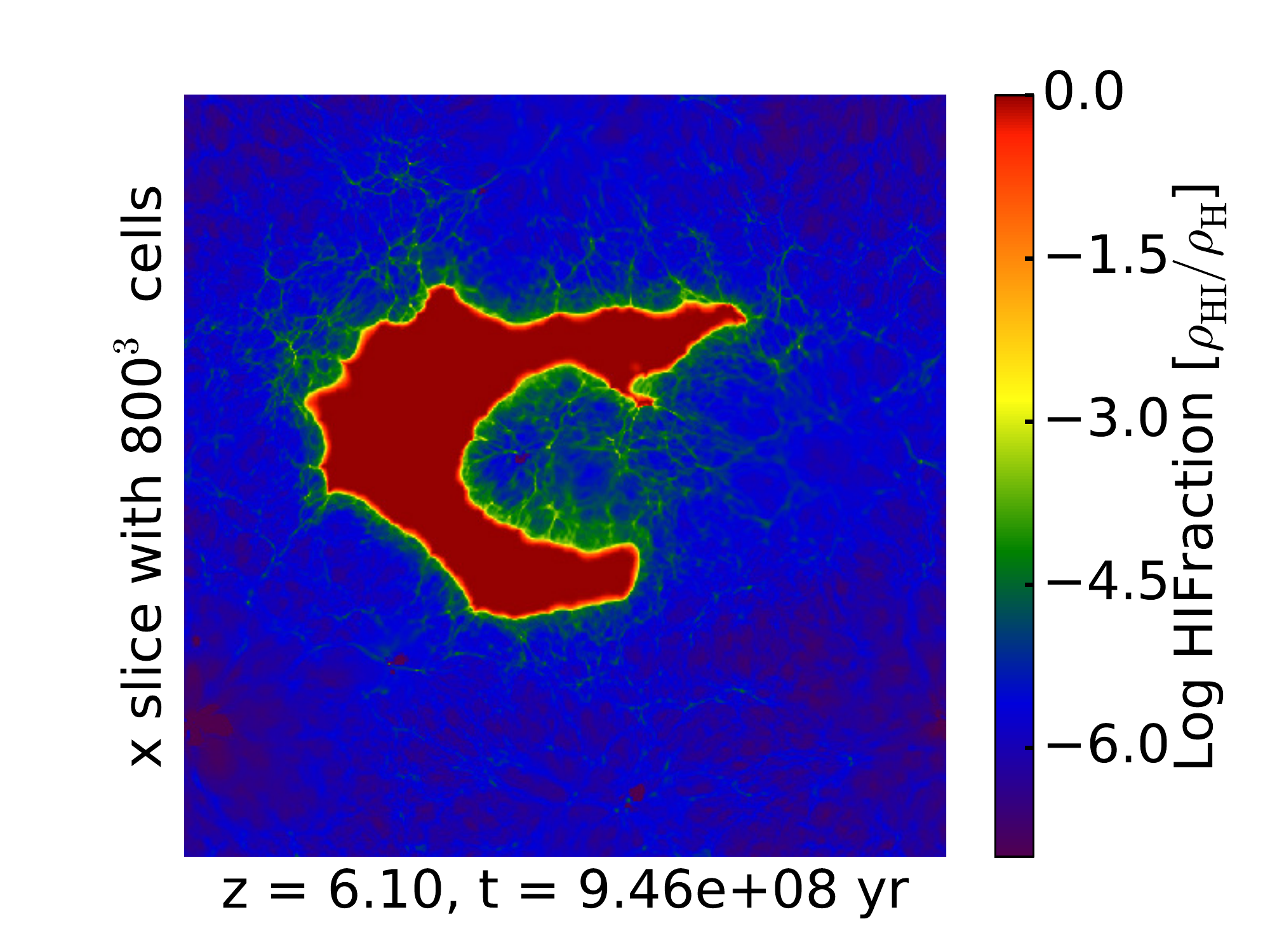}
	\end{minipage}
\vspace*{-2.00mm}\\
	\begin{minipage}[h]{0.33\linewidth}
	\centering
	\includegraphics[trim = 7mm 9mm 1mm 7mm, clip, width=1.0\textwidth]{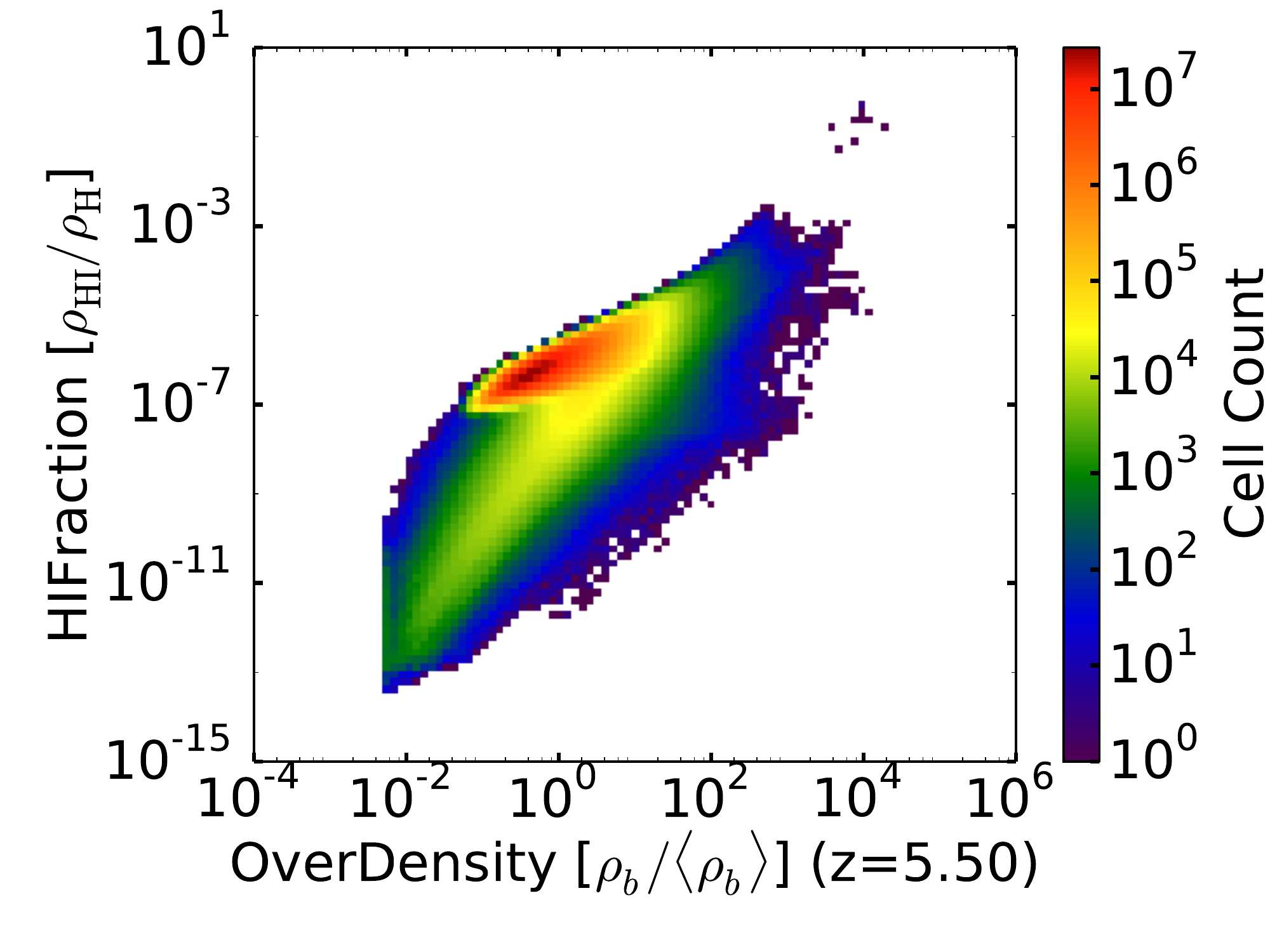}
	\end{minipage}
\hspace*{-2.00mm}
	\begin{minipage}[h]{0.33\linewidth}
	\centering
	\includegraphics[trim = 10mm 0mm 7mm 7mm, clip, width=1.0\textwidth]{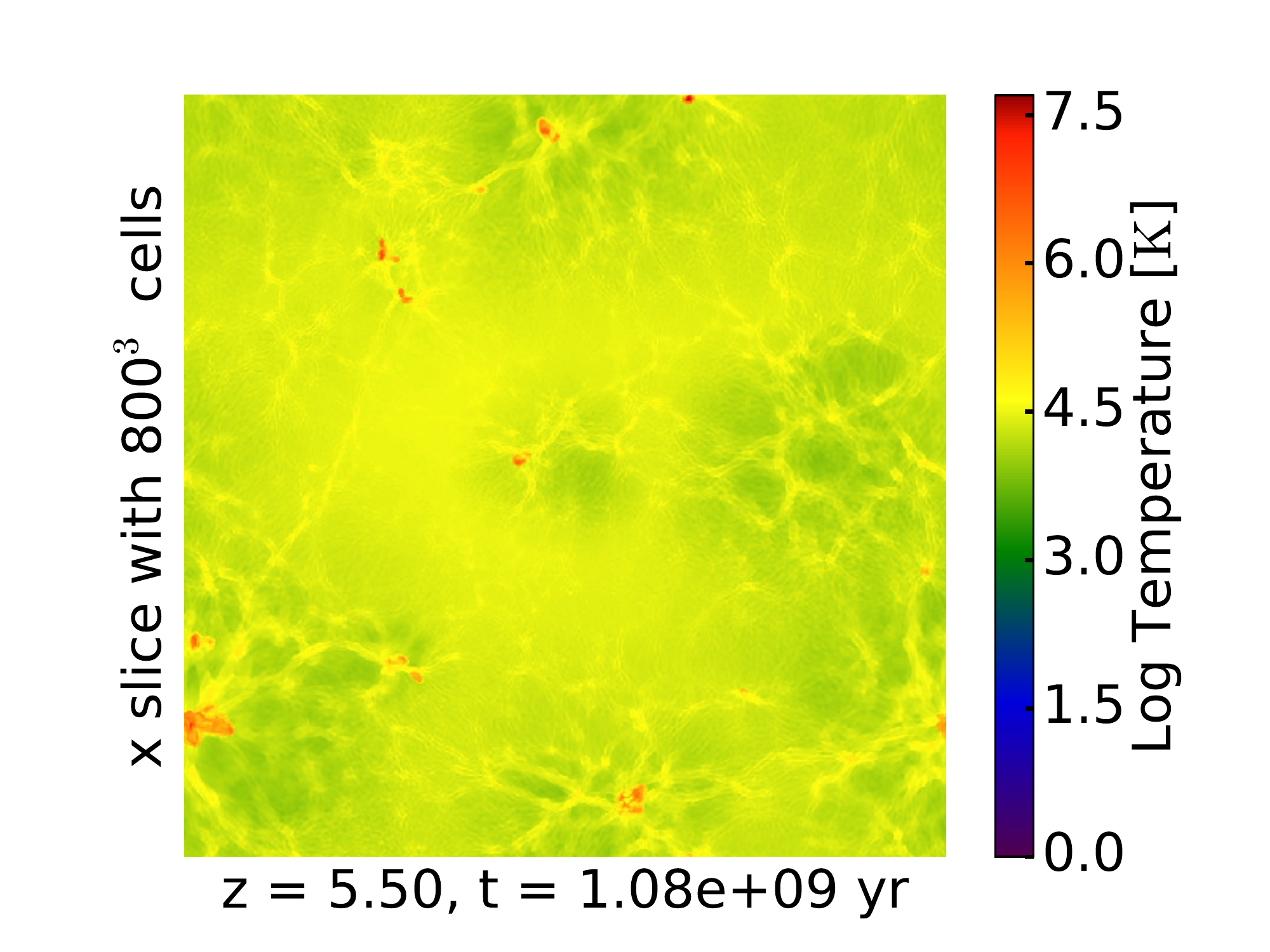}
	\end{minipage}
\hspace*{-2.00mm}
	\begin{minipage}[h]{0.33\linewidth}
	\centering
	\includegraphics[trim = 10mm 0mm 7mm 7mm, clip, width=1.0\textwidth]{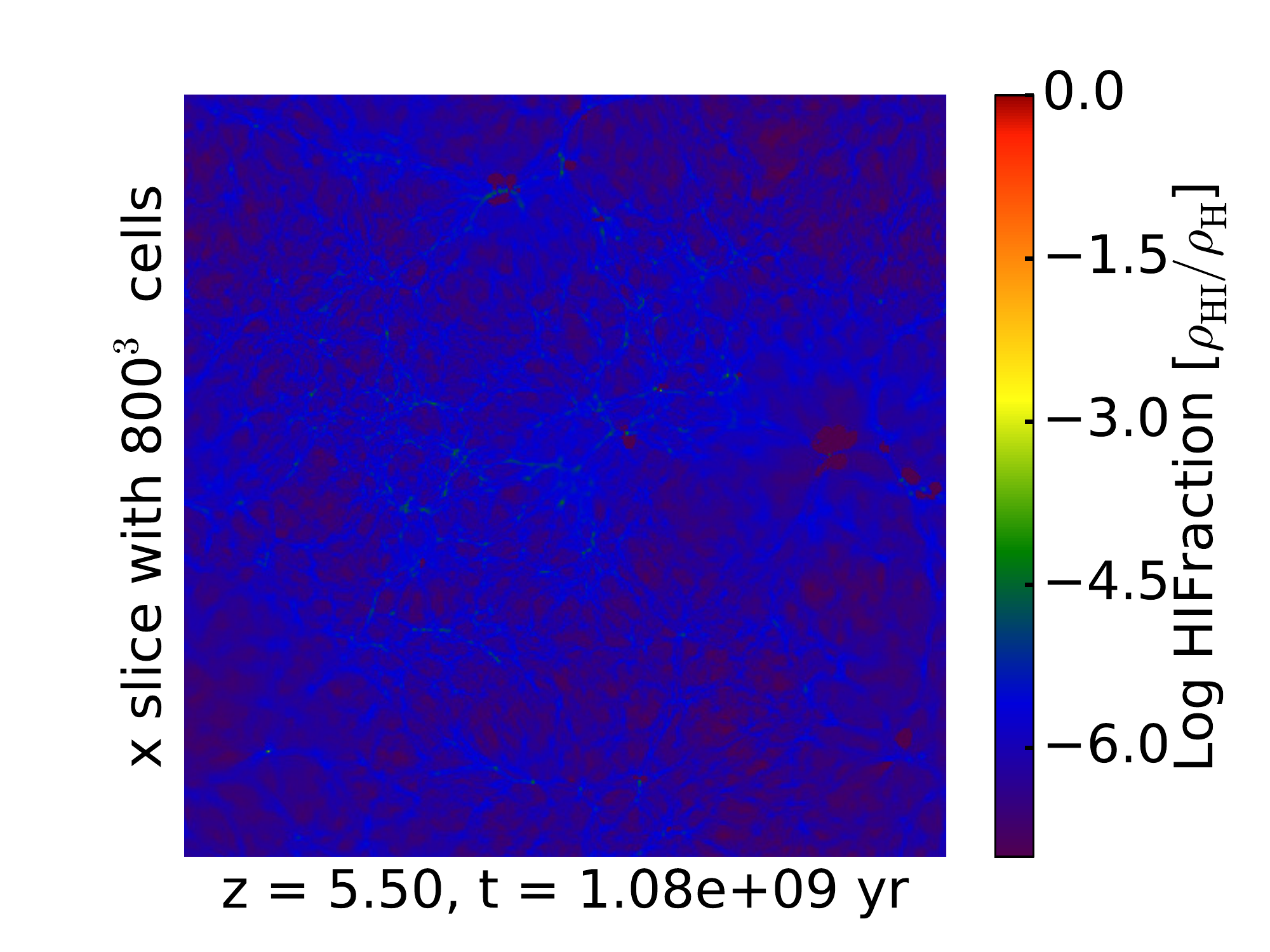}
	\end{minipage}
\vspace*{-2.00mm}\\
	\begin{minipage}[h]{0.33\linewidth}
	\centering
	\includegraphics[trim = 7mm 9mm 1mm 7mm, clip, width=1.0\textwidth]{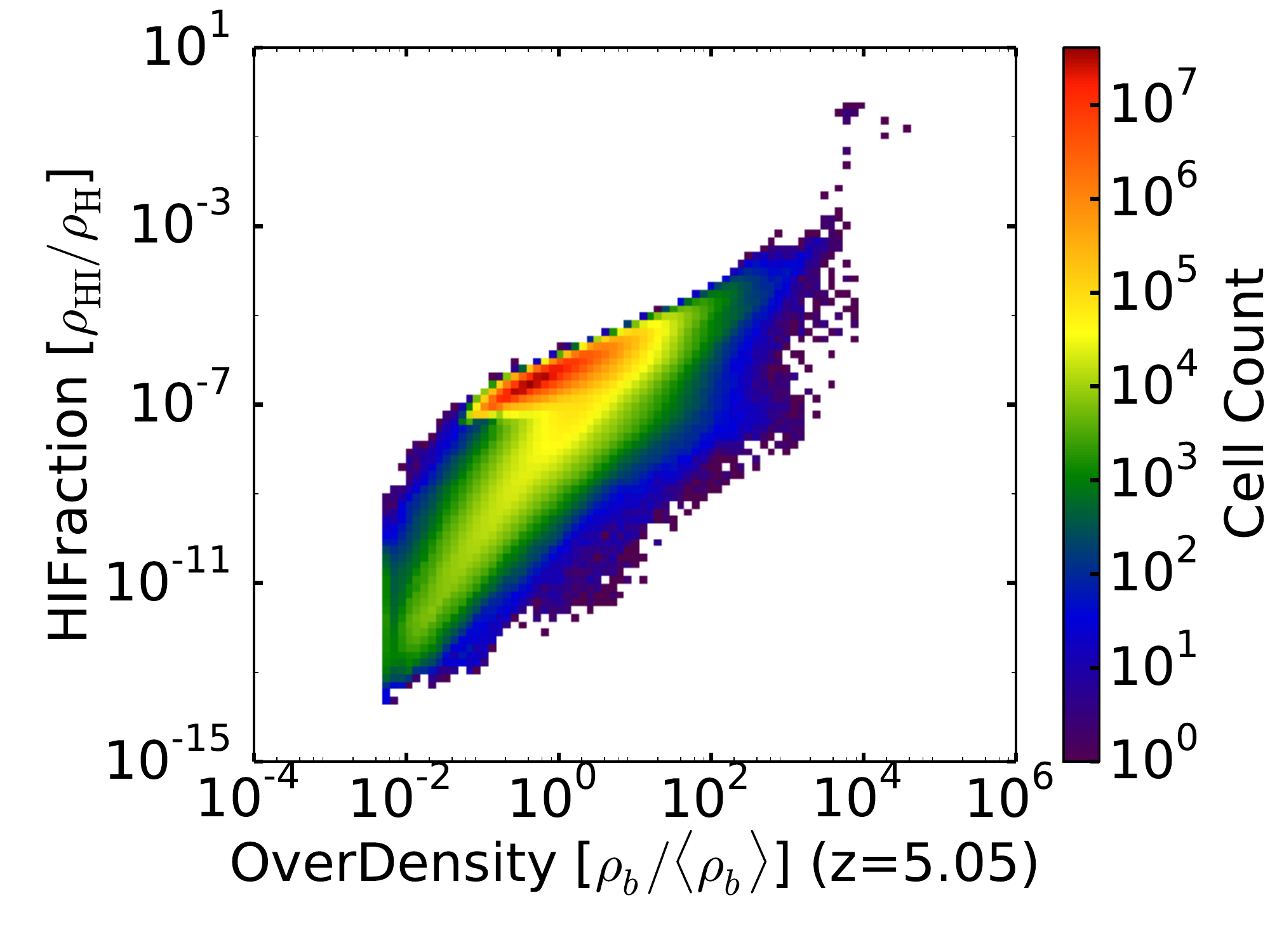}
	\end{minipage}
\hspace*{-2.00mm}
	\begin{minipage}[h]{0.33\linewidth}
	\centering
	\includegraphics[trim = 10mm 0mm 7mm 7mm, clip, width=1.0\textwidth]{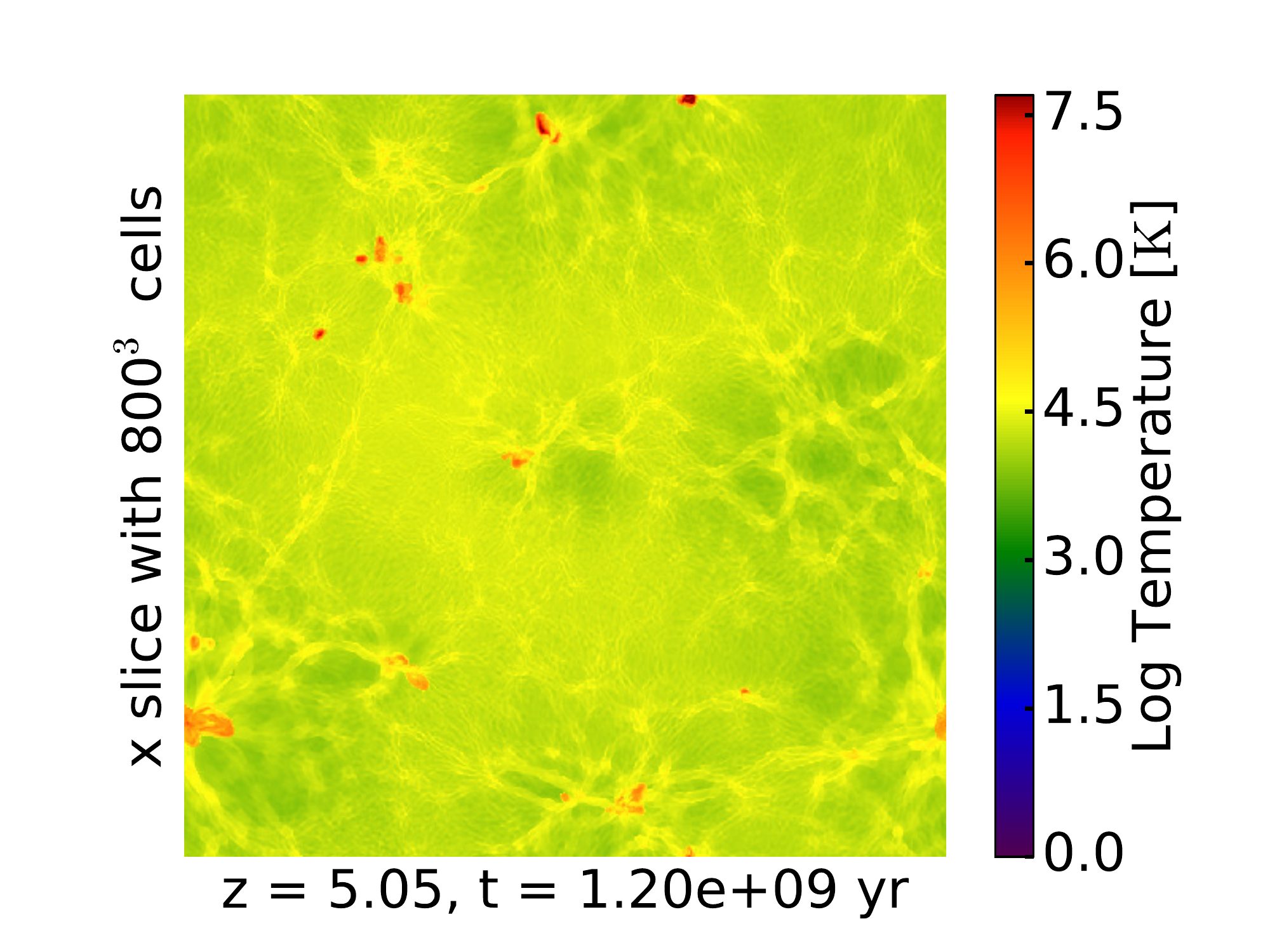}
	\end{minipage}
\hspace*{-2.00mm}
	\begin{minipage}[h]{0.33\linewidth}
	\centering
	\includegraphics[trim = 10mm 0mm 7mm 7mm, clip, width=1.0\textwidth]{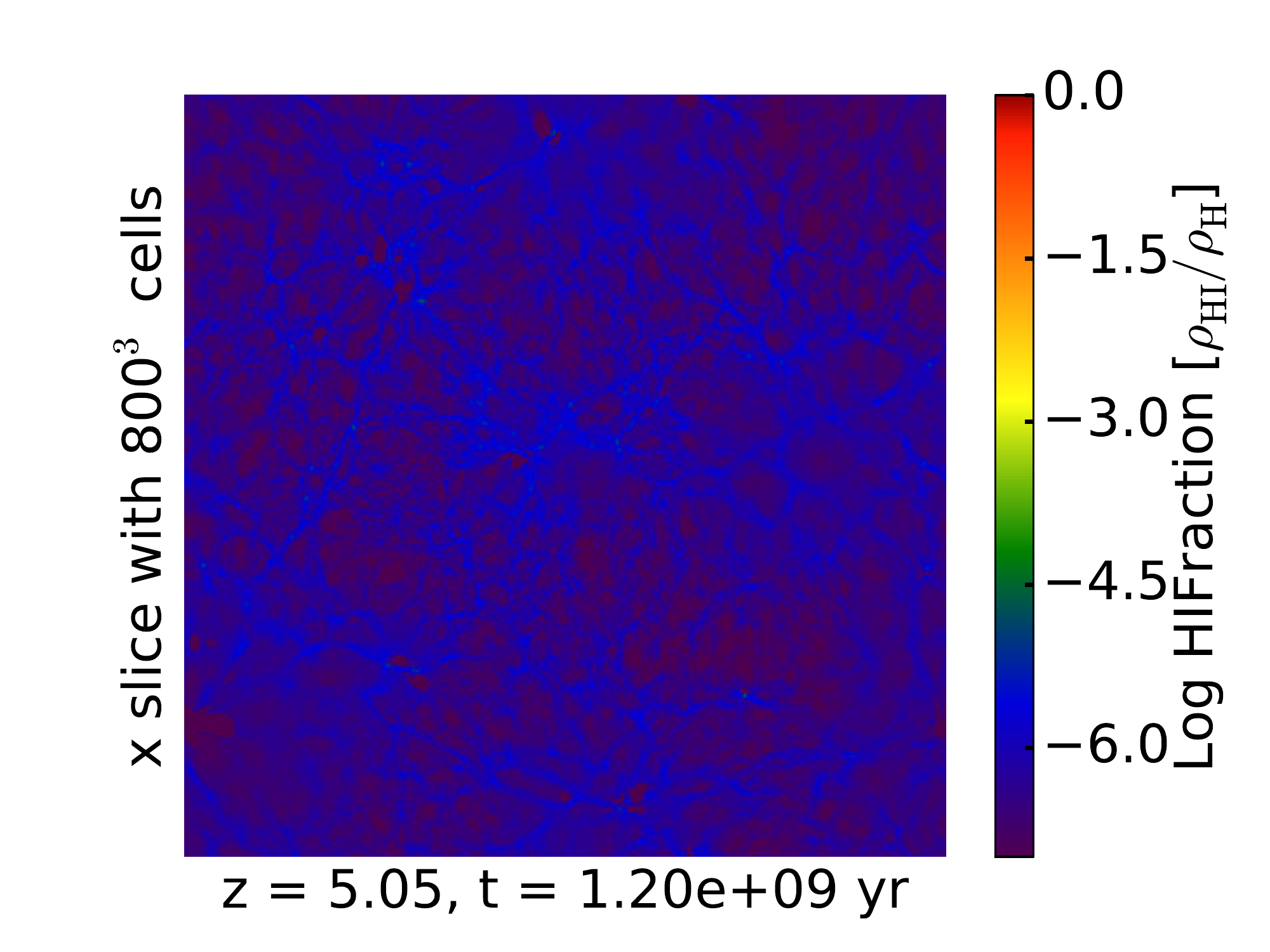}
	\end{minipage}

	\caption{{\em Left}: Phase diagram of neutral hydrogen fraction versus baryon overdensity with decreasing redshift from top to bottom.  {\em Middle}: Slices of Log Temperature [K] through a region that remained mostly neutral until just before overlap at redshift of $\sim$5.8. {\em Right}: Slices of neutral hydrogen fraction through the same region as before.  Please refer to \S\ref{IOOI} for detailed description.}
  \label{NeutralPhase}
\end{figure*}

To investigate how reionization progresses in regions of different density, we plot in Figure \ref{NeutralPhase} the hydrogen neutral fraction ($\rho_\mathrm{H\,I}/\rho_\mathrm{H}$) versus overdensity $\Delta_b\equiv\rho_b/\langle\rho_b\rangle$ in the left column, and in the right column a slice of the gas temperature, with redshift decreasing from top to bottom.  One would expect if inside-out ionization is the case, that the neutral fraction of higher density region should drop down more quickly than lower density regions.  Below, we will describe each row of the figure in more detail.

Looking at the redshift $z=10$ row, we see in the gas temperature slice that two isolated regions of ionization appear due to UV feedback from new stars, indicated by the $\sim$10$^4$K gas . These regions correspond to places on the neutral fraction vs. overdensity phase plot where a small amount of volume emerges around $\Delta_b$ of $10^{-1}-10^1$, reaching Well Ionized to Fully Ionized levels.  The T $\sim$10$^7$K region corresponds to the extended tail of very low neutral fraction gas in the left column, and indicates gas shock heated by supernova feedback.  Although the cell count of shock heated gas will grow, it remains orders of magnitude smaller compared to the photoionized regions that we will emphasize.  Even at this early stage, there are high density regions above $\Delta_b$ of 10$^2$-10$^3$ that are Well Ionized; this is due to their close proximity to the ionizing sources, supporting the Inside-out paradigm.

Looking at the next row of figures at a redshift of $z=7$, we see that the volume of Well Ionized regions has increased greatly, and so has the shock heated region in the phase plot.  We also see that most, but not all the $\Delta_b > 10^2$ cells have reached the Well Ionized level.  Although a large portion of the volume is in the Well Ionized regime, the majority of the volume (the red pixels) is still neutral, as we can see in the corresponding temperature slice plot.  Most of the volume is still well under 10$^4$K, where we expect the temperature to hover around once the ionization front has passed through the region and the gas has had time to come into photoionization thermal eqilibrium. 

By a redshift of $z=6.1$, we see from the left column that the region that is ionized beyond the Fully Ionized level (an irony in terms, which means there is definitely room for improvement in the naming convention), dominates the simulation volume.  There are still some regions not yet consumed by the ionization front, that is seen on the top of the neutral fraction plot and on the right according to the temperature slice.

The next row at redshift of $z=5.5$ is after the entire volume has been swept over by ionization fronts.  Most of the volume is beyond the Well Ionized level, except for a few cells around $\Delta_b \sim 10^2$.  There are also some cells that are still neutral around $\Delta_b \sim 10^4$.  They remain neutral because their densities are so high, leading to high recombination rates.  Over time these cells will shift up and down the neutral fraction plot with waves of star formation and supernova explosions since they are likely close to the source of the radiation and kinetic energy.

The last row of Figure \ref{NeutralPhase} is at redshift $z\sim5$, where we can see that the previous few cells that have yet to reach Well Ionized levels around $\Delta_b \sim 10^2-10^3$ have now disappeared.  The cells that have not reached Well Ionized level before are cells where either the radiation is not strong enough due to shielding effects or the density is so high the gas recombines quickly even after being ionized.  After the ionization front has passed though and highly ionized the IGM, there is little material left to shield against the radiation background and we see all but the densest few cells become Well Ionized. The high density region reaching the same ionization level after the under dense void, would fit well with the description for the Outside-in model.  Note, that the remaining cells that finally reached Well Ionized levels, are orders of magnitude smaller in total volume compare to the rest of the cells at the same density.  So if we call cells of $\Delta_b \sim 10^2$ filaments, not all dense filaments get Well Ionized until late in the EoR.  Before the volume is filled with radiation, these dense filaments are able to remain relatively neutral.

Unfortunately, the evolution of these redshift panels is not enough to capture the propagation of radiation fronts from the initial sources, but they do convey the overall ionization history of the universe.  The panels suggest that the region surrounding the ionization sources, whether they are dense cores, filaments, or voids, are all affected by the radiation on roughly the same time scale.  However, the degree to which they are ionized is different.  It is this difference, that is the key to answering the original question, whether the universe ionize inside-out or outside-in.  

When focusing on the ionization of the IGM, lets for a moment neglect the $\Delta_b \sim10^4$ cells that shift ionization level with waves of star formation which comprise a tiny fraction of the volume.  If we use the ``Ionized'' level to characterize something as completely ionized and draw the line for neutral fraction at 10\%, then the universe reaches end of EoR before $z \sim 5.5$.  Since radiation propagates from sources outward, that would correspond to the Inside-out picture.  If we were to instead draw the completely ionized line at ``Well Ionized'' level, then we can see that even at $z \sim 5.5$, there is a small peak in the dense region of the phase diagram ($\Delta_b \sim 2.4\times10^2$) that has yet to reach below the line to be considered completely ionized.  This would correspond to the Outside-in picture which reaches end of EoR sometime before $z \sim 5$ (or Inside-outside-middle if one uses the \cite{FinlatorEtAl2009} terminology and considers the neutral peak to be a part of the filaments).  And finally, if we were to draw the line at the ``Fully Ionized'' level, the universe has yet to ionize even for regions that are only 10$\times$ over dense.  Thus the ionization history is a story with many perspectives, and it really depends on how the story teller draws the line as to whether Inside-out, Outside-in, or Inside-outside-middle is a better qualitative description. 

%To conclude this arduous and complex tale told by the few panels shown, is that reionization can be
%told with varying accuracy and focus.  The accuracy and focus can and will change the narrative that is finally distilled.  We would like to claim that the previous 
%authors all told a side of a story on this in, out, middle issue.  But, at the risk of sounding clich\'e, the whole story indeed is a combination of those pieces.  So, 
%unless we use the same quantitative language to describe these simulations, we may seem like disagreeing, while trying to tell a side of the same story.

%\begin{figure}
%	\includegraphics[width=0.7\textwidth]{color_phase_slice.png}
%	\caption{Left column: Plot of hydrogen neutral fraction versus over density with decreasing redshift from top to bottom.  Right column: Plot of Log temperature [K] projections through the simulation centered on a region that remained mostly neutral right before the ionization fronts engulf it at redshift of $\sim$6, with decreasing redshift from top to bottom.  Please refer to \S\ref{IOOI} for detailed description.}
%      \label{NeutralPhase}

%\end{figure}

\section{Clumping Factors and the Photon Budget for Reionization}
\label{sec:ClumpingFactors}

\subsection{Clumping Factor Analysis of Madau}
\label{Madau}
In this section we begin our examination of Equation \eqref{eq:ndot} from \cite{MadauEtAl1999} as an accurate predictor of when reionization completes, focusing on the clumping factor. While  it is true that the Madau-type analysis was not designed to predict the precise redshift for reionization completion, only the ionization rate density needed to maintain the IGM in an ionized state after reionization has completed, it is effectively being used in this way when it is applied to galaxy populations at increasingly higher redshifts $z=6-7$  (cf. \cite{FanEtAl2006, RobertsonEtAl2013}).
Our methodology is the following. The simulation supplies $\dot{N}_{sim}(z)$ ionizing photons, which increases with decreasing redshift because the SFRD increases with decreasing redshift.  Equation \eqref{eq:ndot} poses a minimum requirement on the ionizing emissivity to maintain the IGM in an ionized state at given redshift z. This requirement decreases with decreasing redshift due to the strong z dependence.  We look to see if the box becomes fully ionized when these two curves cross; i.e., when $\dot{N}_{sim} \geq \dot{\mathcal{N}}_{ion} $. In subsequent sections we do this for more recent definitions of the clumping factor that have been introduced by various authors, in roughly chronological order. 

%\subsubsection{Original Methods}
%\label{OriginalMethods}

The way the clumping factor is introduced and used, is to estimate the amount of recombination that radiation has to overcome, in order to keep the universe ionized \citep{GnedinOstriker1997,ValageasSilk1999,MadauEtAl1999,FanEtAl2006}.  In a homogeneous universe, the hydrogen recombination rate is also homogeneous, and is a simple function of the mean density, ionization fraction, and temperature. The clumping factor is a correction factor to account for density inhomogeneities induced by structure formation, although in principle inhomogeneties in ionization fraction and temperature are also important.   
%In its original definition, the clumping factor is defined when only hydrogen information is available.  Therefore, when the universe is 100\% ionized, the free %electron number density equals that of the hydrogen number density denoted by $n_\mathrm{H\,II}$, multiplied by $(1+2\chi)$, where $\chi$ is the cosmic fraction of helium to hydrogen.  
The most common definition for the clumping factor is:
\begin{equation}
	C=\frac{\langle n_\mathrm{H\,II}^2 \rangle}{\langle n_\mathrm{H\,II} \rangle^2}
	\label{eq:clumpingfactor}
\end{equation}
Where the $\langle\rangle$ brackets denotes an average over the simulation volume. To see where this comes from lets look at the change of $n_\mathrm{H\,II}$ with respect to time due to recombinations:
%Paraphrasing the original authors, if we only consider the non-expanding cosmological box, 
%To see where this comes from look at the change of $n_\mathrm{H\,II}$ with respect to time of 
% recombination $t_{rec}$, using the recombination portion of Equation \eqref{eq:chemical_ionization},
\begin{align}
	\label{eq:recombtime}
	\frac{\partial n_\mathrm{H\,II}}{\partial t} 
&= -n_\mathrm{e} n_\mathrm{H\,II}\alpha_B(T)\notag\\
	\frac{\partial n_\mathrm{H\,II}}{n_\mathrm{H\,II}} 
&= -\partial t n_\mathrm{e}\alpha_B(T)\notag\\
	\int^{n_f}_{n_i}\frac{\partial n_\mathrm{H\,II}}{n_\mathrm{H\,II}} 
&= -\int^{t_f}_{t_i}\partial t n_\mathrm{e}\alpha_B(T)\notag\\
	ln\left(\frac{n_f}{n_i}\right) 
&= -(t_f-t_i)n_\mathrm{e}\alpha_B(T), \notag\\
	\frac{n_f}{n_i} 
&= exp(-t_{rec}n_\mathrm{e}\alpha_B)
\end{align}
In the last step, we have set $(t_f-t_i)$ to be $t_{rec}$.  This leads to 
\begin{equation}
    t_{rec} = [n_\mathrm{e}\alpha_B(T)]^{-1}
    \label{recombtime}
\end{equation}
being the characteristic time when the fraction $n_f/n_i =  1/e$.  Using this expression for the recombination time, one can rewrite the right hand side of the equation as
\begin{align}
	\label{eq:trec}
	\frac{\partial n_\mathrm{H\,II}}{\partial t} 
&= - n_\mathrm{H\,II}n_\mathrm{e}\alpha_B(T) = - n_\mathrm{H\,II} / t_{rec}\notag\\
&= - n_\mathrm{H\,II} (1+2\chi) n_\mathrm{H\,II} \alpha_B(T) \notag\\
&= - n_\mathrm{H\,II}^2 (1+2\chi) \alpha_B(T) \notag\\
\end{align}
where in the last two steps, following \cite{MadauEtAl1999}, we replace $n_\mathrm{e}$ with $(1+2\chi)n_\mathrm{H\,II}$ assuming helium is fully ionized. Here $\chi$ is the cosmic fraction of helium. Taking the volume average we have:
\begin{align}
\label{eq:trecpart2}
	\langle \frac{\partial n_\mathrm{H\,II}}{\partial t} \rangle
&= - \langle n_\mathrm{H\,II}^2 (1+2\chi) \alpha_B(T) \rangle \notag\\
&= - \langle n_\mathrm{H\,II}^2 \rangle (1+2\chi) \alpha_B \notag\\
&= - \langle n_\mathrm{H\,II} \rangle^2 (1+2\chi) \alpha_B C\notag\\
&= - \langle n_\mathrm{H\,II} \rangle /\bar{t}_{rec}
\end{align}
%When all the helium is ionized, one is justified to replace $n_\mathrm{e}$ by %$(1+2\chi)n_\mathrm{H\,II}$.  
In the above we have made the oft-used assumption of a uniform IGM temperature of $10^4$K, making the Case B recombination coefficient, $\alpha_B$ a constant.  Note this is not physically justified, but since the temperature of the IGM is not well determined observationally, it is a useful approximation, and one that is embedded in Equation \eqref{eq:ndot}. With this simplifying assumption, when taking the volume average on both sides of the equation, we may rewrite the result in the same form as the first line in Equation \eqref{eq:trec}.  Therefore, the effective recombination time can be written as 
\begin{equation}
	\bar{t}_{rec} = t_\mathrm{Madau} \equiv [(1+2\chi)\langle n_\mathrm{H\,II} \rangle \alpha_B C]^{-1}
	\label{eq:tmadau}
\end{equation}
%We are hesitant to label the recombination time $t_\mathrm{Madau}$ as a mean quantity here, %as oppose to \cite{MadauEtAl1999}, which we will discuss in more detail in \S\ref{Discussion}.
This expression is the same as Equation (20) of \cite{MadauEtAl1999} if we substitute $\langle n_\mathrm{H\,II} \rangle$ for $\bar{n}_\mathrm{H}$. In the case of a fully ionized universe these two quantities are equivalent. We note that $t_\mathrm{Madau}$ is not at all the volume average of $t_{rec}$ but is $\langle t_{rec}^{-1} \rangle ^{-1}C^{-1}$, which weights regions
with the {\em shortest} recombination times; i.e. regions at the mean density and above. If we now make the 
{\em ansatz} $\dot{\mathcal{N}}_{ion} \times \bar{t}_{rec} = \bar{n}_\mathrm{H}(0)$, we may derive Equation (26) in \cite{MadauEtAl1999}, updated by \cite{FanEtAl2006}, repeated here for convenience:
%If the clumping factor is calculated from an accurate simulation, then the clumping factor can 
%be used in the above work's Equation (24) in semi-analytical models to predict the ionization 
%rate.  One will arrive at the equation
\begin{equation}
	\label{eq:updatedNdot}
	\dot{\mathcal{N}}(z)=10^{51.2}s^{-1}Mpc^{-3}\left(\frac{C}{30}\right)\left(\frac{\Omega_\mathrm{b} h^2}{0.02}\right)^{2}\left(\frac{1+z}{6}\right)^{3}.
\end{equation}

%and updated in \citep{FanEtAl2006} with appropriate assumptions.  
This equation gives an estimate of the ionizing photon production rate density (in units of s$^{-1}$Mpc$^{-3}$comoving) that is needed to balance the recombination rate density (the right-hand-side of Equation \eqref{eq:updatedNdot}) in a completely ionized universe.  Values for $C$ ranging $\sim$10-30 are often quoted from earlier hydrodynamical simulations such as \cite{GnedinOstriker1997}, and $\sim 3$ for more recent work following \cite{PawlikEtAl2009, RaicevicTheuns2011, ShullEtAl2012, FinlatorEtAl2012} and the methods there.
\begin{figure}
	\includegraphics[width=0.5\textwidth]{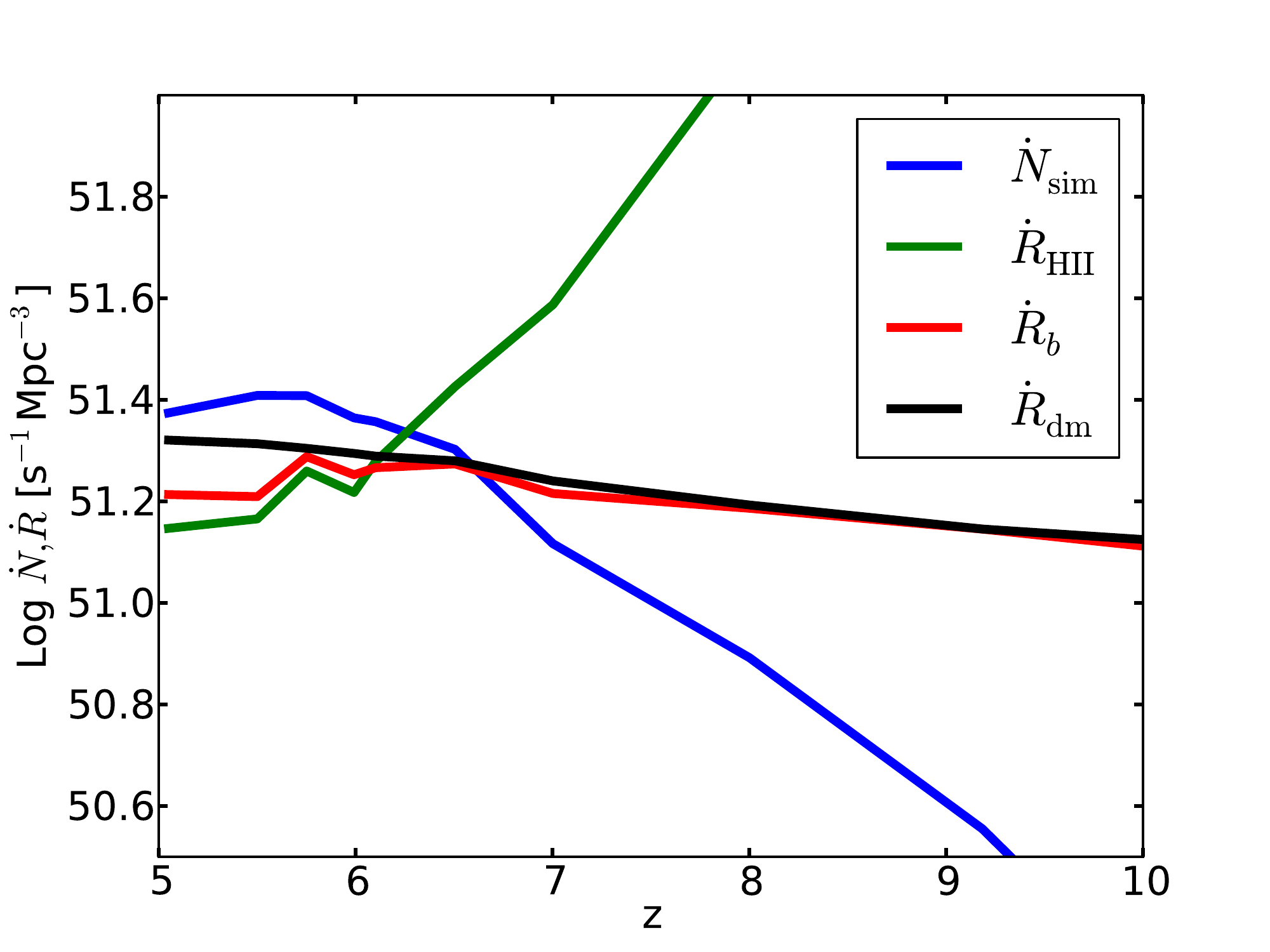}
	\caption{Ionizing photon production rate density and various estimates of the recombination rate density versus redshift. The blue curve labeled ``$\dot{N}_{sim}$'' is the measured photon production rate density averaged over the entire simulation volume. The green curve labeled ``$\dot{R}_\mathrm{H\,II}$'' is the recombination rate density estimate from using the clumping factor calculated with Equation \eqref{eq:clumpingfactor} substituted in Equation \eqref{eq:updatedNdot}. The red curve labeled ``$\dot{R}_b$'' is Equation \eqref{eq:updatedNdot} evaluated using a clumping factor calculated from the baryon density. The black curve labeled ``$\dot{R}_\mathrm{dm}$'' is using a clumping factor calculated with dark matter density.}
	\label{unthresholded}
\end{figure}

\begin{figure}
	\includegraphics[width=0.5\textwidth]{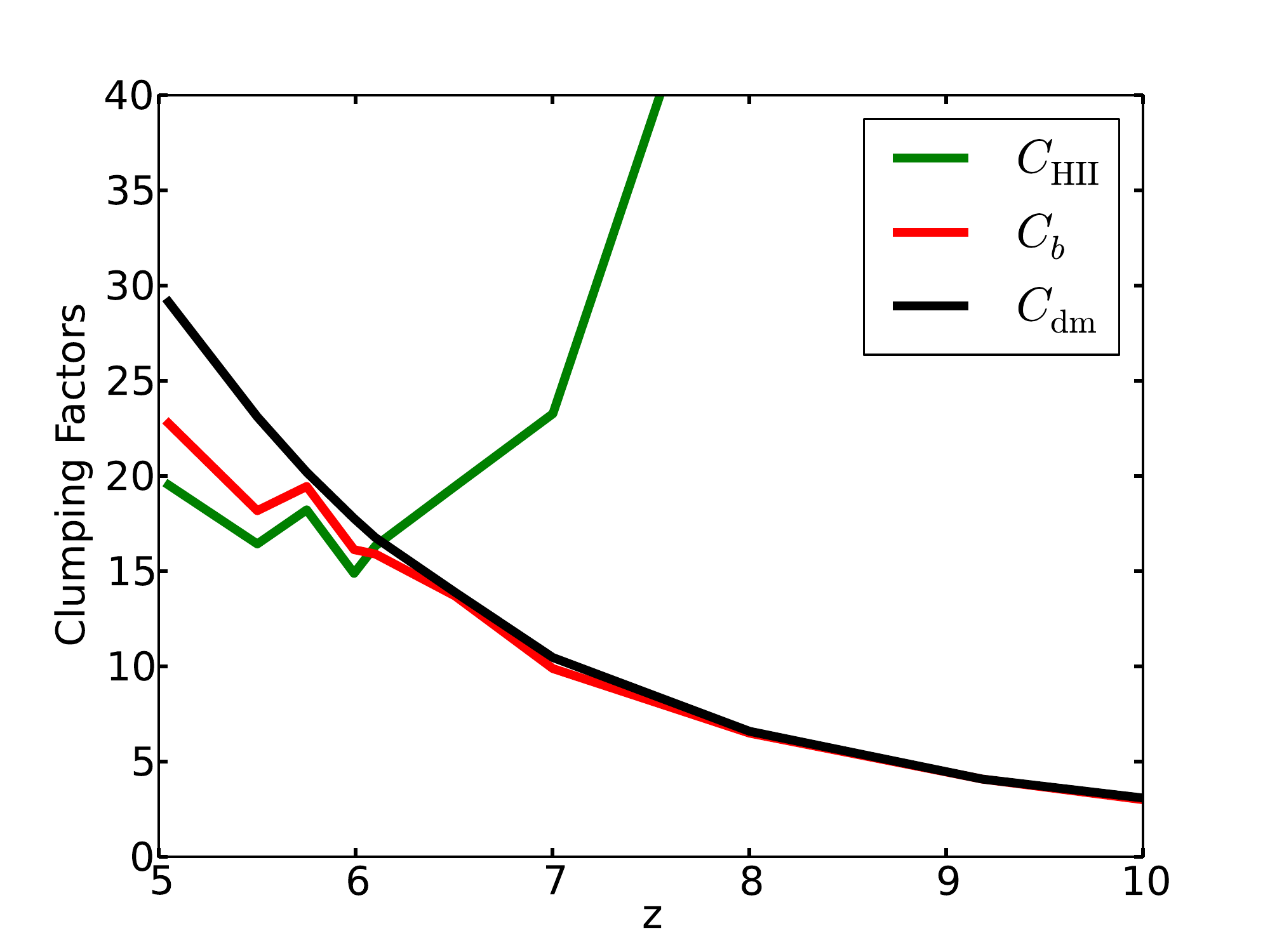}
	\caption{Unthresholded clumping factors used in Fig. \ref{unthresholded}. $C_{HII}, C_b, C_{dm}$ are calculated from the unthresholded H \footnotesize{II}, baryon, and dark matter densities, respectively.}
	\label{unthreshclumping}
\end{figure}

We follow these earlier studies using our own simulation data.  In Figure \ref{unthresholded} we plot the ionizing photon production rate density and recombination rate density from our fiducial simulation.  The curve in blue labeled $\dot{N}_{sim}$ is the photon production rate density from the simulation, calculated using a time average of the volume integrated ionizing emissivity $\eta$ (Equation \eqref{eq:emissivity}) divided by the average energy per photon which we obtain directly from the SED.  The other three curves plot Equation \eqref{eq:updatedNdot} for three methods for calculating $C$: green uses the H {\footnotesize II} density directly (Equation \eqref{eq:clumpingfactor}); red uses the baryon density $C=\langle \rho^2_b \rangle / \langle \rho_b \rangle^2$; and black uses the dark matter density $C=\langle \rho^2_{dm} \rangle / \langle \rho_{dm} \rangle^2$. In all cases no thresholding is being applied (the effect of threholding is examined in the next section); the averages are done over every cell in the simulation including those inside the virial radii of galaxies. The H {\footnotesize II} curve drops sharply with decreasing redshift because $C$ is large when the H {\footnotesize II} distribution is patchy. The baryon and dark matter curves track one another for $z > 6$ because the clumping factors are nearly the same, but begin to separate after overlap as the baryon clumping factor drops due to Jeans smoothing.  

Where the ionization and recombination rate density lines cross is roughly when we expect the universe to become highly ionized.
If we define the end of the EoR as when 99.9\% of the volume has  reached the Well Ionized level, then our simulation reaches that point around $z\sim5.8$ according to Figure \ref{linearIonized}.  The $\dot{N}_{sim}$ curve crosses the $\dot{R}_\mathrm{H\,II}$ curve at $z\sim6.2$. This is somewhat reassuring since we are counting every ionizing photon emitted and every recombination, at lease insofar as Equation \eqref{eq:updatedNdot} provides a good estimate of that. The recombination rate density curves using clumping factors computed from the baryon and dark matter densities curves cross the $\dot{N}_{sim}$ curve at a somewhat higher redshift of $z \approx 6.6$. By following the original methodology of using the clumping factor to estimate recombinations, we find that the clumping factor calculated with the H {\footnotesize II} density field to be the closest predictor for the end of EoR in our simulation. 

\begin{figure}
	\includegraphics[width=0.5\textwidth]{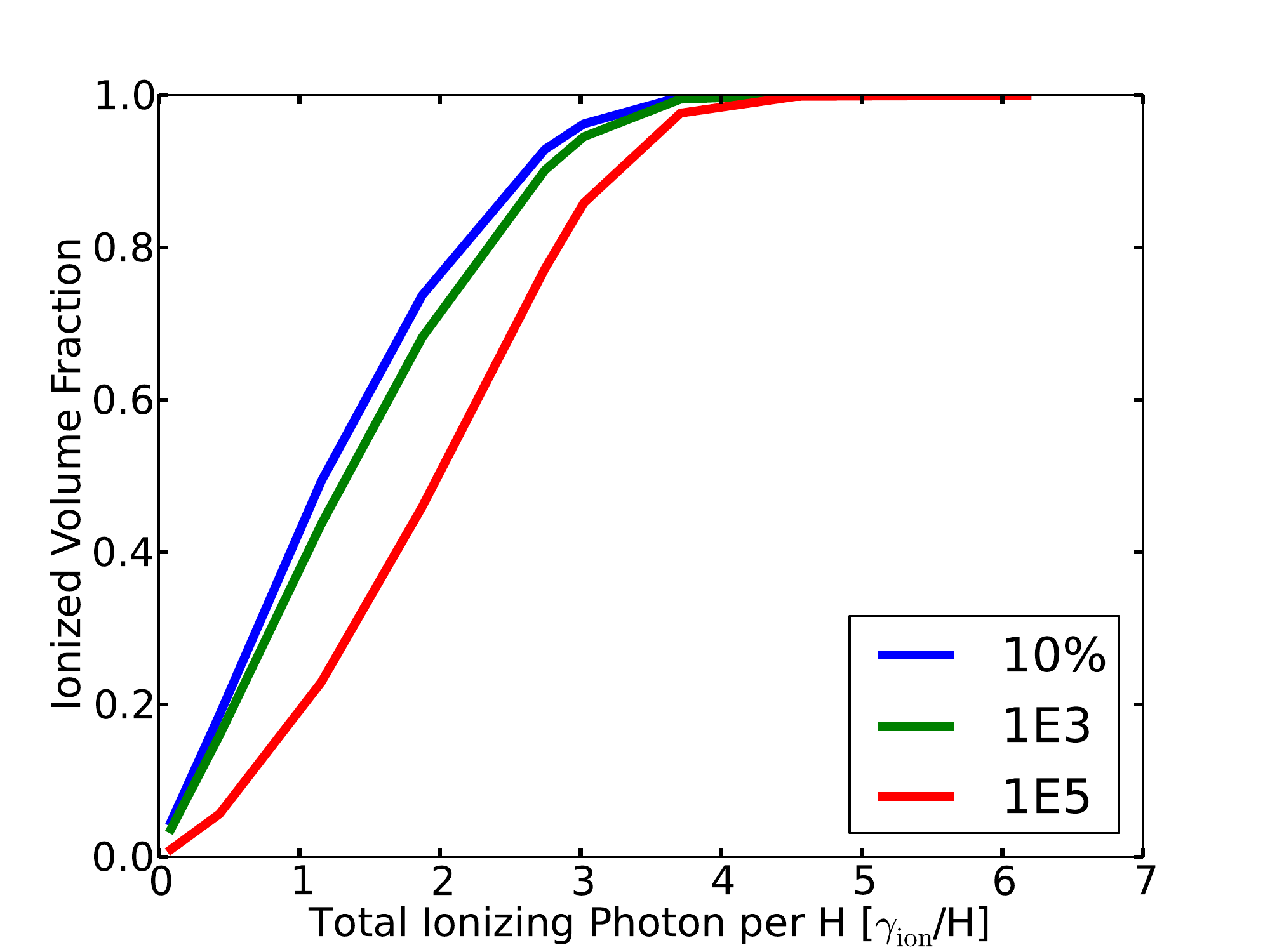}
	\caption{Ionized volume fraction as a function of the number of ionizing photons emitted per H atom averaged over the entire simulation volume (including inside halos) for three different ionization levels: $f_i \geq 0.1$ (blue line);  $f_i \geq 0.999$ (green line); $f_i \geq 0.99999$ (red line). Compare with Fig. \ref{threshphotonbudget} which excludes gas inside halos.}
	\label{unthreshphotonbudget}
\end{figure}

The photon budget that enabled us to reach different levels of ionization is plotted in Figure \ref{unthreshphotonbudget}.  
Here we plot the evolution of the ionized volume fraction versus $\gamma_{ion}/H=\int dt \dot{N}_{sim} / \bar{n}_\mathrm{H}(0)$. So, for the same definition for the end of EoR, we see that we need $\sim$4 photons per hydrogen atom to  achieve. This cannot be considered a converged result because this estimate includes the dense gas inside galaxies, which is not well resolved in our simulation. Even though a small fraction of the baryons reside inside galaxies, due to the short recombination time many ionizing photons are required to keep the gas ionized.  Since we have not resolved the internal structure of galaxies, and higher resolution would likely result in higher density gas, we must consider $\gamma_{ion}/H=4$ a lower bound. We eliminate this issue in the next subsection by excluding the dense gas in halos from the calculation. 
%By the same logic, as the definition of EoR changes, so will the photon per hydrogen number.

%\subsubsection{Investigating Thresholding Method}
%\label{InvestigatingThresholdingMethod}
\subsection{Quantitative Analysis of Recombinations}

As the clumping factor method grew in popularity, various authors have applied thresholds of one form or another to improve upon its accuracy in predicting the recombination rate density needed to maintain an ionized universe. When thresholds are applied, parts of the volume are excluded from the photon counting analysis.   \cite{PawlikEtAl2009, RaicevicTheuns2011} and others, limit the calculation of the clumping factor to the low density IGM by using $\Delta_b$ thresholds, usually set at 100.  They threshold out gas in virialized halos and the self-shielded collapsed objects, because radiation does not penetrate these objects, or they recombine too fast, which leaves them neutral and not contributing to recombinations in the IGM. More recently \cite{ShullEtAl2012} has also thresholded out void regions ($\Delta_b < 1$), arguing that they do not contribute appreciably to the total recombinations due to their long recombination times. 

\begin{figure*}[!tp]
     \begin{minipage}[h]{0.33\linewidth}
        \centering
        \includegraphics[trim = 5mm 8mm 0mm 0mm, clip, width=1.0\textwidth]{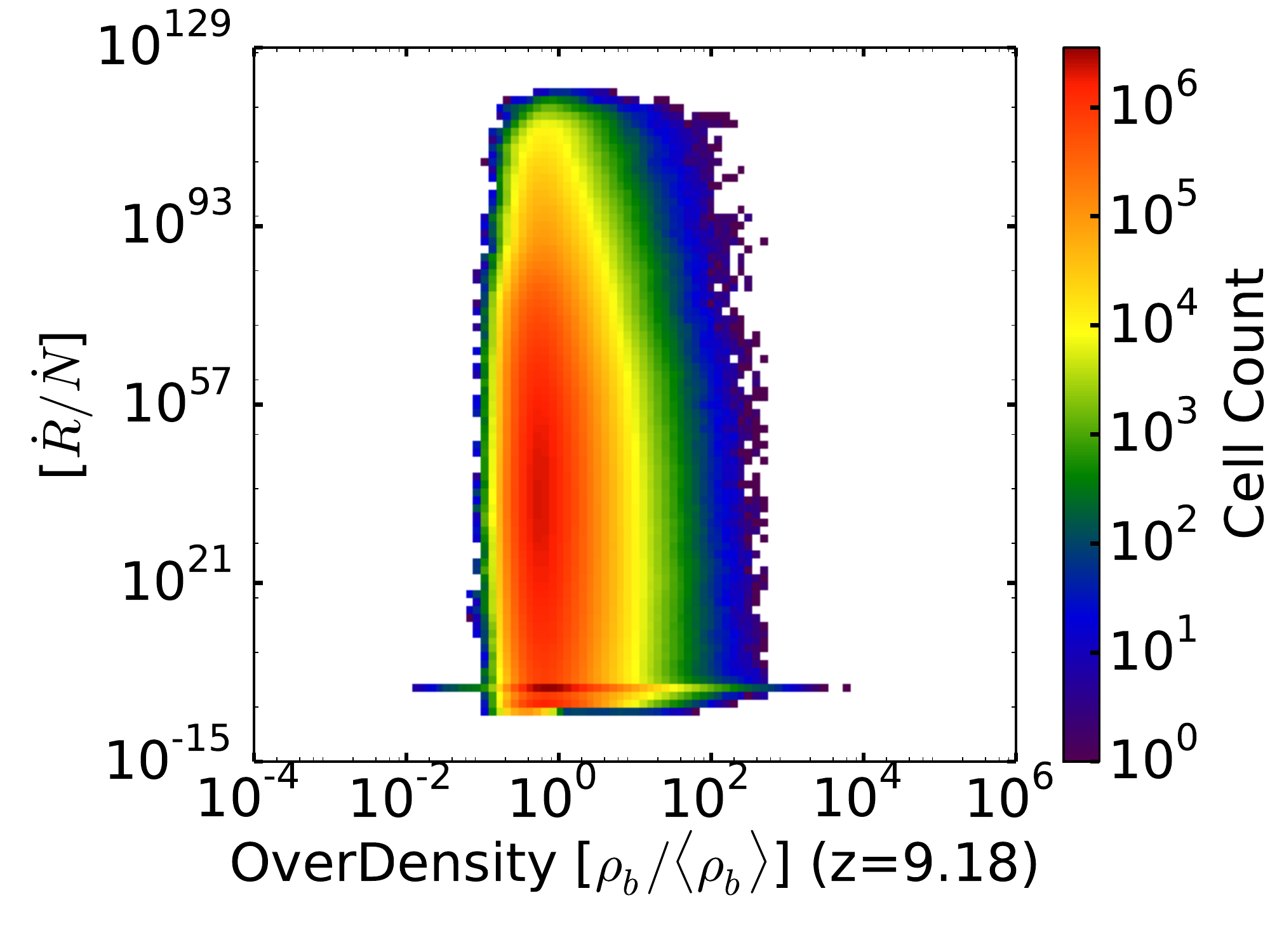}
     \end{minipage}
\hspace*{-2.00mm}
    \begin{minipage}[h]{0.33\linewidth}
       \centering
       \includegraphics[trim = 5mm 8mm 0mm 0mm, clip, width=1.0\textwidth]{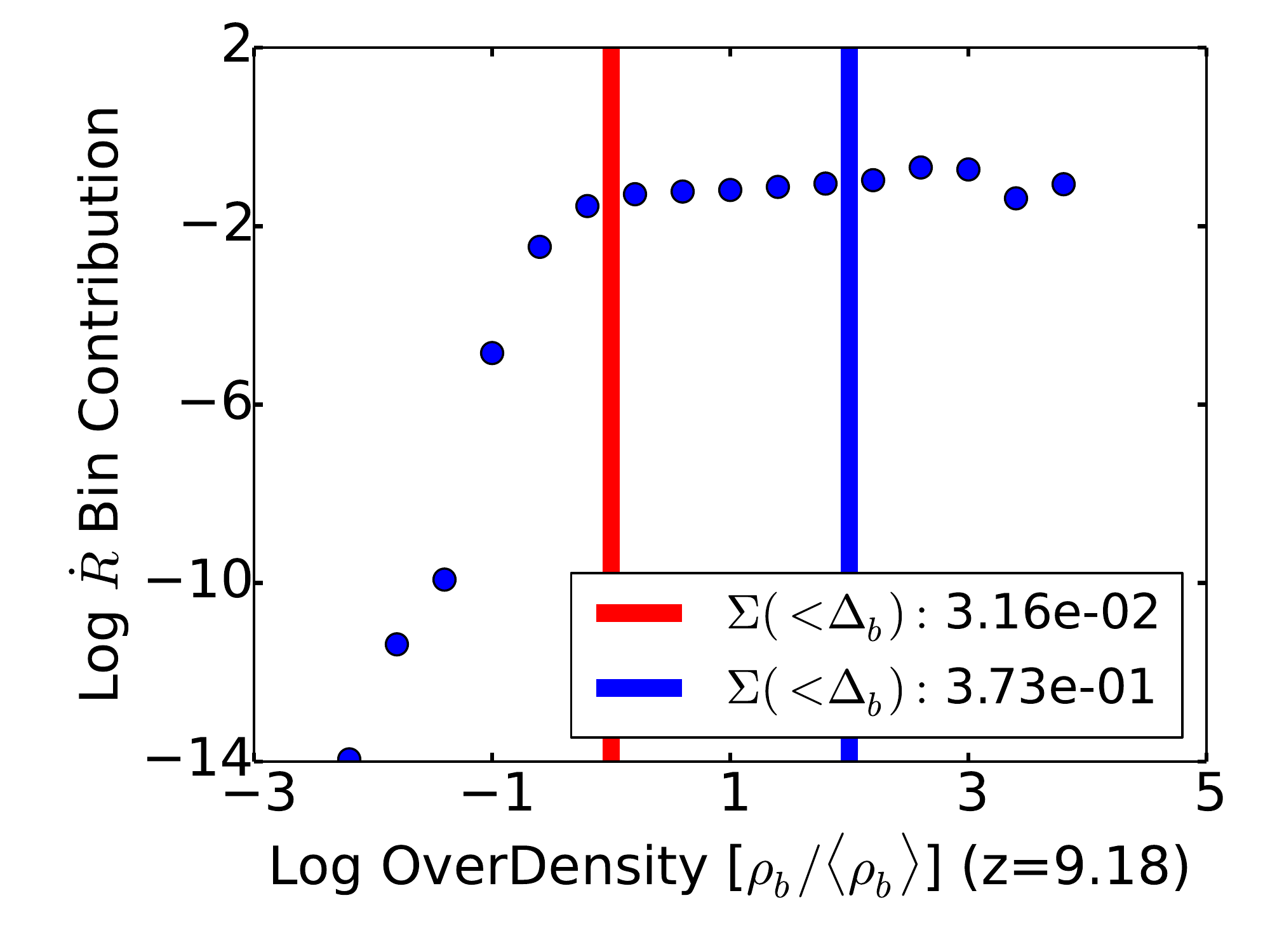}
     \end{minipage}
\hspace*{-4.00mm}
    \begin{minipage}[h]{0.33\linewidth}
       \centering
       \includegraphics[trim = 5mm 8mm 0mm 0mm, clip, width=1.0\textwidth]{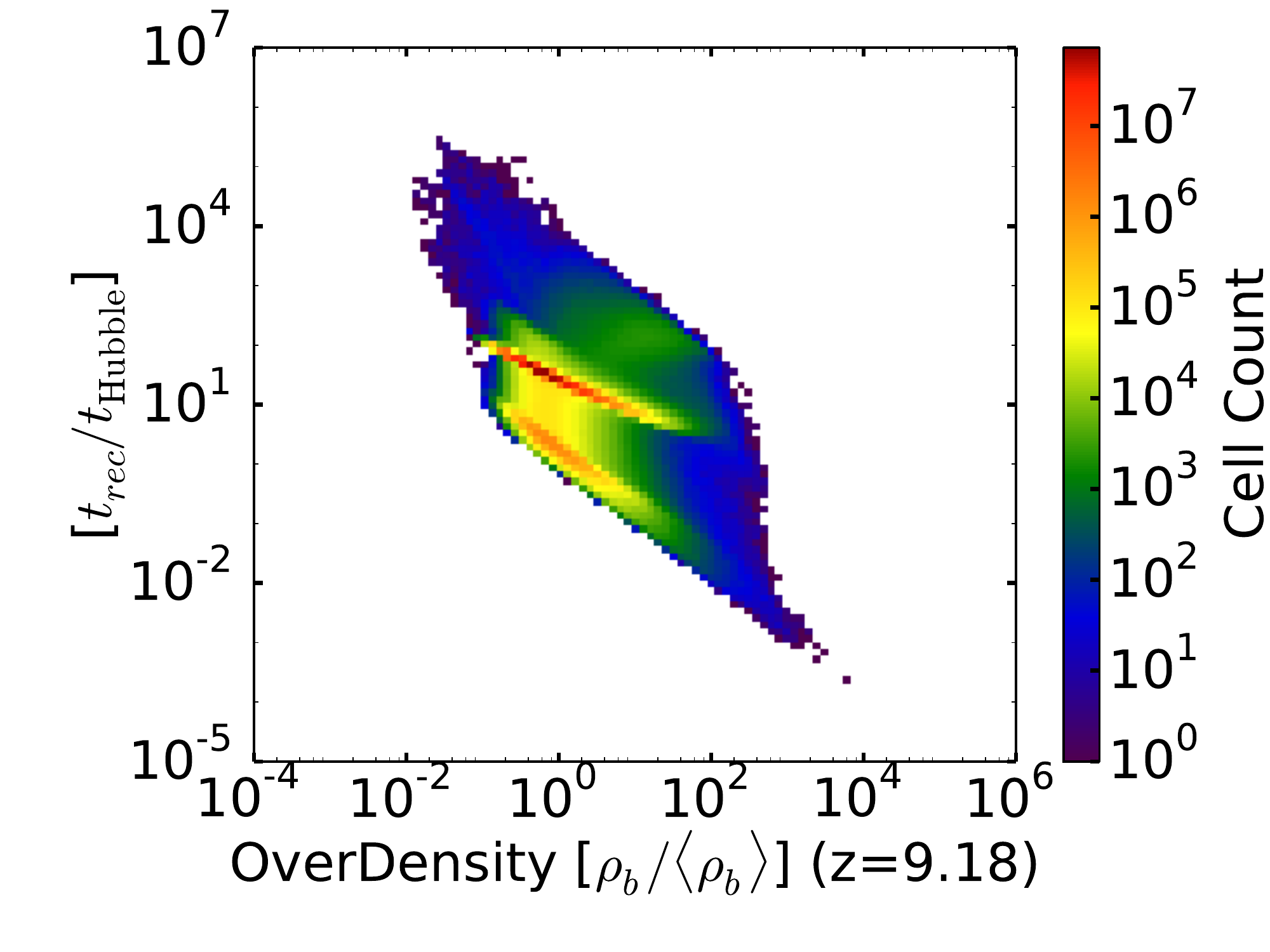}
    \end{minipage}
\\
     \begin{minipage}[h]{0.33\linewidth}
        \centering
        \includegraphics[trim = 5mm 8mm 0mm 0mm, clip, width=1.0\textwidth]{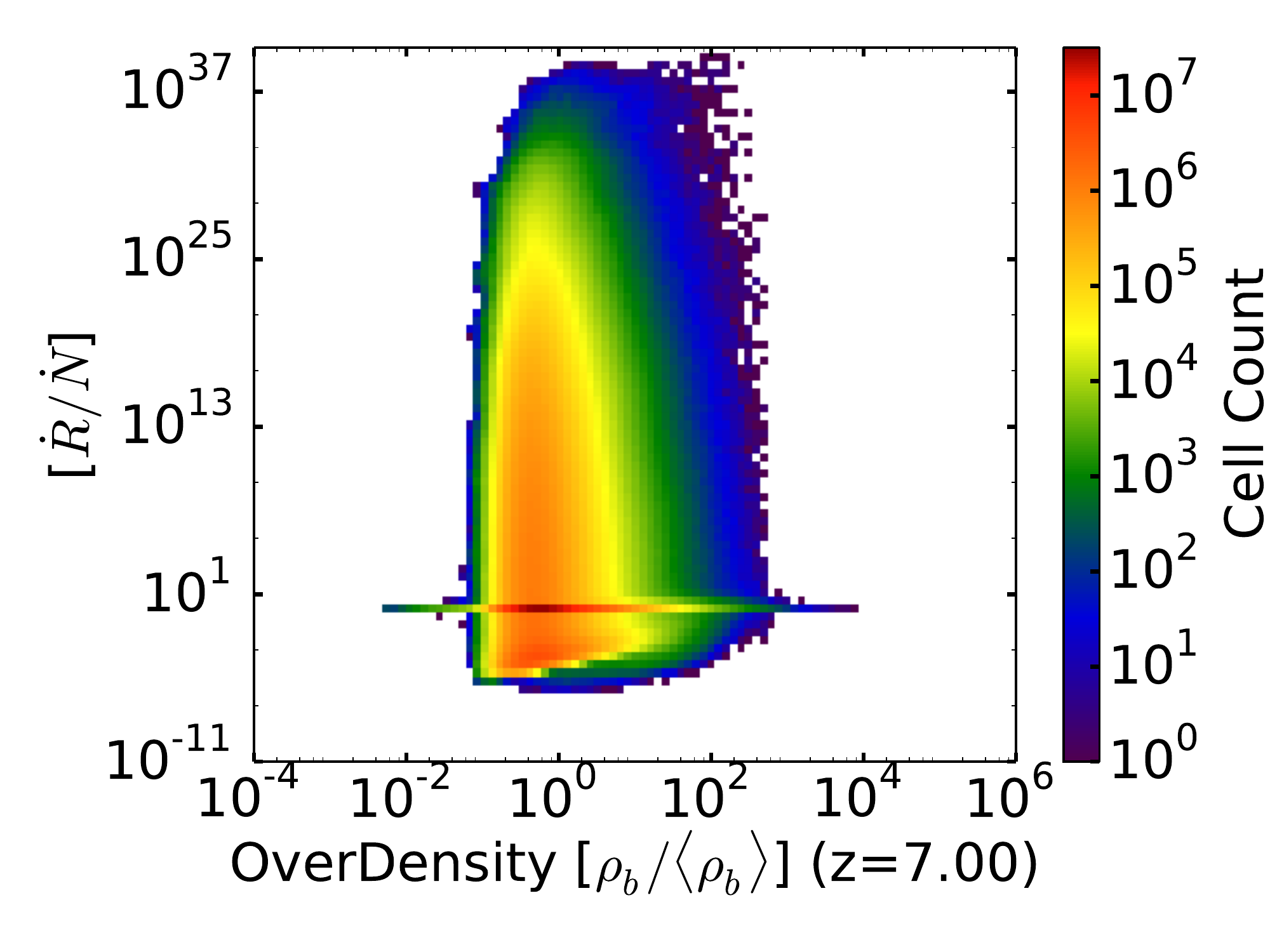}
     \end{minipage}
\hspace*{-2.00mm}
    \begin{minipage}[h]{0.33\linewidth}
       \centering
       \includegraphics[trim = 5mm 8mm 0mm 0mm, clip, width=1.0\textwidth]{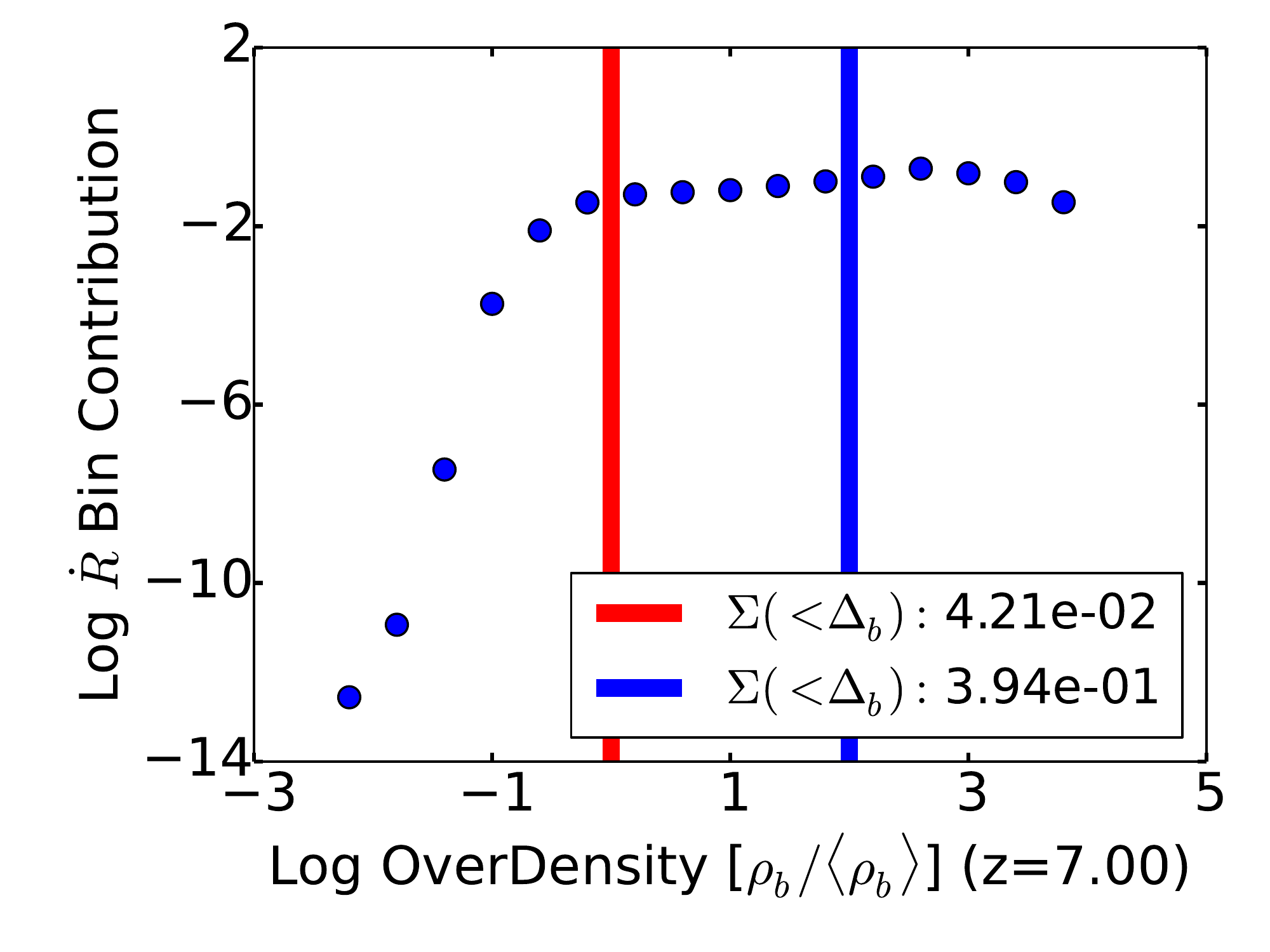}
     \end{minipage}
\hspace*{-4.00mm}
    \begin{minipage}[h]{0.33\linewidth}
       \centering
       \includegraphics[trim = 5mm 8mm 0mm 0mm, clip, width=1.0\textwidth]{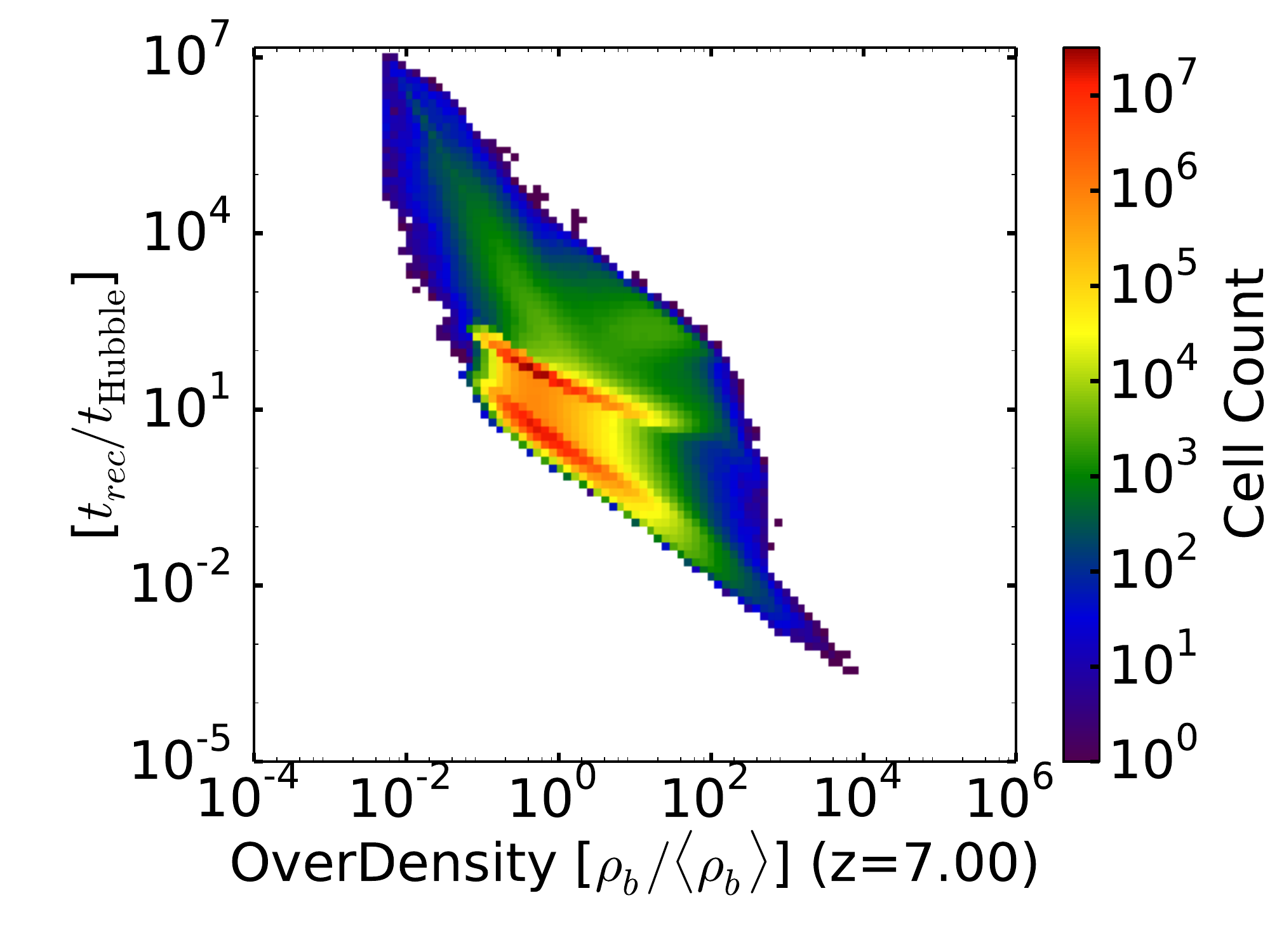}
    \end{minipage}
\\
     \begin{minipage}[h]{0.33\linewidth}
        \centering
        \includegraphics[trim = 5mm 8mm 0mm 0mm, clip, width=1.0\textwidth]{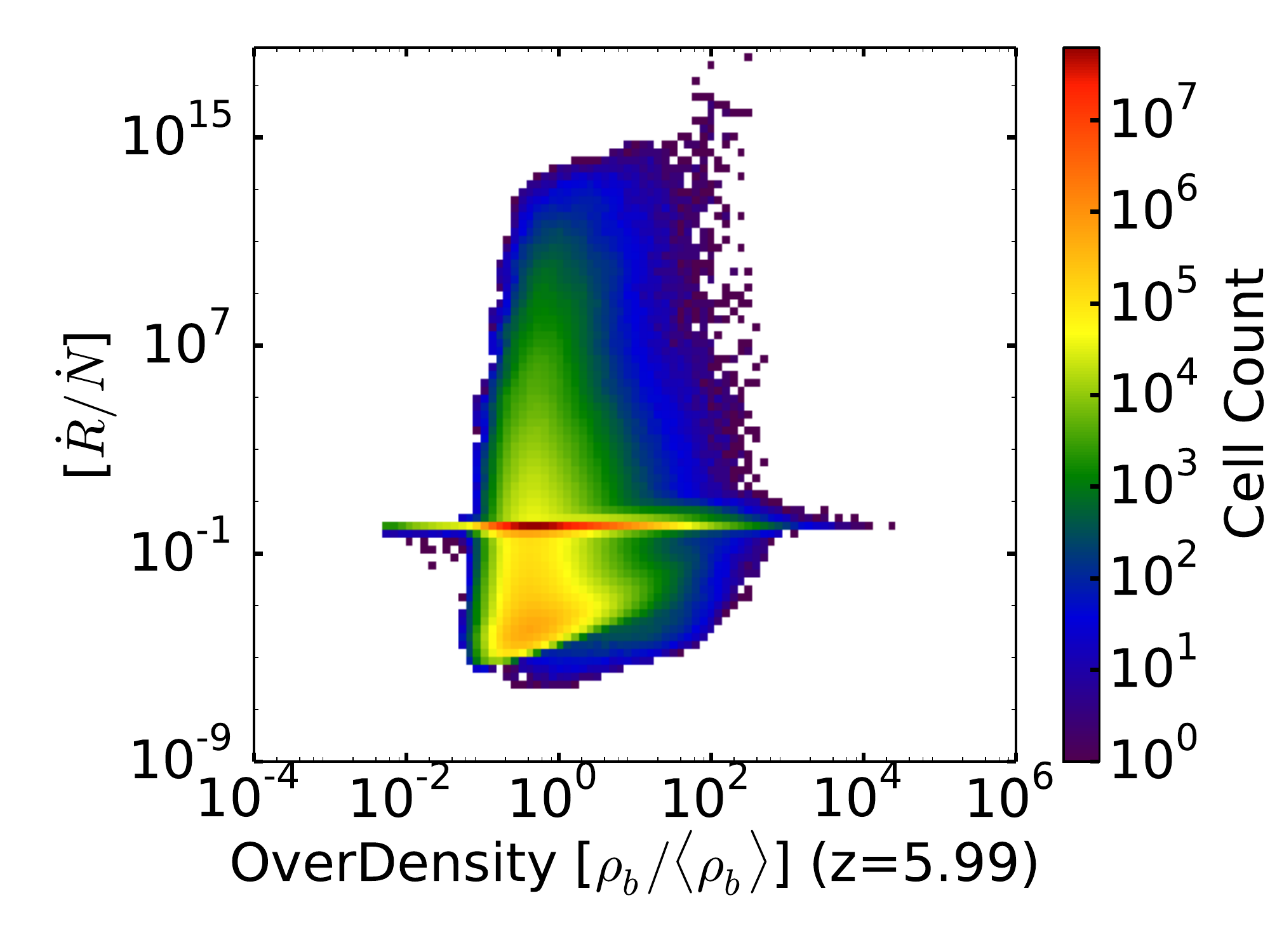}
     \end{minipage}
\hspace*{-2.00mm}
    \begin{minipage}[h]{0.33\linewidth}
       \centering
       \includegraphics[trim = 5mm 8mm 0mm 0mm, clip, width=1.0\textwidth]{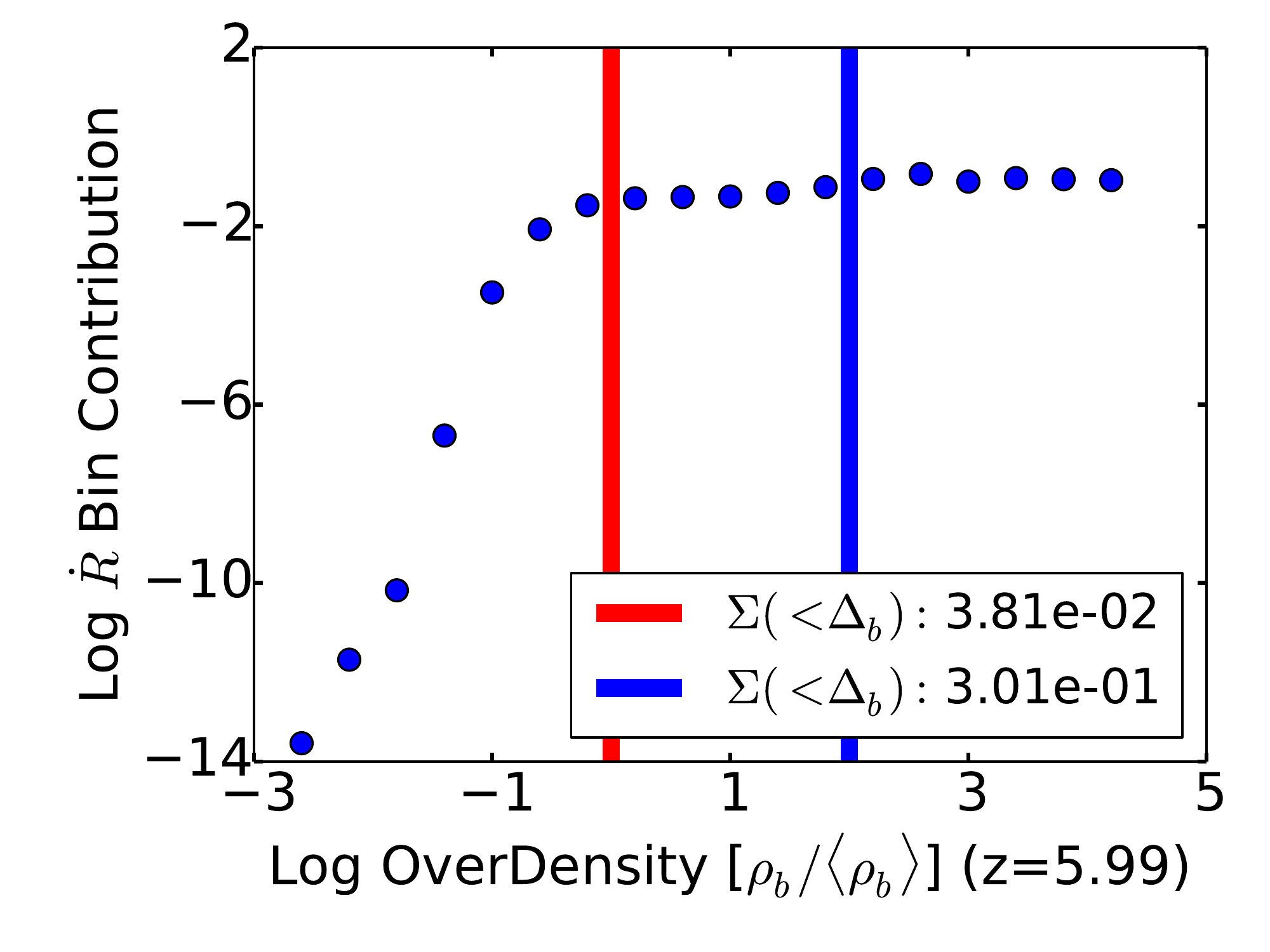}
     \end{minipage}
\hspace*{-4.00mm}
    \begin{minipage}[h]{0.33\linewidth}
       \centering
       \includegraphics[trim = 5mm 8mm 0mm 0mm, clip, width=1.0\textwidth]{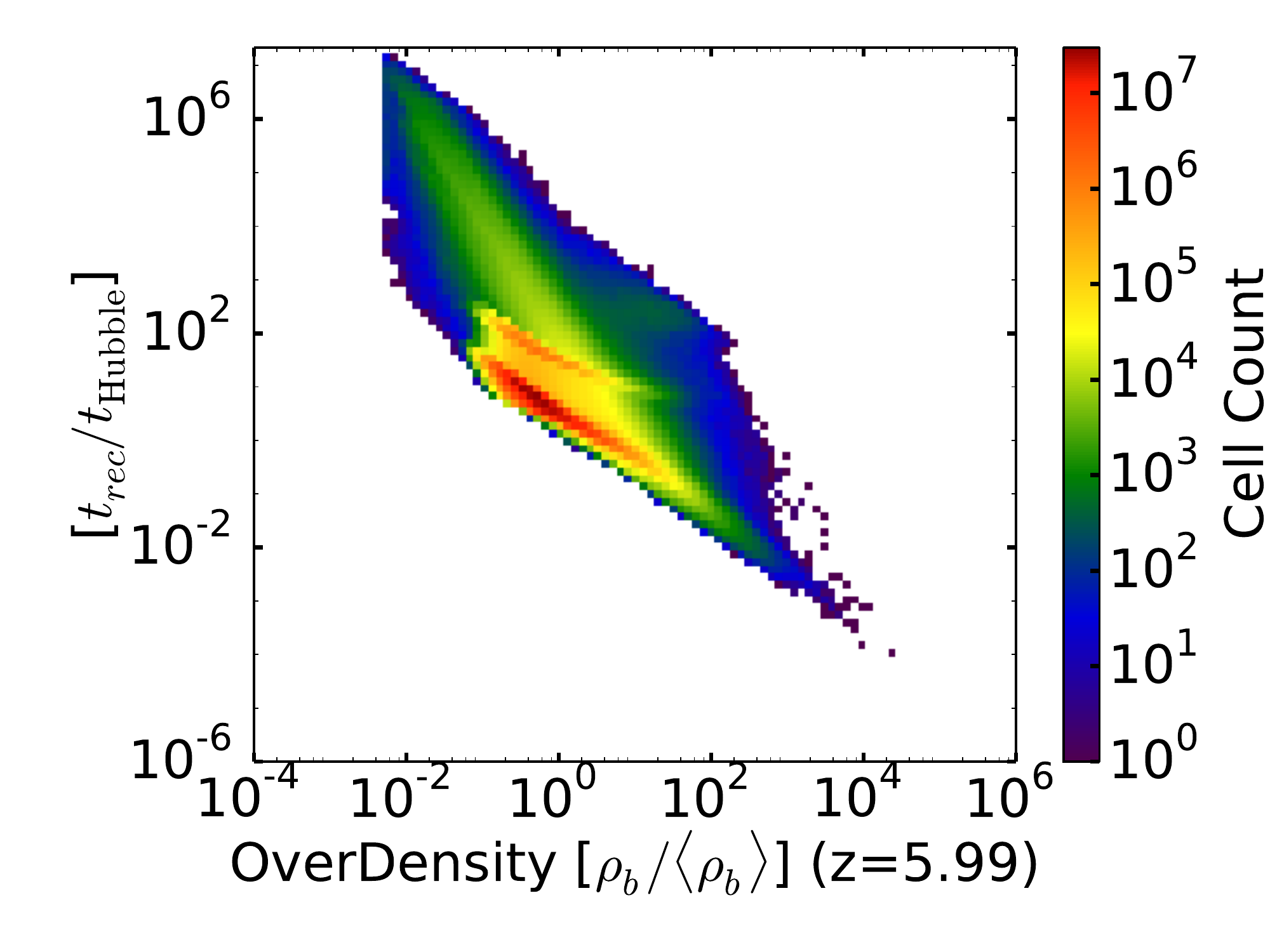}
    \end{minipage}
\\
     \begin{minipage}[h]{0.33\linewidth}
        \centering
        \includegraphics[trim = 5mm 8mm 0mm 0mm, clip, width=1.0\textwidth]{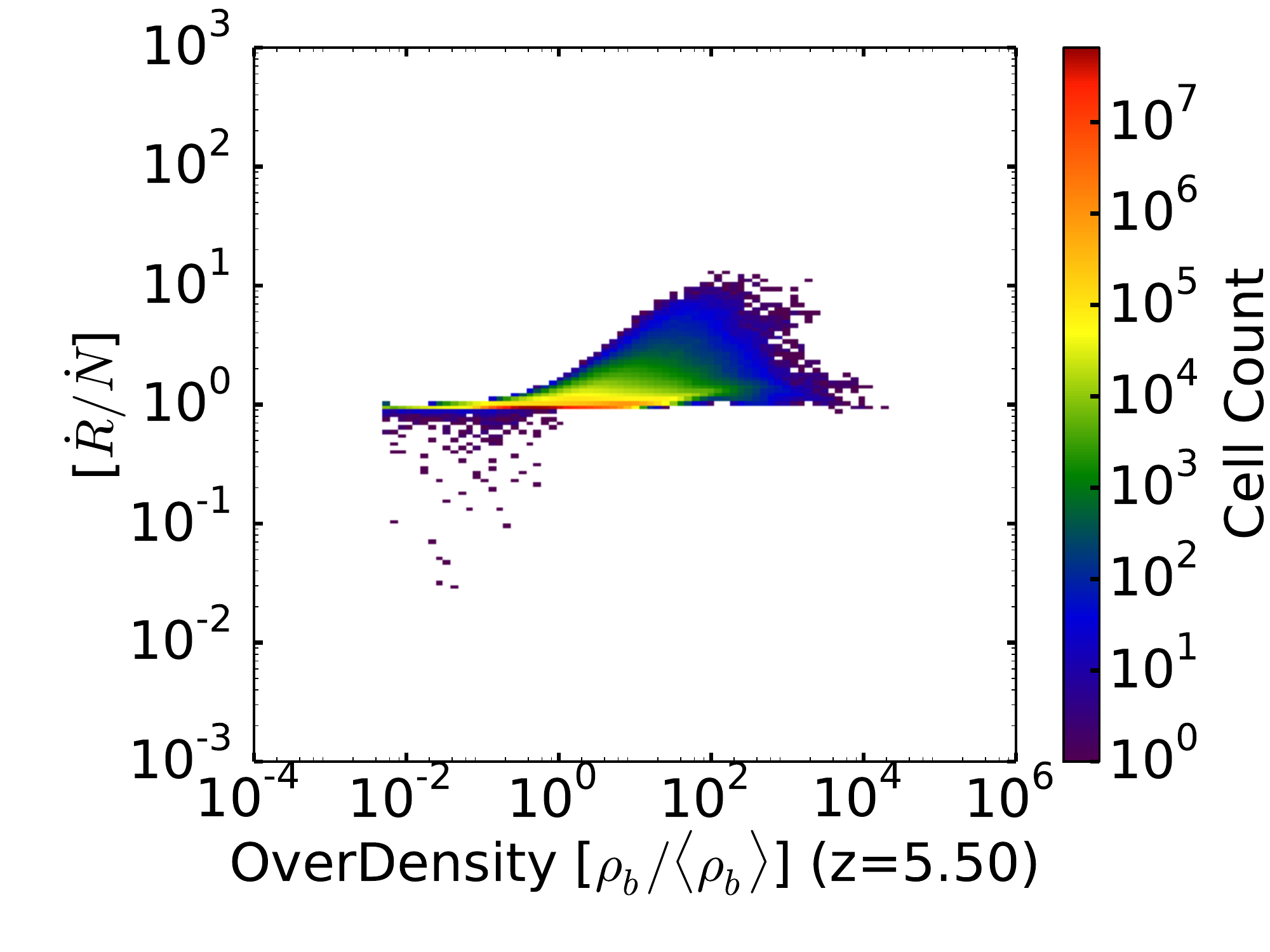}
     \end{minipage}
\hspace*{-2.00mm}
    \begin{minipage}[h]{0.33\linewidth}
       \centering
       \includegraphics[trim = 5mm 8mm 0mm 0mm, clip, width=1.0\textwidth]{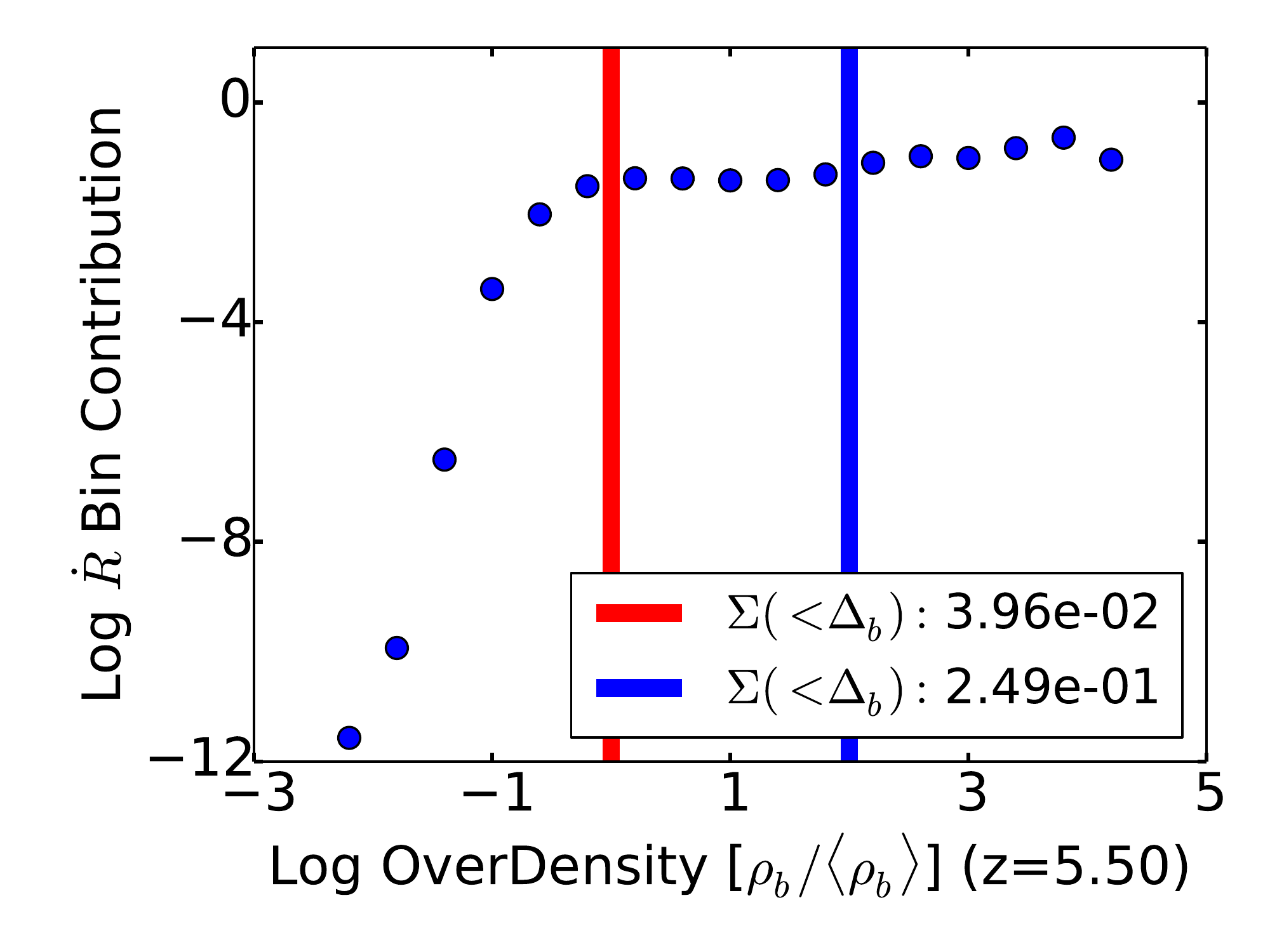}
     \end{minipage}
\hspace*{-4.00mm}
    \begin{minipage}[h]{0.33\linewidth}
       \centering
       \includegraphics[trim = 5mm 8mm 0mm 0mm, clip, width=1.0\textwidth]{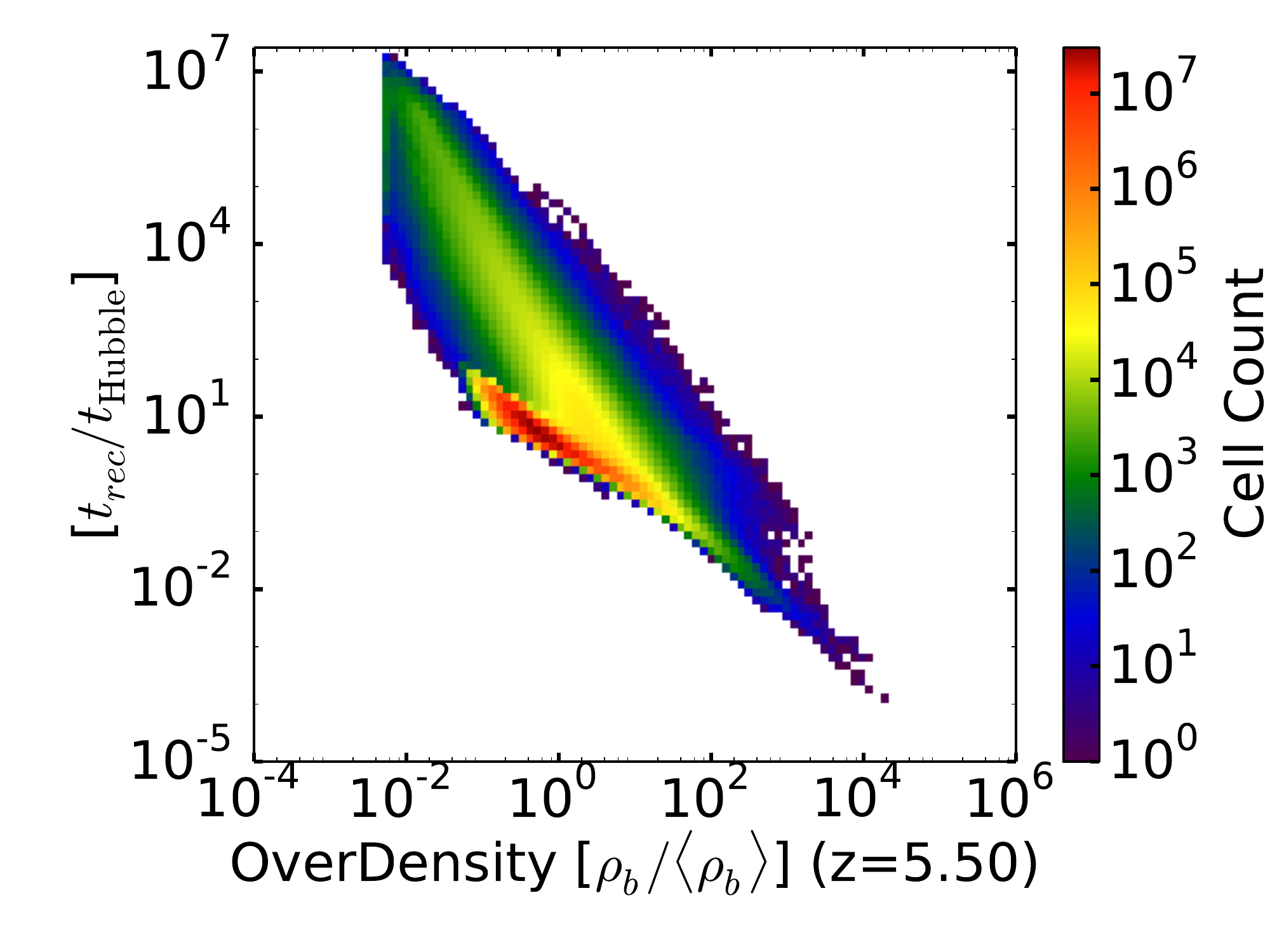}
    \end{minipage}
\\ 
     \begin{minipage}[h]{0.33\linewidth}
        \centering
        \includegraphics[trim = 5mm 8mm 0mm 0mm, clip, width=1.0\textwidth]{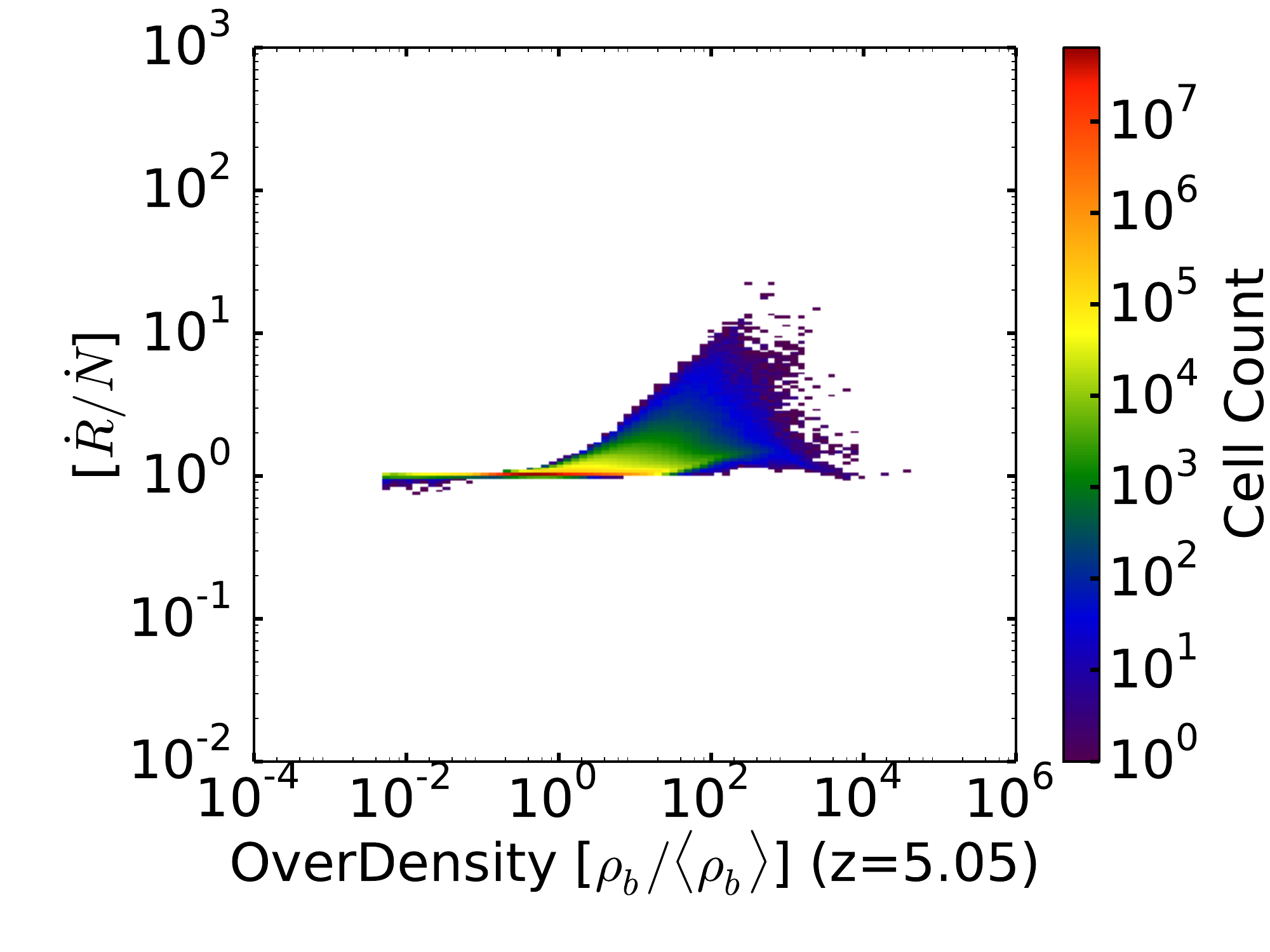}
     \end{minipage}
\hspace*{-2.00mm}
    \begin{minipage}[h]{0.33\linewidth}
       \centering
       \includegraphics[trim = 5mm 8mm 0mm 0mm, clip, width=1.0\textwidth]{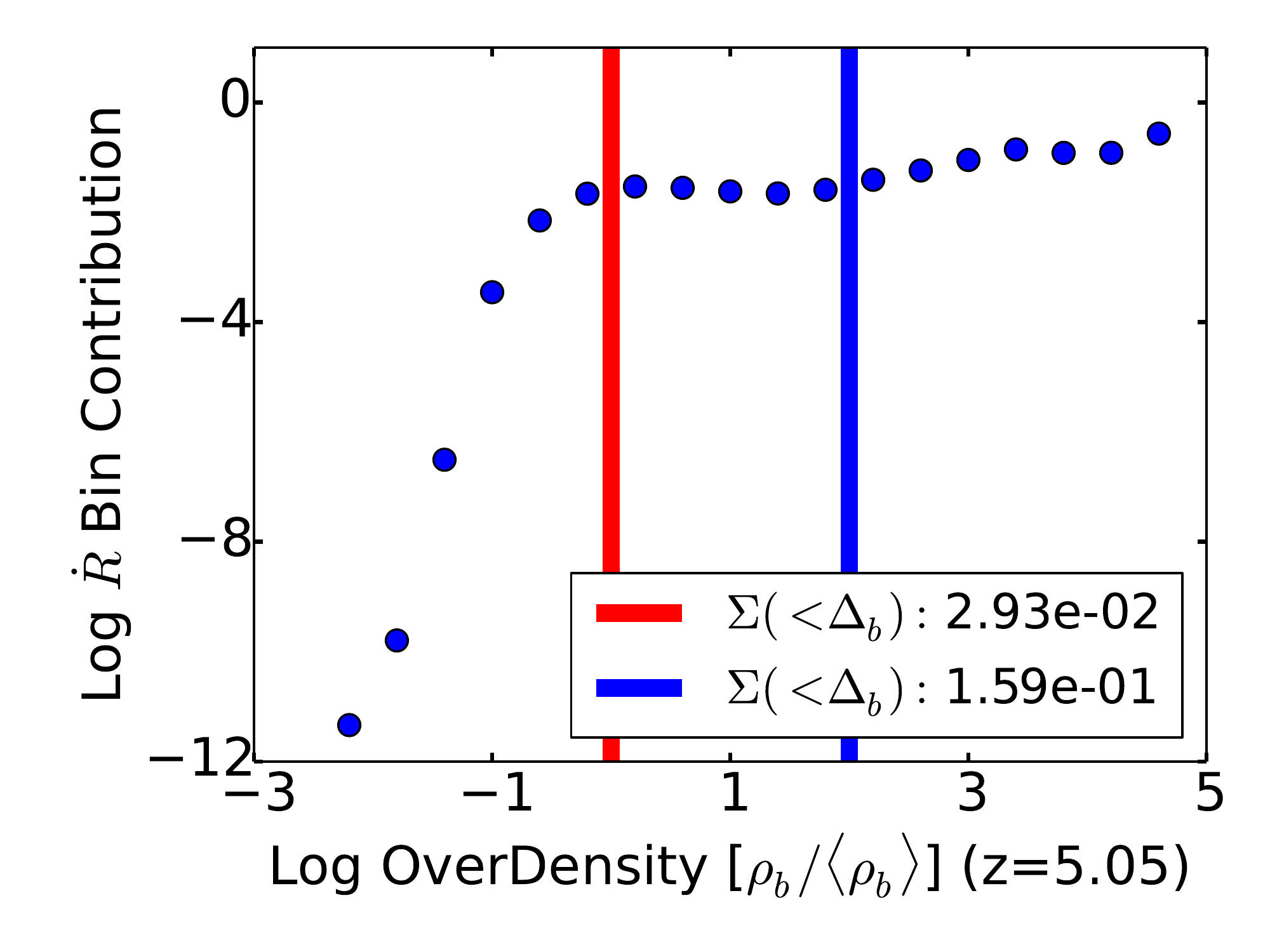}
     \end{minipage}
\hspace*{-4.00mm}
    \begin{minipage}[h]{0.33\linewidth}
       \centering
       \includegraphics[trim = 5mm 8mm 0mm 0mm, clip, width=1.0\textwidth]{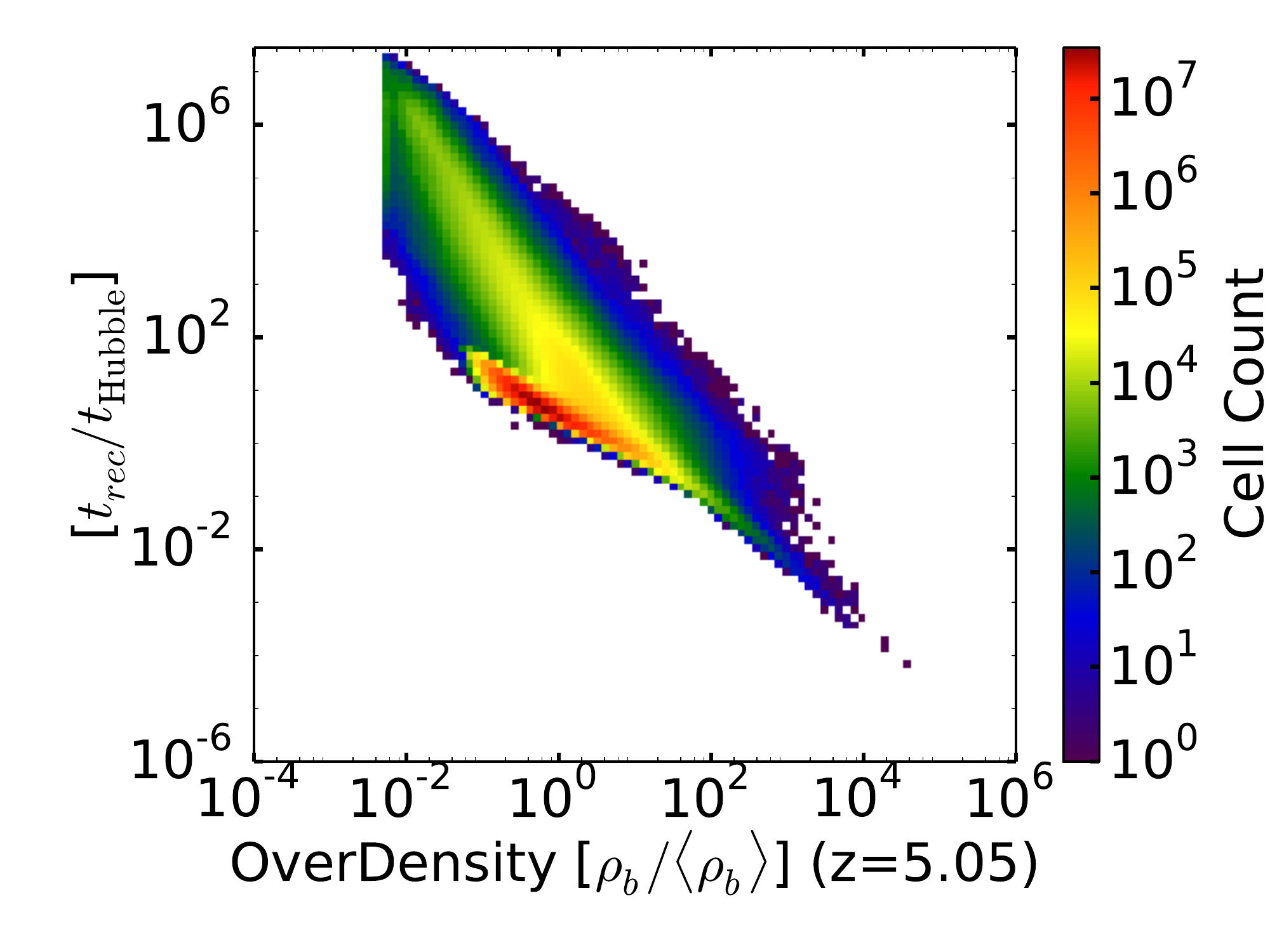}
    \end{minipage}
\\ 
    \caption{Quantifying recombination information.  Left column is a 2D distribution of recombination rate density divided by ionization rate density versus overdensity.  Middle column is plot relative bin contribution to the total recombination rate density versus overdensity bins.  The lines show the cumulative of all previous bins.  Blue line is at $\Delta_b$=100, red line is at $\Delta_b$=1.  Right column is plot of recombination time divide by Hubble time versus overdensity.  All three columns evolve with decreasing redshift from top to bottom.}
    \label{recomb}
\end{figure*}

To investigate the contribution of gas of different density to the total recombination rate density, we plot in Figure \ref{recomb}, three quantities dealing with recombinations in our simulation.  In the left column we have a 2D distribution plot of recombination rate density $\dot{R} = n_\mathrm{H\,II}n_e\alpha_B(T)$ divided by ionization rate density $\Gamma_\mathrm{H\,I}^{ph}n_\mathrm{H\,I}$ versus baryon overdensity $\Delta_b$, where  
\begin{align}
  \label{eq:photoionization}
  \Gamma_\mathrm{H\,I}^{ph} &= \frac{c E}{h} \left[\int_{\nu_\mathrm{H\,I}}^{\infty}
    \frac{\sigma_\mathrm{H\,I}(\nu) \chi_E(\nu)}{\nu}\,\mathrm d\nu \right]  \bigg /
  \left[\int_{\nu_\mathrm{H\,I}}^{\infty} \chi_E(\nu)\,\mathrm d\nu\right].
\end{align}
Here, $\sigma_\mathrm{H\,I}(\nu)$ and $\nu_\mathrm{H\,I}$ are the ionization cross section and ionization threshold for H {\footnotesize I}, respectively, and 
$h$ is Planck's constant (Paper I). In the middle column we plot the relative bin contribution to the total recombination rate density versus $\Delta_b$.  We draw vertical lines at $\Delta_b$=1 and 100, and in the legend box calculate the cumulative contribution to total reionizations to those thresholds.  In the right column, we plot the cell recombination time divided by the Hubble time versus $\Delta_b$.  All three columns evolve with descreasing redshift from top to bottom.

% ZahnEtAl2009 bubble threshold at x=0.9, ionized is when photon number > hydrogen number
% ShinEtAl2008 ionized when x>0.5

%\begin{figure*}[p]
%	\includegraphics[width=1.0\textwidth]{recomb.png}
%	\caption{Plot of recombination information.  Left column is a 2D distribution of recombination rate %density divide by ionization rate density versus over density.  Middle column is plot relative bin %contribution to the total recombination rate density versus over density bins.  The lines show the %cumulative of all previous bins.  Blue line is at $\Delta_b$=100, red line is at $\Delta_b$=1.  Right %column is plot of recombination time divide by Hubble time versus over density.  All three columns %evolve with descreasing redshift from top to bottom.}
%	\label{recomb}
%\end{figure*}

At $z\sim9$, in the left column of Figure \ref{recomb}, we see that even though there are regions of the volume that are in approximate ionization equilibrium (indicated by the horizontal distribution near 10$^0$), there is a wide distribution of cells far out of equilibrium, some even off by $\sim120$ orders of magnitude.  The middle column shows that about 37\% of all recombinations happen below a $\Delta_b$ of 100, and about 3.2\% happen below $\Delta_b$ of 1.  The phase diagram in the right column shows that there is a bimodal distribution of cells in terms of their recombination time normalized by Hubble time.  The top concentration of cells are more neutral, having long recombination times, and the lower concentration of cells are photoionized, having smaller recombination times.  The recombination time is lower for the ionized cells simply because there are more free electrons available to recombine with protons.  The blue cloud at low $\Delta_b$ and high $t_{rec}/t_\mathrm{Hubble}$ are the small number of cells that are shock heated to $T >$10$^6$K by supernova feedback.  Due to this high temperature, even though there are more free electrons their recombination times remain long.  

At $z\sim7$, more of the volume has reached the Well Ionized level, and we see the size of the out of equilibrium distribution shrink in the left column.  Now the maximum is only $\sim37$ orders of magnitude higher compared to equilibrium.  The middle column shows about 40\% of total recombinations are happening below $\Delta_b$ of 100, and about 4.2\% happens below $\Delta_b$ of 1.  In the right column, we see roughly equal numbers of cells in the upper (more neutral) distribution as compared to the lower (more ionized) distribution, whereas the top was much greater in numbers before.  As more cells become ionized to a high degree, their recombination time will decrease and their cell counts will shift to the lower distribution.

At $z\sim6$, looking at the left column, most of the cells are now in equilibrium.  This is indicated by the peak of the distribution in red, being near zero on the y-axis.  The maximum of the distribution is now less than 19 orders of magnitude apart from equilibrium.  The middle column showing 30\% to 3.8\% recombinations below $\Delta_b$ of 100 and 1, respectively.  The right column shows that the majority of the cells are now in the more ionized distribution and have a low recombination time.  This can be verified by looking at the same redshift in Figure \ref{NeutralPhase}, where most of the cells are at the Well Ionized level compared to fewer before.

At $z\sim5.5$, after the entire volume has become Well Ionized, and the vertical spread of the distribution has collapsed to about an order of magnitude away from equilibrium with the vast majority of the cells in equilibrium.  The fraction of recombinations are 25\% and 4\% below $\Delta_b$ of 100 and 1, respectively.  Looking at the recombination time to Hubble time, we no longer see the bimodal distribution of neutral cells and highly ionized cells, we only see the bottom distribution of highly ionized cells now.  The small distribution of shock heated gas is still present, but now seem more prominent with the absence of the neutral distribution.

%{\bf added revision and subsequent concluding paragraph, please check} 
At $z\sim5$, on the left column, the few cells that are in the low density void, which were recombining slower than ionizing are now all near equilibrium.  Cells that are higher in $\Delta_b$ are more likely to be above equilibrium.  In the middle column, we see the fraction of recombinations are 16\% and 2.9\% for region below $\Delta_b$ of 100 and 1, respectively.  Not much has  changed in the recombination time column except there are fewer cells above the $\Delta_b$ of 10$^4$, possibly due to effect of Jeans smoothing.

We see that there is no real one-to-one correspondence between overdensity and the quantities we show on the y-axis.  That is because in a given panel, we are only seeing two dimensions of a multidimensional physical process that depends on locality to sources of radiation, the behavior of said sources at a given moment, the local density of neutral and ionized gas, temperature, among others.  It is helpful to speak about the average behavior in any given overdensity as we have done, but we should always keep in mind that the average may not be as representative of the wider distribution as we may think.

%\subsubsection{Thresholded Values}
%\label{ThresholdedValues}
\subsection{Investigating Thresholded Clumping Factor Analyses}
\subsubsection{Excluding Halos}
\label{ExcludingHalos}

We saw in \S\ref{Madau} that using the unthresholded H {\footnotesize II} density field to calculate $C$ via Equation \eqref{eq:clumpingfactor} yields a reasonably good estimate of when reionization completes (Figure \ref{unthresholded}). This is perhaps not surprising since we count every ionizing photon emitted and every recombination to the accuracy of Equation \eqref{eq:updatedNdot}. Possible sources of disagreement between theory and simulation are: (1) inaccuracies in estimating the recombination rate density using Equation \eqref{eq:updatedNdot}; (2) breakdown of the ``instantaneous approximation'' used to derive Equation \eqref{eq:updatedNdot} due to history-dependent effects; (3) finite propagation time for I-fronts to cross voids; and (4) numerical inaccuracies. Regarding possibility (4) we note that our mathematical formalism is photon conserving, and that our I-front tests in Paper I show that I-fronts propagate at the correct speed, which is an indication that numerical photon conservation is good. 

To investigate whether improved estimates of the recombination rate density will improve the agreement, we follow the practice of some recent investigators \citep{PawlikEtAl2009, RaicevicTheuns2011} and threshold out dense gas bound to halos, leaving only the diffuse IGM to consider. The motivation for this is that since we are only interested in the photon budget required to maintain the diffuse IGM in an ionized state, by excluding the complicated astrophysics within halos we have a simpler problem to model and resolve numerically. 

\begin{figure}
	\includegraphics[width=0.5\textwidth]{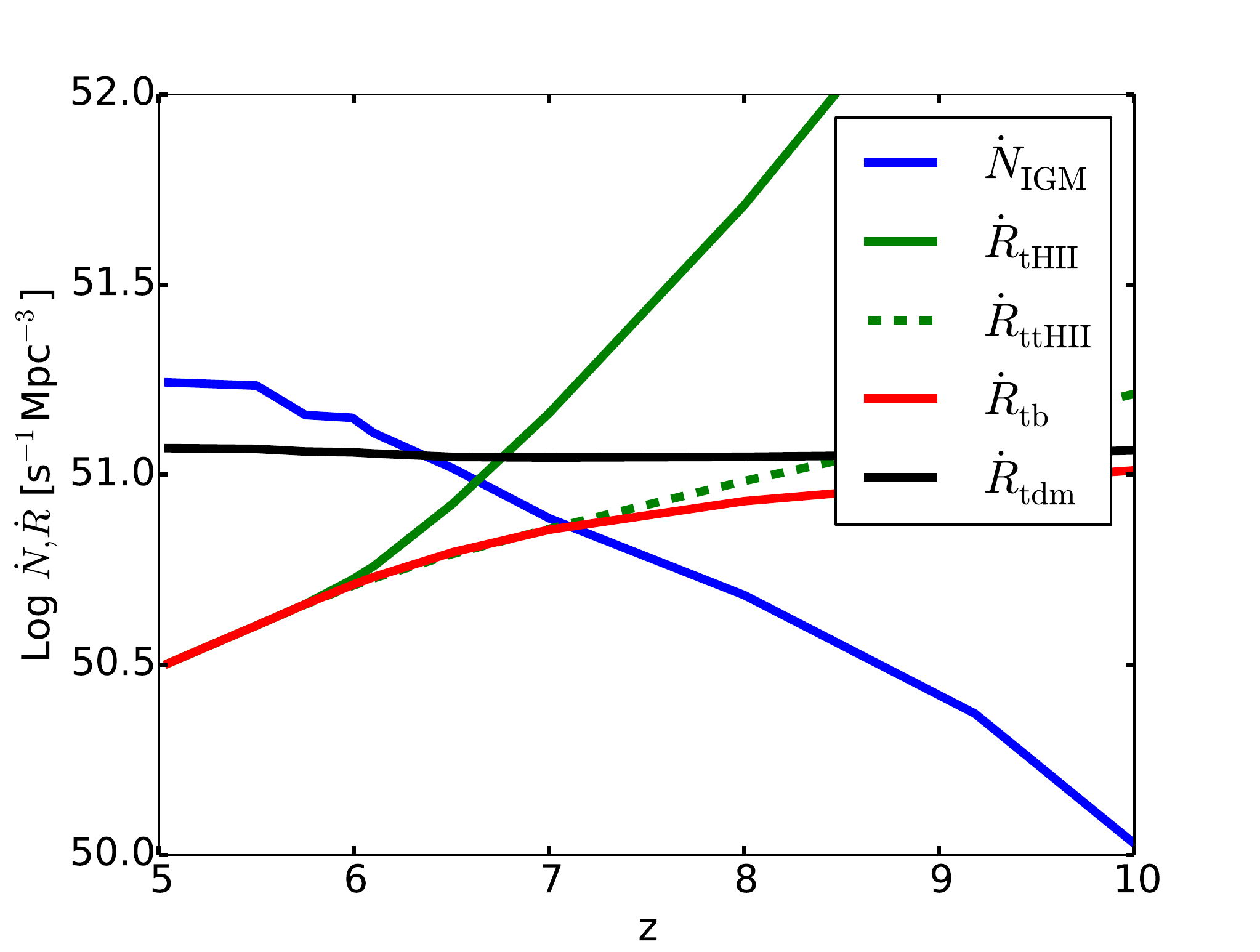}
	\caption{Same quantities as Figure \ref{unthresholded}, except now the ``$\dot{N}_\mathrm{IGM}$'' curve is the number of ionizing photons which escape into the IGM (see \S\ref{escape}). The recombination rate densities with a subscript that begins with ``t" are calculated as described in the caption for Figure \ref{unthresholded}, except that the clumping factors are computed excluding regions satisfying $\Delta_b > 100$. The curve labelled $\dot{R}_{ttHII}$ is calculated from Equation (22) using the doubly-thresholded clumping factor $C_{ttHII}$ defined in Figure \ref{threshclumping}.}
	\label{thresholded}
\end{figure}

%We wondered, as the people that apply thresholding in calculating the clumping factor, what would change if the same thresholding is applied to Figure \ref{unthresholded}?  
To proceed we must calculate the ionization and recombination rate densities outside of collapsed objects. We estimate the number of ionizing photons escaping halos by multiplying  $\dot{N}_{sim}(z)$ by a global escape fraction $\bar{f}_{esc}(z)$ derived in \S\ref{escape} and plotted in Figure \ref{RadEscFraction}:
\begin{equation}
	\dot{N}_{IGM}(z)=\bar{f}_{esc}(z)\dot{N}_{sim}(z)
	\label{eq:ndot_igm}
\end{equation}
\noindent
The recombination rate density outside of halos is calculated using Equation \eqref{eq:updatedNdot} where now the clumping factor is thresholded such that only cells for which $\Delta_b < 100$ contribute to the sum. As in Figure \ref{unthresholded} we plot three curves for the recombination rate density calculated using Equation \eqref{eq:updatedNdot} using H {\footnotesize II}, baryons, and dark matter density fields. These are plotted in Figure \ref{thresholded} as green, red, and black curves, respectively. We see that the recombination rate density based on the singly thresholded H {\footnotesize II} (labeled $\dot{R}_\mathrm{tH\,II}$) and on the thresholded dark matter (labeled $\dot{R}_\mathrm{tdm}$) curve cross the ionizing emissivity curve labeled ``$\dot{N}_\mathrm{IGM}$'' at $z \approx 6.7$ in Figure \ref{thresholded}, whereas the thresholded baryon density curve (labeled $\dot{R}_{tb}$) crosses ``$\dot{N}_\mathrm{IGM}$'' at $z \sim 7.2$. Taking the doubly-thresholded H {\footnotesize II} curve as the best estimate for the recombination rate density, we find that restricting the analysis to only IGM gas yields poorer agreement than the simpler, global model of Madau, which at first blush is a perplexing result. By thresholding out the gas in galaxies we have isolated the thing we care about: the ionization balance of the IGM. Why then should the implied redshift of reionization completion become worse compared to the analysis in \S\ref{Madau}? We defer addressing this question until later sections. 

%. we plot the fraction of ionization rate density in the simulation that happens at $\Delta_b<100$ region (outside of collapsed, self shielded region) as the curve ``tIon'' in Figure \ref{EscFraction} (we will discuss Figure \ref{EscFraction} in more detail in \S\ref{ConsistencyCheck}).  This redshift dependent curve ``tIon'', we refer to as our escape fraction.  Second, we take the original ``FromSim'' curve from Figure \ref{unthresholded}, which describes the total amount of ionizing photon production rate density, and we multiply that by the escape fraction curve, the result is ``$\dot{N}_\mathrm{IGM}$'' curve in Figure \ref{thresholded}.  We calculate the ionization escape fraction using the same methodology as the recombination fraction (the same way as blue line values in the middle column of Figure \ref{recomb}).

\begin{figure}
	\includegraphics[width=0.5\textwidth]{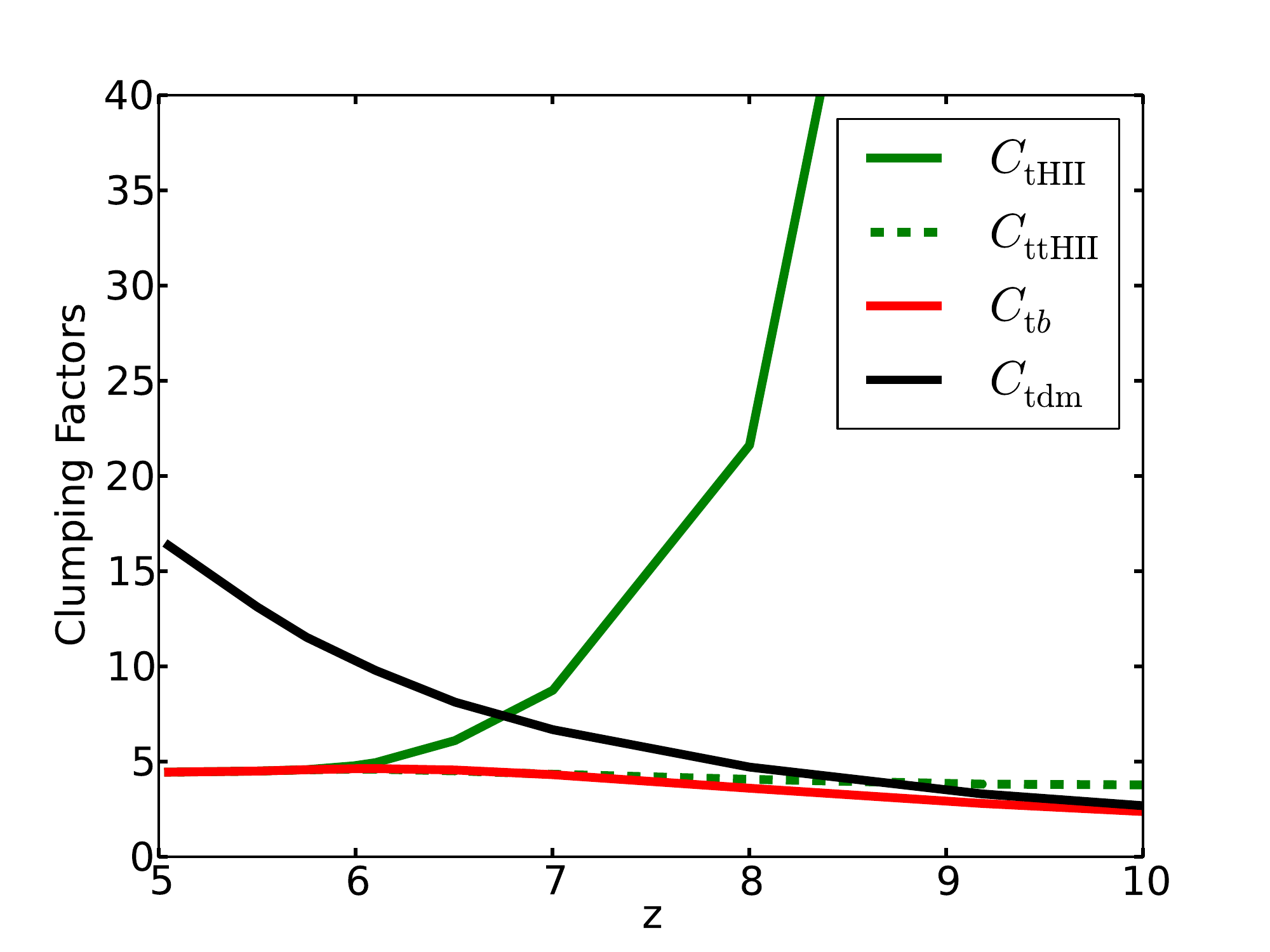}
	\caption{Thresholded clumping factors used in Fig. \ref{thresholded}. $C_{tHII}, C_{tb}, C_{tdm}$ are calculated using thresholded H II, baryon, and dark matter density fields, respectively, where only cells satisfying $\Delta_b < 100$ contribute. $C_{ttHII}$ is calculated from the H II density where only cells satisfying $\Delta_b < 100$ and $f_i > 0.1$ contribute.}
	\label{threshclumping}
\end{figure}

\begin{figure}
	\includegraphics[width=0.5\textwidth]{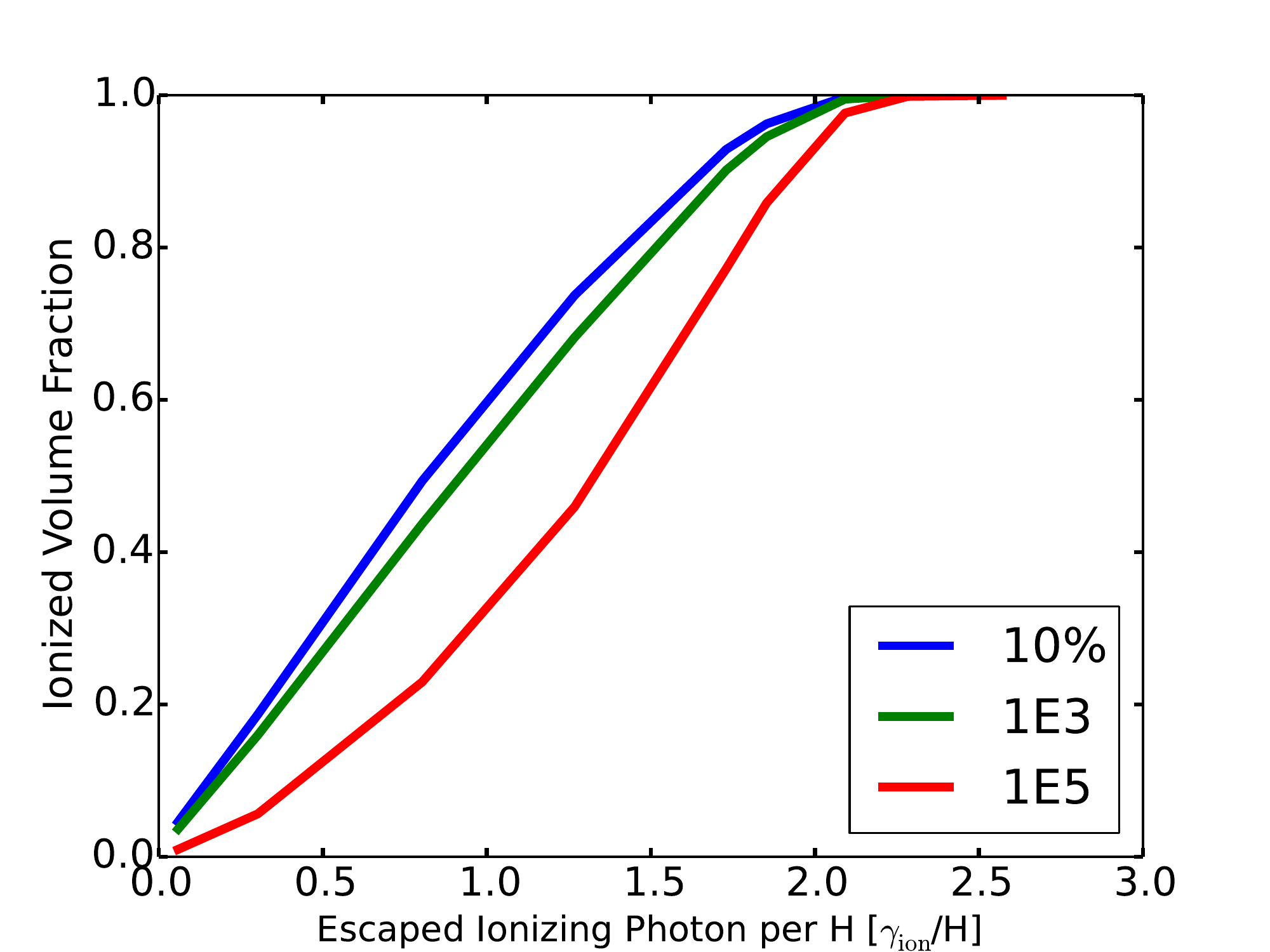}
	\caption{Ionized volume fraction as a function of the number of ionizing photons emitted per H atom averaged over the entire simulation volume (excluding gas inside halos) for three different ionization levels: $f_i \geq 0.1$ (blue line);  $f_i \geq 0.999$ (green line); $f_i \geq 0.99999$ (red line). Compare with Fig. \ref{unthreshphotonbudget} which includes gas inside halos.}
	\label{threshphotonbudget}
\end{figure}

Finally, we ask how many ionizing photons per H atom are required to convert the neutral gas residing outside halos to a well ionized state. We repeat the analysis of Figure \ref{unthreshphotonbudget} and show the result in Figure \ref{threshphotonbudget}.  We see that the effect of counting only escaped photons on the photon budget is significant.  Previously, we summed $\dot{N}_{sim}(z)$ and divided by the total number of hydrogen atoms in the simulation volume, and used that as our progress variable. 
%Now we multiply the photon production by the rate escape fraction toon each time interval, to consider only the photons escaped, and divide by the number of hydrogen atoms in the simulation.  
In Figure \ref{threshphotonbudget} we sum $\dot{N}_{IGM}(z)$ and divide by the number of hydrogen atoms in the thresholded volume, and use that as our progress variable. 
Instead of needing $\sim$4 to ionize the IGM, now we only need $\sim$2 photons per hydrogen atom for 99.9\% of the universe to reach Well Ionized level. This result supports the ``photon starved'' reionization scenario discussed by \cite{BoltonHaehnelt2007}.  

%\begin{figure}
%	\includegraphics[width=0.5\textwidth]{fig14.eps}
%	\caption{}
%	\label{}
%\end{figure}

%\subsubsection{Restricting Methods}
%\label{RestrictingMethods}
\subsubsection{Including Temperature Corrections}
\label{IncludingTemperatureCorrections}

During the preparation of this paper, a new way of estimating the recombinations in the IGM appeared in the literature.  The authors \citep{ShullEtAl2012,FinlatorEtAl2012} reformulated the expression for the clumping factor taking the temperature dependence of the recombination rate into account. We briefly investigate their methods here.  In order for the calculation of the clumping factor to take only IGM gas that is ionized but recombining, several additional thresholds were applied.  Equation (15) in \cite{ShullEtAl2012} is a new expression for the clumping factor, similar in form to \cite{Gnedin2000},
\begin{equation}
	C_\mathrm{RR}=\frac{\langle n_e n_\mathrm{H\,II}\alpha_B(T) \rangle}{\langle n_e \rangle \langle n_\mathrm{H\,II} \rangle \langle \alpha_B(T) \rangle}
	\label{eq:CRR}
\end{equation}
with the following thresholds applied: 1$<\Delta_b<$100, 300K$<$$T$$<10^5$K, Z$<10^{-6}$Z$_\odot$, $x_e$$>$0.05.  Here,  Z is metalicity and $x_e$ is the ionized fraction.  The reason that a lower limit threshold is applied to the baryon overdensity, the authors argued, is because very little recombinations happen there, due to the low density.  \cite{ShullEtAl2012} also provide a new formulation for ionizing photon rate density that uses this definition of the clumping factor, in their Equation (10),
\begin{align}
	\frac{dN}{dt}=4.6\times 10^{50}\mathrm{s}^{-1}\mathrm{Mpc}^{-3}\notag\\
\times \(\frac{(1+z)}{8}\)^3 T_4^{-0.845}\(\frac{C}{3}\)
	\label{eq:ShullNdot}
\end{align}
Here, T$_4$ is mean IGM temperature measured in units of 10$^4$K.  

\begin{figure}
	\includegraphics[width=0.5\textwidth]{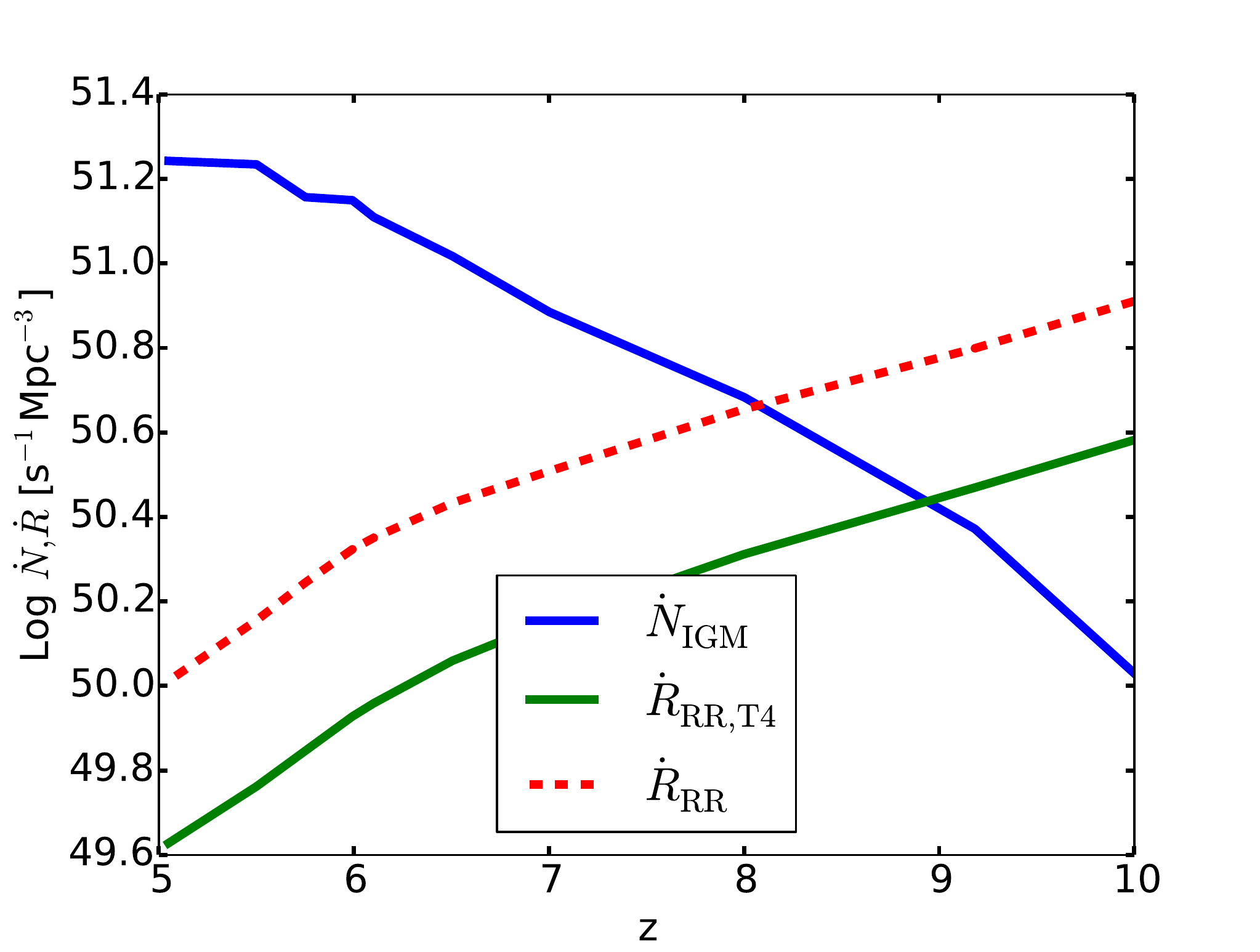}
	\caption{Ionizing photon injection rate density in the IGM from the simulation $\dot{N}_{IGM}$ versus the predictions of Equation \eqref{eq:ShullNdot}, evaluated with two choices for the clumping factor which take temperature corrections into account.   The curve labeled ``$\dot{R}_\mathrm{RR,T4}$'' is from Equation \eqref{eq:ShullNdot}, with T$_4$ being the average temperature in C$_\mathrm{RR}$ region in units of 10$^4$K.  The curve ``$\dot{R}_\mathrm{RR}$'' is calculated the same way as $\dot{R}_\mathrm{RR,T4}$ except now T$_4$ is set to 1 in Equation \eqref{eq:ShullNdot}, for an effective IGM temperature of 10$^4$K.}
	\label{Shull}
\end{figure}

Equation \eqref{eq:ShullNdot} is proposed as an improvement over Equation \eqref{eq:ndot}. To see if this is the case we used our data to evaluate the clumping factor C$_\mathrm{RR}$ and then used Equation \eqref{eq:ShullNdot} to calculate ionizing photon rate density versus redshift needed to maintain an ionized IGM. The result is shown in Figure \ref{Shull}. The curve labeled $\dot{R}_\mathrm{RR,T4}$ in green uses the average temperature, in units of 10$^4$K, of the region that satisfies the C$_\mathrm{RR}$ thresholds for T$_4$ in Equation \eqref{eq:ShullNdot}.  The curve $\dot{R}_\mathrm{RR}$ uses 1 in place of T$_4$ in Equation \eqref{eq:ShullNdot}, essentially fixing the IGM temperature to a constant 10$^4$K.  The green curve is lower than the red curve because the average temperature in the simulation is higher than $10^4$K. The blue curve labeled $\dot{N}_{IGM}$  is as defined previously. We see that Equation \eqref{eq:ShullNdot} predicts that reionization completes at significantly higher redshifts than exhibited by the simulation, calling into question the validity of the analysis. 

\begin{figure}
	\includegraphics[width=0.5\textwidth]{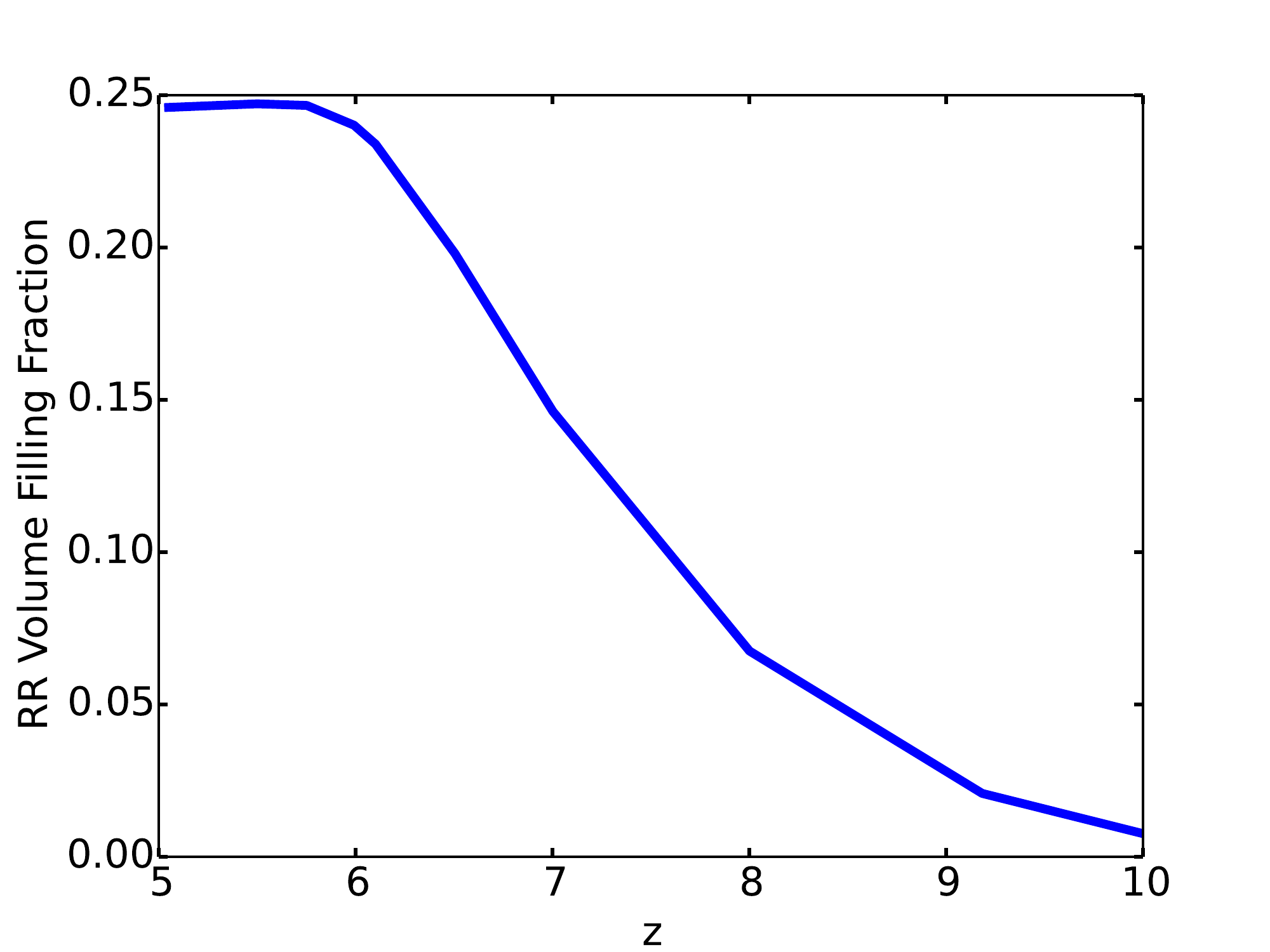}
	\caption{Evolution of the volume filling fraction with redshift of regions satisfying the C$_\mathrm{RR}$ thresholding criteria.}
	\label{volumefracCRR}
\end{figure}

We find it curious that as the clumping factor analysis is refined through physically well-motivated modifications, it yields predictions for the redshift of reionization completion that become worse and worse, moving to higher redshift rather than lower redshift. This suggests that there is something fundamentally wrong with the whole approach, and that the seemingly good agreement found in \S\ref{Madau} was fortuitous. One worrisome aspect about the utility of Equation \eqref{eq:ShullNdot} is that the fraction of simulation volume included in the C$_\mathrm{RR}$ thresholds is actually quite small. This is illustrated in Figure \ref{volumefracCRR}. The included volume grows from 3\% at $z=9$ to only 23\% of the simulation volume by overlap. One wonders about the validity of making global statements about reionization based on such a restricted sample of the IGM. It is also unclear how we should interpret the redshift at which lines across in Figure \ref{Shull}. Should we interpret it as the redshift below which an ionization rate given by Equation \eqref{eq:ShullNdot} can keep the whole volume ionized, or only the fraction of the volume satisfying the thresholds? If it is the former, how do we account for the time it takes for I-fronts to cross neutral voids?

At this point the reader may rightfully claim that the Madau-type analysis was never meant to predict the precise redshift for reionization completion, only the ionization rate density needed to maintain the IGM in an ionized state after reionization has completed. We would agree with that. However it is effectively being used in this way when it is applied to galaxy populations at increasingly higher redshifts $z=6-7$  (cf. \cite{FanEtAl2006, RobertsonEtAl2013}). Our investigations indicate that formulae such as Equation \eqref{eq:ndot} and \eqref{eq:ShullNdot} are not reliable estimates of when reionization completes. In \S\ref{Discussion} we examine whether they can be usefully applied at lower redshifts, as originally intended.

%In Figure \ref{Shull}, we show the results of using their clumping factor definition (except we do not have metalicity field in our data, therefore, there is a slight difference in the thresholding done, but it is not expected to change our results much), with most of the thresholds are applied.  The curve ``IRRFromSim''  is taking the photon production rate (``FromSim'' in Figure \ref{unthresholded}) and multiplying it by a different ionization fraction.  This time, the ionization fraction is calculated by summing the ionization from regions that satisfy the C$_\mathrm{RR}$ thresholds (1$<\Delta_b<$100, 300K$<$$T$$<10^5$K, $x$$>$0.05), and dividing it by the total ionization in the entire simulation.  The curve ``C\_RRT4'' in green uses the average temperature, in units of 10$^4$K, of the region that satisfies the C$_\mathrm{RR}$ thresholds for T$_4$ in Equation \eqref{eq:ShullNdot}.  The curve ``C\_RR'' uses 1 in place of T$_4$ in Equation \eqref{eq:ShullNdot}, essentially fixing the IGM temperature to a constant 10$^4$K.  The use of 10$^4$K in \citep{FinlatorEtAl2012} for the denominator of C$_\mathrm{RR}$ will be discussed in \S\ref{Discussion}.

%We can see that while setting the the temperature to 10$^4$K helps get the crossing with ``IRRFromSim'' closer to our expected redshift of $\sim$6, it is still further than the original or the single threshold methods.  Since this result is very sensitive on the IGM temperature, we have to beware the conclusion drawn, which we will tackle more in-depth in \S\ref{Discussion} of the article.

\subsection{Comparing Clumping Factors}
\label{ClumpingFactorEvolution}

For ease of comparison we collect into one plot all the H {\footnotesize II} clumping factors used in the previous sections. The unthresholded H {\footnotesize II} calculated using Equation \eqref{eq:clumpingfactor} is denoted C$_\mathrm{H\,II}$. The singly thresholded clumping factor is denoted C$_\mathrm{tH\,II}$, in which the threshold $\Delta_b<100$ is being applied. The curve labeled C$_\mathrm{RR}$ plots the evolution of Equation \eqref{eq:CRR} with the following thresholds: 1$<\Delta_b<$100, 300K$<$$T$$<10^5$K, $x_e$$>$0.05. For comparison we also plot a doubly thresholded H {\footnotesize II} clumping factor denoted C$_\mathrm{ttH\,II}$ with thresholds $\Delta_b<100$ and $x_e>0.05$, which can be thought of as the clumping factor inside H {\footnotesize II} regions excluding the dense gas in halos. 
%We now show the clumping factors we calculated and used in Equation \eqref{eq:updatedNdot}, \eqref{eq:ShullNdot}.  The two equations estimate the amount of photon production rate density required to keep the universe ionized.  Figure \ref{ClumpingFactors} shows various clumping factors that is calculated using the H {\footnotesize II} field.  The first three clumping factors are calculated using Equation \eqref{eq:clumpingfactor} with H {\footnotesize II} number density.  The prefix ``t'' indicates the threshold of $\Delta_b<100$ being applied.  The ``tt'' prefix indicates two thresholds are applied, they are: $1<\Delta_b$ and $x>0.05$.  The curve ``ttC$_\mathrm{H\,II}$'' is not used in our other analysis, we show it here only to illustrate the effects of applying different thresholds.  The last curve is the C$_\mathrm{RR}$ clumping factor from Equation \eqref{eq:CRR} with the following thresholds: 1$<\Delta_b<$100, 300K$<$$T$$<10^5$K, $x$$>$0.05.

We see a clear trend that as more thresholds are applied the lower the value of the clumping factor goes.  This is because as more regions of the volume are excluded from the averaging process the remaining regions are more homogeneous exhibiting less variations.  If no thresholds are applied, the H {\footnotesize II} clumping factor starts around 200 at $z\sim9$ (Figure \ref{unthresholded}).  Such high values arise because when the first couple of ionizing sources created high H {\footnotesize II}, they are localized and spread far apart, making the H {\footnotesize II} density very clumpy.  As more of the universe is ionized, the H {\footnotesize II} density becomes more homogeneous.  We see the single and double thresholded H {\footnotesize II} clumping factors become the same after overlap with a value of $\sim 4.5$  because the second threshold $x_e>0.05$ is satisfied everywhere. 

The clumping factor that is not based on the H {\footnotesize II} density alone is C$_\mathrm{RR}$.  We see from Equation \eqref{eq:CRR}, C$_\mathrm{RR}$ depends on electron number density, H {\footnotesize II} number density, and the case B hydrogen recombinationation coefficient $\alpha_B(T)$, which is itself dependent on the gas temperature T (fit to Table 2.7 in \cite{OsterbrockFerland2006} implemented in Enzo).  $\alpha_B(T)$ depends on T to a negative power and this causes Equation \eqref{eq:CRR} to sometimes have a very low numerator compared to the denominator.  This as well as the exclusion of gas in the voids leads to the low clumping factor value of $\sim 2$ we see in the graph.  It is very possible to have a value that is smaller than unity, which can lead to even more confusion with the original definition of the clumping factor in Equation \eqref{eq:clumpingfactor}.  There, the clumping factor can only have a value of greater than 1, and 1 occurs only in the case of homogeneous distribution of the gas number density.

\begin{figure}
	\includegraphics[width=0.5\textwidth]{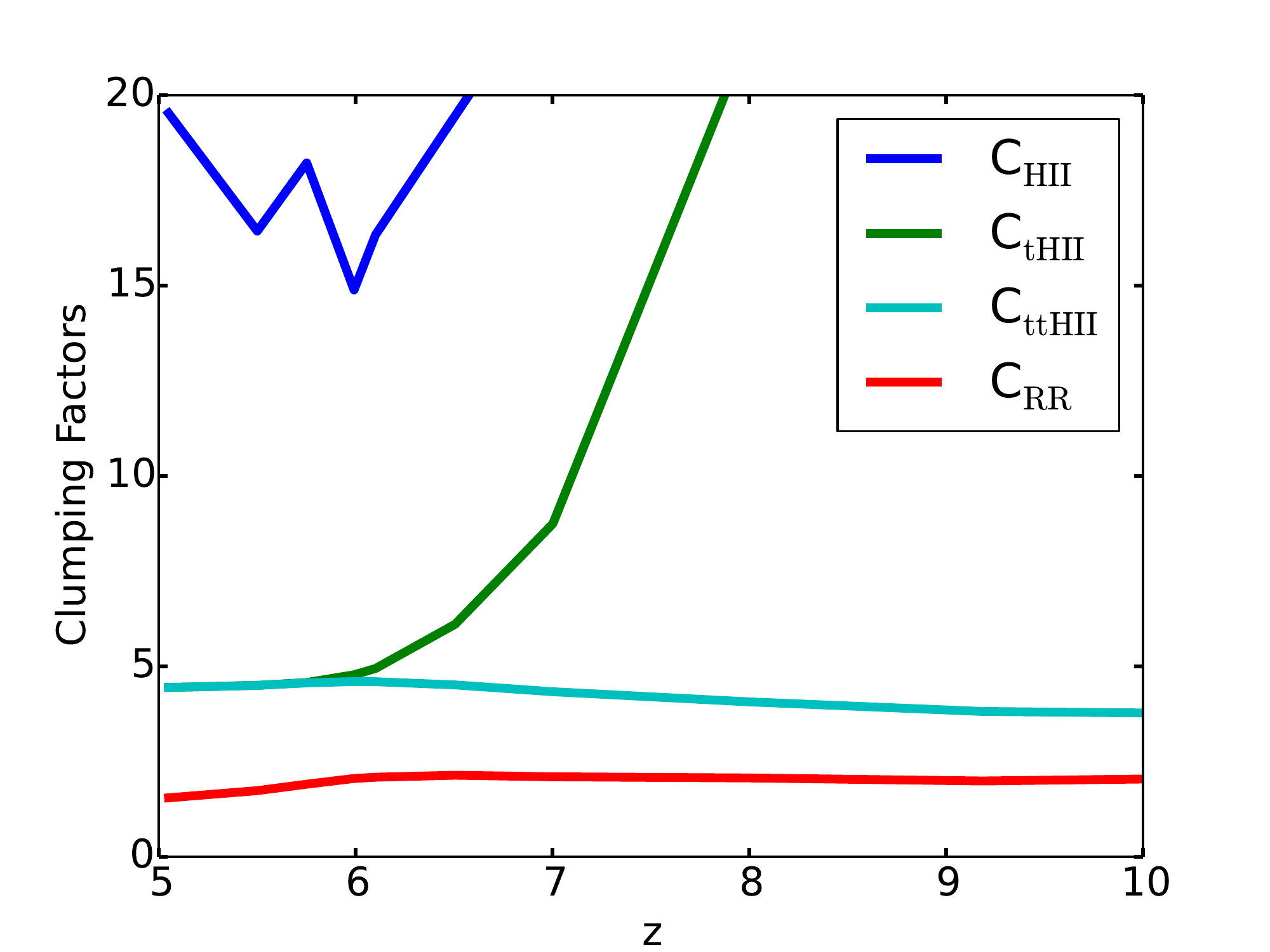}
	\caption{Various clumping factors versus redshift.  C$_\mathrm{H\,II}$ is Equation \eqref{eq:clumpingfactor} used in $\dot{R}_\mathrm{H\,II}$ curve in Figure \ref{unthresholded}, C$_\mathrm{tH\,II}$ is used in $\dot{R}_\mathrm{tH\,II}$ curve in Figure \ref{thresholded}, C$_\mathrm{ttH\,II}$ is clumping factor with two thresholds applied, $\Delta_b < 100$ and $f_i>0.1$, shown here solely for comparison.  C$_\mathrm{RR}$ is the value of recombination rate clumping factor from Equation \eqref{eq:CRR} with the 5 thresholds applied.}
	\label{ClumpingFactors}
\end{figure}

\section{A Global Estimate for Circumgalactic Absorption of Ionizing Radiation}
\label{escape}

The ionizing escape fraction from galaxies is an important parameter in models of reionization. Typically, one thinks about the escape fraction as a property of individual galaxies, determined by the absorption of ionizing radiation on small scales in the ISM. However it is interesting to ask whether there is significant absorption in the denser Circumgalactic Medium (CGM) surrounding galaxies. If we write the total escape fraction as the product of escape fractions, then $f_{esc}=f_{esc}(ISM)f_{esc}(CGM)$.  Here we use our simulation to derive an estimate of the globally averaged escape fraction as a function of redshift  due to the circumgalactic medium $\bar{f}_{esc}(CGM)$. 

Recall from Sec. \S\ref{Method} that the halo escape fraction is not a model input parameter, but is rather an ouput since the equation of radiative transfer is solved throughout the computational domain. Our halos are not well resolved internally, and so we are underestimating the amount absorption of ionizing radiation on galaxy ISM scales. However if significant absorption  occurs on scales of the virial radius or larger, then that would be simulated reasonably accurately. In the following we assume this is the case, and present results that can be taken to be an upper limit on the total escape fraction (ISM+CGM). 

Rather than measure the escape fraction halo by halo and take the average over all halos, we use a simpler method. Since we know every ionization requires an ionizing photon, and we have the ionization rate density as a field defined at every grid cell, then we can estimate $\bar{f}_{esc}(CGM)$ as follows (hereafter we drop the CGM modifier with the reader's understanding that this is what we are estimating):
\begin{equation}
\bar{f}_{esc}(I_t) = \int_{V_t}  n_\mathrm{H\,I}\Gamma_\mathrm{H\,I}^{ph} d^3x \bigg / \int_{V}  n_\mathrm{H\,I}\Gamma_\mathrm{H\,I}^{ph} d^3x  ,
\label{eq:fesc}
\end{equation}
\\where $\Gamma_\mathrm{H\,I}^{ph}$ is evaluated cell by cell via Equation \eqref{eq:photoionization}, $V$ is the simulation volume and $V_t$ denotes the integration includes only cells which satisfy $\Delta_b < 100.$ In other words, $\bar{f}_{esc}$ is the ratio of the number of ionizations in the IGM, as defined by the overdensity threshold, to the total number of ionizations in the volume. The modifier $I_t$ refers to this method of estimating $\bar{f}_{esc}$ (a superior method is presented below).

\begin{figure}
	\includegraphics[width=0.5\textwidth]{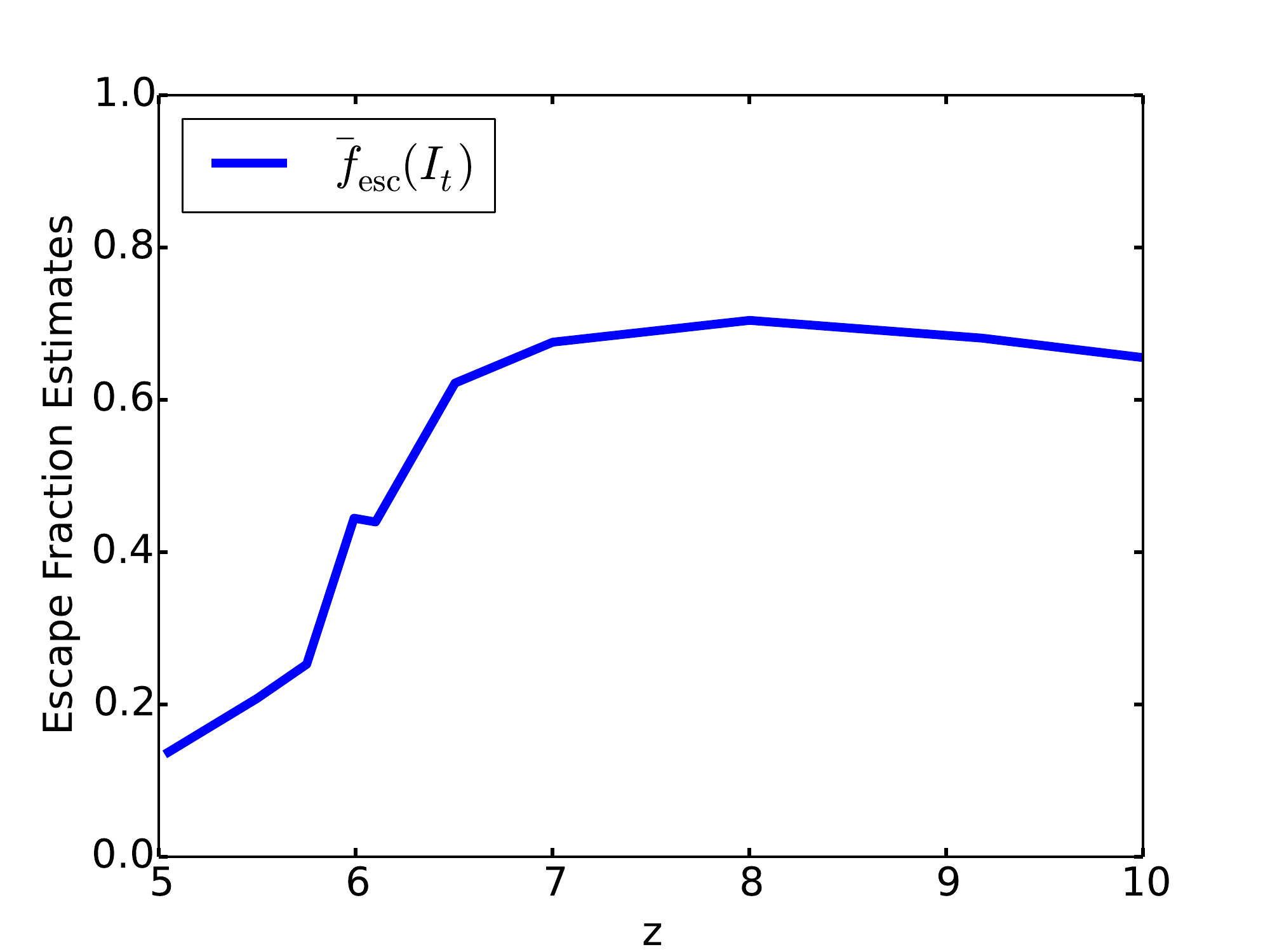}
	\caption{Estimate of the globally averaged ionizing radiation escape fraction due to circumgalactic absorption $\bar{f}_{esc}(I_t)$ computed as the ratio of the volume integrated ionization rate in the IGM ($\Delta_b < 100$) divided by the total ionization rate (Eq. \eqref{eq:fesc}). }
	\label{EscFraction}
\end{figure}

\begin{figure}
	\includegraphics[width=0.5\textwidth]{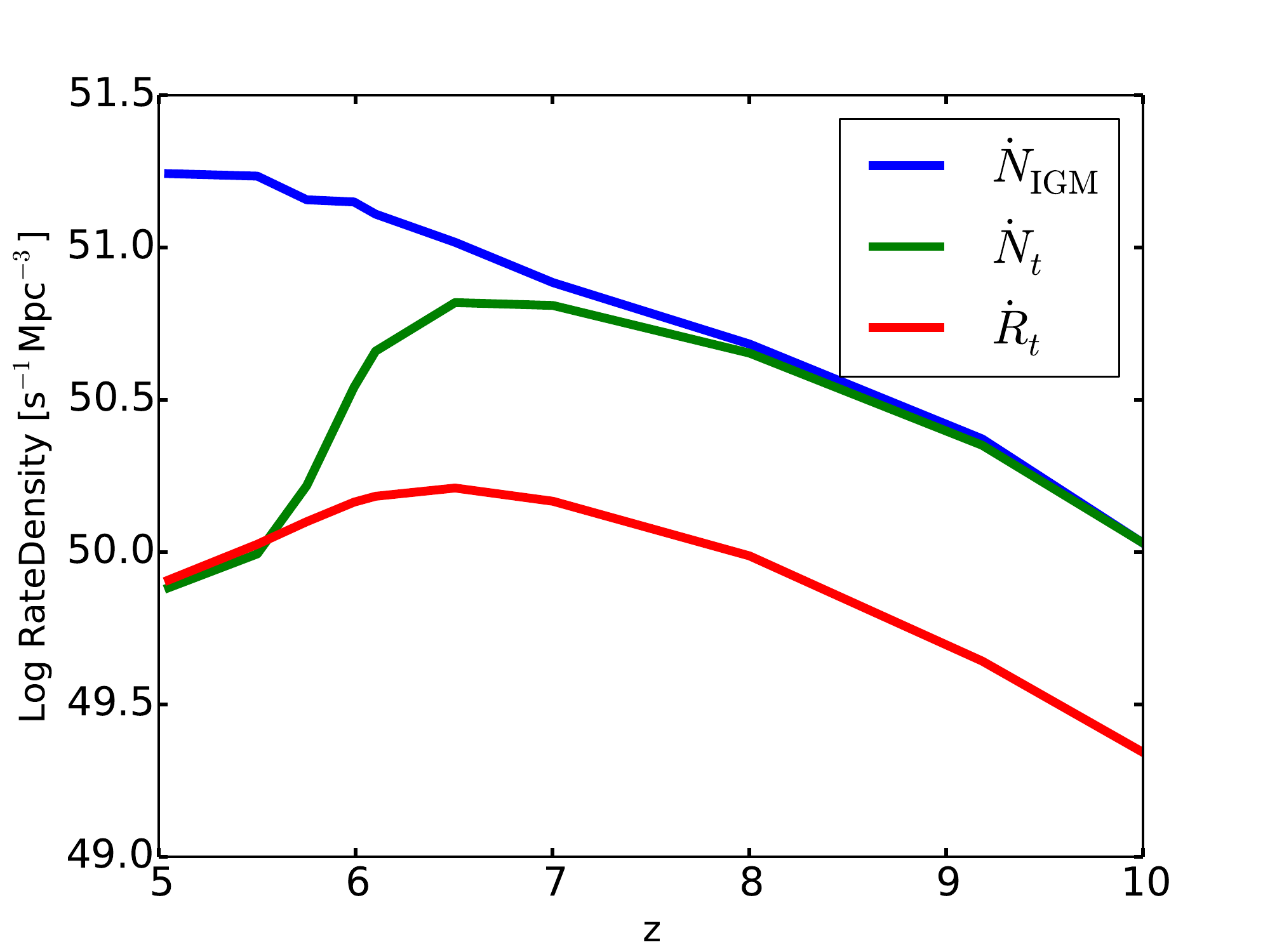}
	\caption{Evolution of the volume averaged rate densities for: (1) ionizing photons injected into the IGM ($\dot{N}_\mathrm{IGM}$), (2) gas photoionization ($\dot{N}_t$), and (3) gas recombination  ($\dot{R}_t$) integrated over the singly thresholded volume $V_t$ defined as $\Delta_b<100$. The ionization rate density curve tracks the photon injection rate density curve in the photon starved regime at high redshifts, but begins to fall below it as the globally averaged ionization parameter approaches unity (Fig. \ref{IP}). After overlap, in the photon abundant regime, the ionization rate density is $\sim 20\times$ the photon injection rate density, but comes into balance with the recombination rate density.}
	\label{sanitycheckrate}
\end{figure}

The result is plotted in Fig. \ref{EscFraction}.  At high redshifts the escape fraction is high and relatively constant at $\bar{f}_{esc} \sim 0.65-0.7$. As overlap is approached $\bar{f}_{esc}$ drops considerably, reaching values of $\sim 0.2$ by $z=5.$ There is no obvious reason why the escape fraction should drop so dramatically at the epoch of overlap. To investigate this properly would require a statisical analysis of individual halo escape fractions, which we defer to a subsequent paper. Perhaps this is an artifact of how we are estimating $\bar{f}_{esc}$. While it is true that every ionization requires and ionizing photon in the photon starved regime (i.e., before overlap), after overlap the volume becomes optically thin to ionizing radiation, and it is not true that every ionizing photon causes an ionization in the box. This is illustrated in Fig. \ref{sanitycheckrate}. 

%The blue dashed line in Fig. \ref{sanitycheckrate} is $\dot{N}_{IGM}$ defined as $\bar{f}_{esc}\dot{N}_{sim}$. 
The curve labeled $\dot{N}_t$ is the actual ionization rate density measured in the simulation averaged over the entire 20 Mpc cubic volume satisfying the overdensity threshold $\Delta_b < 100$; i.e. precisely the numerator of Eq. \eqref{eq:fesc} divided by 20$^3$. The curve labeled $\dot{R}_t$ is the recombination rate density averaged over the same volume; i.e.
\begin{equation}
\dot{R}_t = \int_{V_t}  n_e n_\mathrm{H\,II}\alpha_B(T) d^3x  .
\label{eq:totrecombst}
\end{equation}

We see that ionization rate density $\dot{N}_t$ grows with redshift and reaches a maximum at $z \approx 6.5$, and then drops by roughly 0.8 dex by overlap completion at $z=5.8$. It continues to decrease thereafter. The reason for this sudden drop is that after overlap there are very few neutral atoms left to ionize ($n_\mathrm{H\,I}/n_\mathrm{H} \sim 10^{-5}$).  
%After overlap the large disparity between the $\dot{N}_{IGM}$ and $\dot{N}_t$ curves can then be understood as saying that the IGM becomes photon abundant. 

\begin{figure}
	\includegraphics[width=0.5\textwidth]{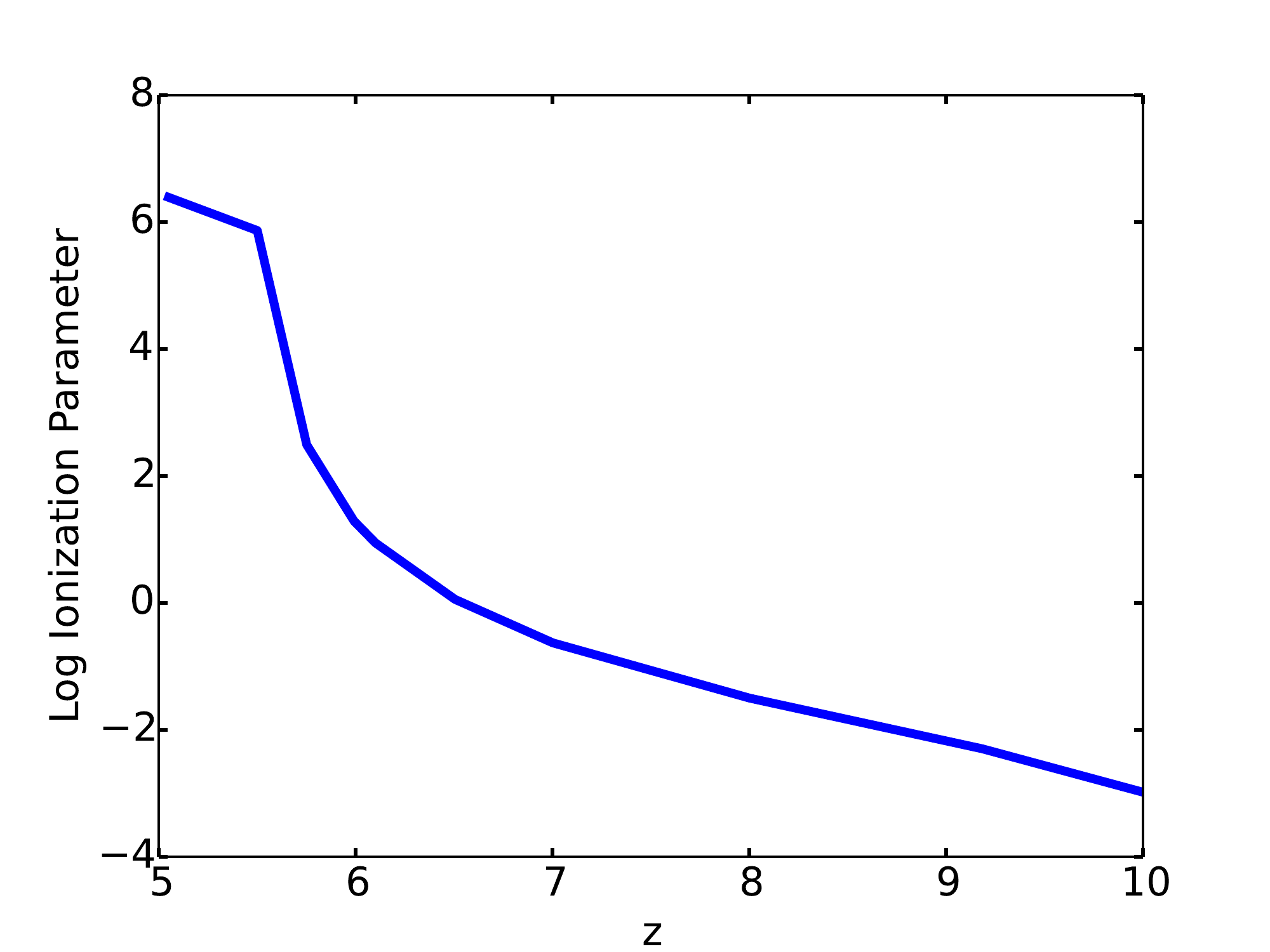}
	\caption{Redshift evolution of the global \hi ionization parameter as defined in Eq. \eqref{eq:IP}. }
	\label{IP}
\end{figure}

This can be illustrated by considering the {\em global ionization parameter}, which is the number of ionizing photons per neutral H atom $\Gamma_{IP} = \langle n_{ph}\rangle/\langle n_{H I} \rangle$ averaged over the entire volume. Specifically, we integrate the grey radiation energy density divided by the mean photon energy $\bar{\epsilon}$ over the singly thresholded volume, and divide by the number of H {\footnotesize I} atoms in the same volume:
\begin{equation}
\Gamma_{IP} =  \int_{V_t}  (E/\bar{\epsilon}) d^3x \bigg / \int_{V_t}  n_{HI} d^3x. 
\label{eq:IP}
\end{equation}
We see from Fig. \ref{IP} that $\Gamma_{IP}$ grows from $\sim 10^{-3}$ at $z=10$ to unity at $z\approx 6.5$ just before overlap. Thereafter $\Gamma_{IP}$ grows very rapidly, reaching a value around $10^5$ at the overlap redshift, and leveling off at around $10^6$ below that.  From the standpoint of the global ionization parameter, reionization begins photon starved but completes photon abundant. 
  
Returning to Fig. \ref{sanitycheckrate} we see that the recombination rate density $\dot{R}_t$ curve tracks the ionization rate density curve to $z\sim 7$, but is about 0.7 dex lower in magnitude, as it must be if the ionized volume filling fraction is to grow. As overlap is approached ionizations and recombinations come into balance, but the recombination rate density has dropped considerably since it reached its maximum value at $z \approx 6.5$. This is also the redshift at which the ionization rate achieves a maximum, and when the global ionization parameter reaches unity.  We also observe that the $f_{esc}$ curve in Fig. \ref{EscFraction} begins its precipitous drop at this redshift. We believe all of these events signal the rapid rise in the global ionization parameter below $z=6.5$, and not some change in the escape fraction of young galaxies. 

Counting the fraction of all ionizations occuring outside halos is not a reliable estimate of the escape fraction for $\Gamma_{IP} \gg 1$ because it does not count the photons in the radiation field that have nothing to ionize. Therefore we need to modify Eq. \eqref{eq:fesc} to include photons which build up of the radiation field: 
\begin{equation}
\bar{f}_{esc} = \int_{V_t}  (n_\mathrm{H\,I}\Gamma_\mathrm{H\,I}^{ph} + \frac{1}{\bar{\epsilon}}\frac{dE}{dt}) d^3x \bigg / \int_{V}  (\eta/\bar{\epsilon}) d^3x . 
\label{eq:fescimproved}
\end{equation}
\\Here the numerator is the rate at which ionizing photons are causing ionizations in the IGM and building up the UV background, and the denominator is volume integrated ionizing photon production rate. 

\begin{figure}
	\includegraphics[width=0.5\textwidth]{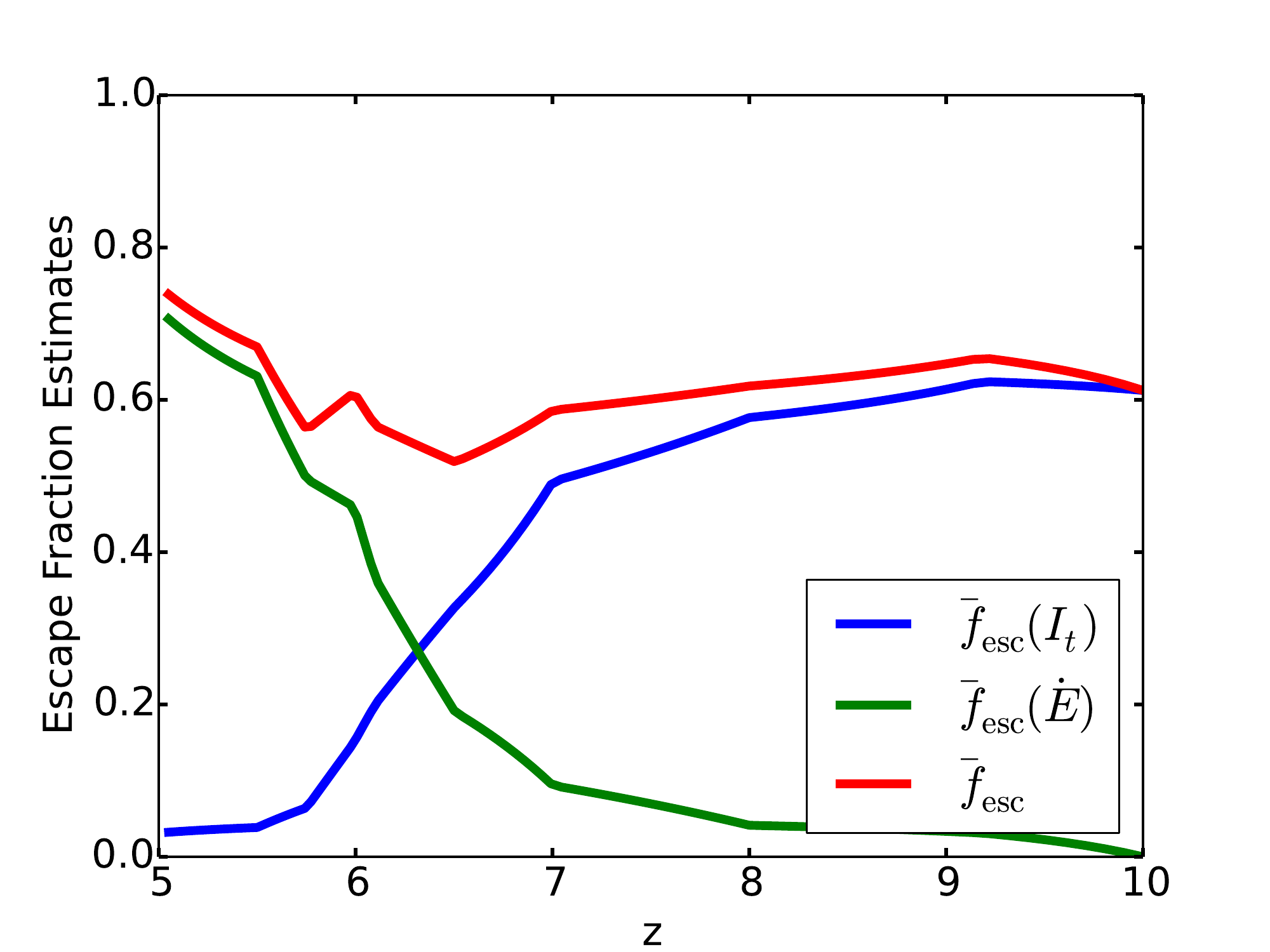}
	\caption{Redshift evolution of the globally averaged escape fraction contribution from circumgalactic absorption as estimated by the number of ionizations occuring in the IGM and the buildup of the ionizing radiation background. The curves labeled $\bar{f}_{esc}(I_t), \bar{f}_{esc}(\dot{E})$ plot the contributions of the first and second terms in Eq. \eqref{eq:fescimproved}, while the curve labeled $\bar{f}_{esc}$ plots their sum.}
	\label{RadEscFraction}
\end{figure}

Fig. \ref{RadEscFraction} plots $\bar{f}_{esc}$ calculated according to Eq. \eqref{eq:fescimproved}. Each contribution to $\bar{f}_{esc}$ is plotted separately, as well as the sum. We see that $\bar{f}_{esc}$ is roughly constant with redshift with a value of around 0.6. We see that as the contribution due to ionizations declines below $z\sim 7$, the contribution due to the change in radiation background intensity increases in a compensating fashion. This confirms our earlier suspicions and gives us a better estimate of the mean circumgalactic attenuation of ionizing radiation from young galaxies. 

To complete the picture we plot in Fig. \ref{sanitycheckrate} the number density of ionizing photons escaping into the IGM, calculated as $\dot{N}_{IGM}=\bar{f}_{esc}\dot{N}_{sim}$, where $\dot{N}_{sim}$ is the ionizing photon production rate in the simulation, and  $\bar{f}_{esc}$ is the improved estimate for the escape fraction calculated using Equation \eqref{eq:fescimproved}. We see that at high redshifts the $\dot{N}_{IGM}$ and $\dot{N}_t$ track each other closely. This tells us two things. First, that reionization at high redshifts when $Q_{\mathrm{H\,II}} \ll 1$ is photon starved, in the sense that every ionizing photon emitted results in an ionization. And second that our estimate of $\bar{f}_{esc}$ is reasonably accurate at these redshifts. However, as redshift decreases, the two curves systematically begin to deviate from one another in the sense that $\dot{N}_t < \dot{N}_{IGM}$. Beginning at $z = 6.5$ the ionization rate density begins to decrease while the ionizing photon production rate into the IGM continues to rise. 
After overlap the large disparity between the $\dot{N}_{IGM}$ and $\dot{N}_t$ curves can then be understood as saying that the IGM becomes photon abundant. 

\begin{figure}
	\includegraphics[width=0.5\textwidth]{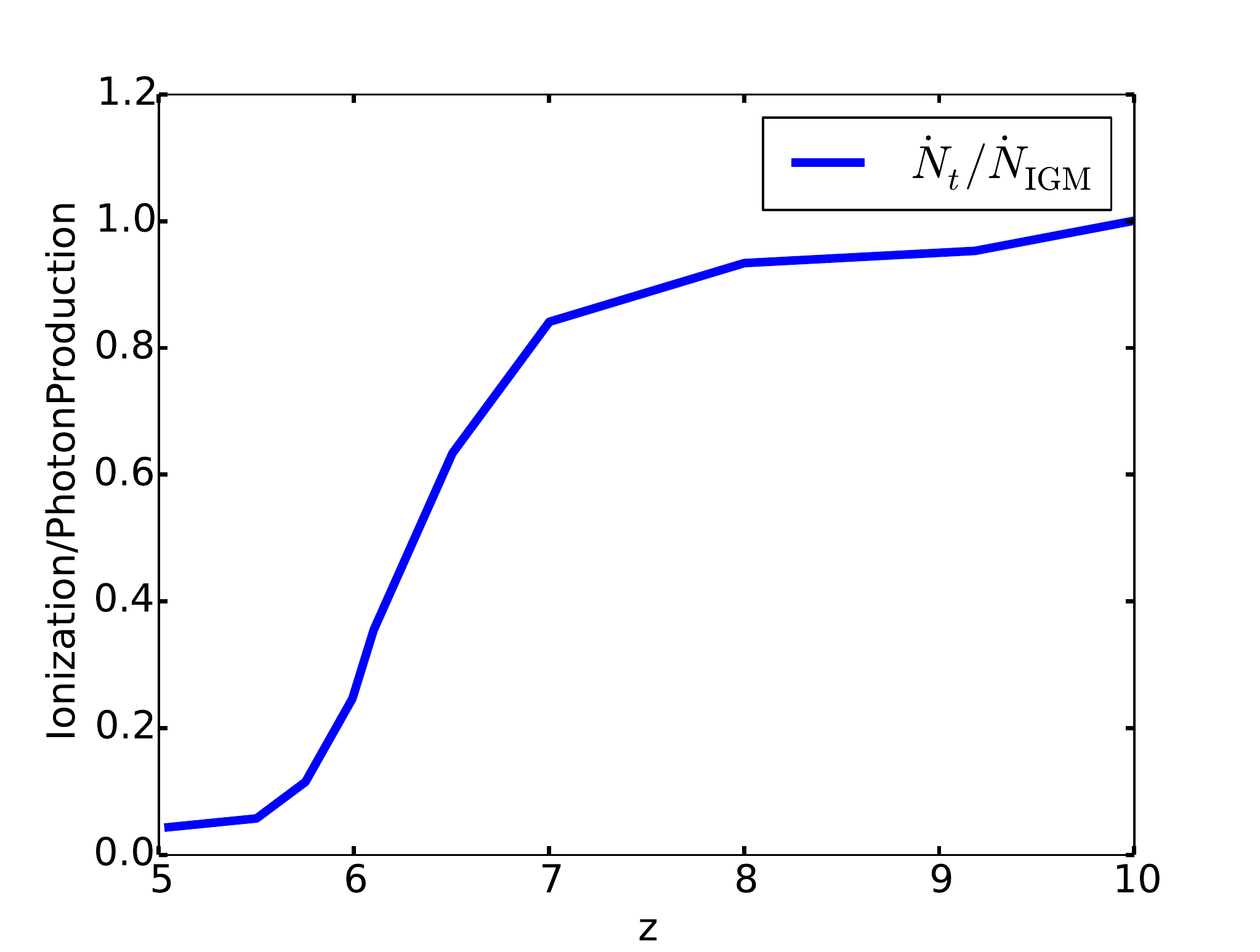}
	\caption{Ratio of the volume integrated photoionization rate in the IGM $\dot{N}_t$ to the integrated photon injection rate into the IGM $\dot{N}_{IGM}$, where the IGM is defined as cells with $\Delta_b < 100$. The ratio is near unity initilly, remains high until $z\approx 7$ ($Q_{HII} \approx 0.5$), and then drops rapidly as overlap is approached and the IGM becomes highly ionized.}
	\label{Ndot_Ratio}
\end{figure}

The ratio of ionization rate density and the photon injection rate into the IGM is plotted in Fig. \ref{Ndot_Ratio}. The ratio is unity initially, and slowly decreases until $z\approx 7$, and then drops rapidly as overlap is approached. After overlap the ratio is about 0.05. In other words, after overlap, the photon production rate is about 20$\times$ the ionization rate in a volume averaged sense. Since the ionization and recombination rates are in balance after overlap, we conclude that the volume averaged photon injection rate is about 20$\times$ the recombination rate.

\section{An Improved Model for the Evolution of $Q_\mathrm{H\,II}$}
\label{Qdot}

In this section we compare the evolution of the ionized volume fraction $Q_\mathrm{H\,II}$ from our simulation with the analytic model introduced by \cite{MadauEtAl1999}. We are motivated to do this because as we have seen from \S\ref{sec:ClumpingFactors}, Equation \eqref{eq:ndot} is not a useful predictor of when $Q_\mathrm{H\,II}$ reaches unity. We therefore want to investigate the accuracy of the time dependent model from which Equation \eqref{eq:ndot} is derived as a limiting case.

 \cite{MadauEtAl1999} derived the following ODE for the evolution of $Q_\mathrm{H\,II}$ (their Equation 20):

%After our initial investigation into the accuracy of Equation \eqref{eq:updatedNdot} for determining the end of EoR, we want to try probe the validity for the equation that determines the volume filling fraction of H {\footnotesize II}.  The algebraic form of Equation (23) in \citep{MadauEtAl1999} should determine the volume filling fration of ionized region, but it is derived with assumptions from the differential form.  Without taking simplifying steps, the differential form is Equation (20) \citep{Madau1999},

\begin{equation}
	\label{eq:dQdt}
	\frac{dQ_\mathrm{H\,II}}{dt} = \frac{\dot{n}_{ion}}{\bar{n}_\mathrm{H}}-\frac{Q_\mathrm{H\,II}}{\bar{t}_{rec}}
\end{equation}
where $\dot{n}_{ion}$ is ionizing photon injection rate, $\bar{n}_\mathrm{H}$ is the mean density of H atoms in the universe, and $\bar{t}_{rec}$ is some characteristic recombination time taking the clumpiness of the IGM into account. For a constant clumping factor and comoving emissivity \cite{MadauEtAl1999} show that 
\begin{equation}
Q_\mathrm{H\,II}(t) \approx \frac{\dot{n}_{ion}}{\bar{n}_\mathrm{H}} \bar{t}_{rec}
\end{equation}
Setting $Q=1$ one arrives at  $\dot{n}_{ion}\bar{t}_{rec}=\bar{n}_\mathrm{H}$, the basis for deriving Equation \eqref{eq:ndot}. \cite{MadauEtAl1999} state that this relation should still be valid provided the clumping factor and comoving emissivity are slowly varying on a timescale of $\bar{t}_{rec}$. We utilize the differential form for our comparison because our emissivity is not a constant value, nor is it slowly varying on a recombination time as $Q \rightarrow 1$, as we show below.  %so the subsquent simplification to get the differential form to the algebraic form is invalid for us.  
%The differential form is shown as Equation \eqref{eq:dQdt}, where $Q$ is the volume filling fraction, $t$ time, $\dot{n}_{ion}$ the number of ionizing photon production rate density, $\bar{n_\mathrm{H}}$ the volume averaged number density of hydrogen atoms, and $t_{rec}$ is the recombination time, which we now investigate.

A practical issue when testing Equation \eqref{eq:dQdt} is how $\bar{t}_{rec}$ should be evaluated when $Q<1$, and in particular when $Q\ll 1$. In the limit $Q\ll 1$ one is dealing with isolated H {\footnotesize II} regions evolving under the influence of local conditions. Yet the definition for $\bar{t}_{rec}$ in Equation \eqref{eq:tmadau} invokes {\em global} values for $C$ and $\langle n_\mathrm{H\,II} \rangle$. Should these quantitles be evaluated locally only within ionized regions? Or are global estimates good enough? In particular, since \cite{MadauEtAl1999}'s Equation (20) uses $\bar{n}_\mathrm{H}$ as a proxy for $\langle n_\mathrm{H\,II} \rangle$, what is the appropriate value for $C$ to use?

A second practical issue is what to take for $\dot{n}_{ion}$. This is commonly understood to be the rate at which ionizing photons are injected into the IGM (e.g., Haardt \& Madau 2012, \S9.3), which in our parlance is $\dot{N}_{IGM}$. Or should we take the actual ionization rate density measured in the simulation $\dot{N}_t$? As we saw in the previous section, these two rates diverge as overlap is approached, and differ by more than an order of magnitude after overlap (Fig. \ref{Ndot_Ratio}). 

\begin{figure}
    \includegraphics[width=0.5\textwidth]{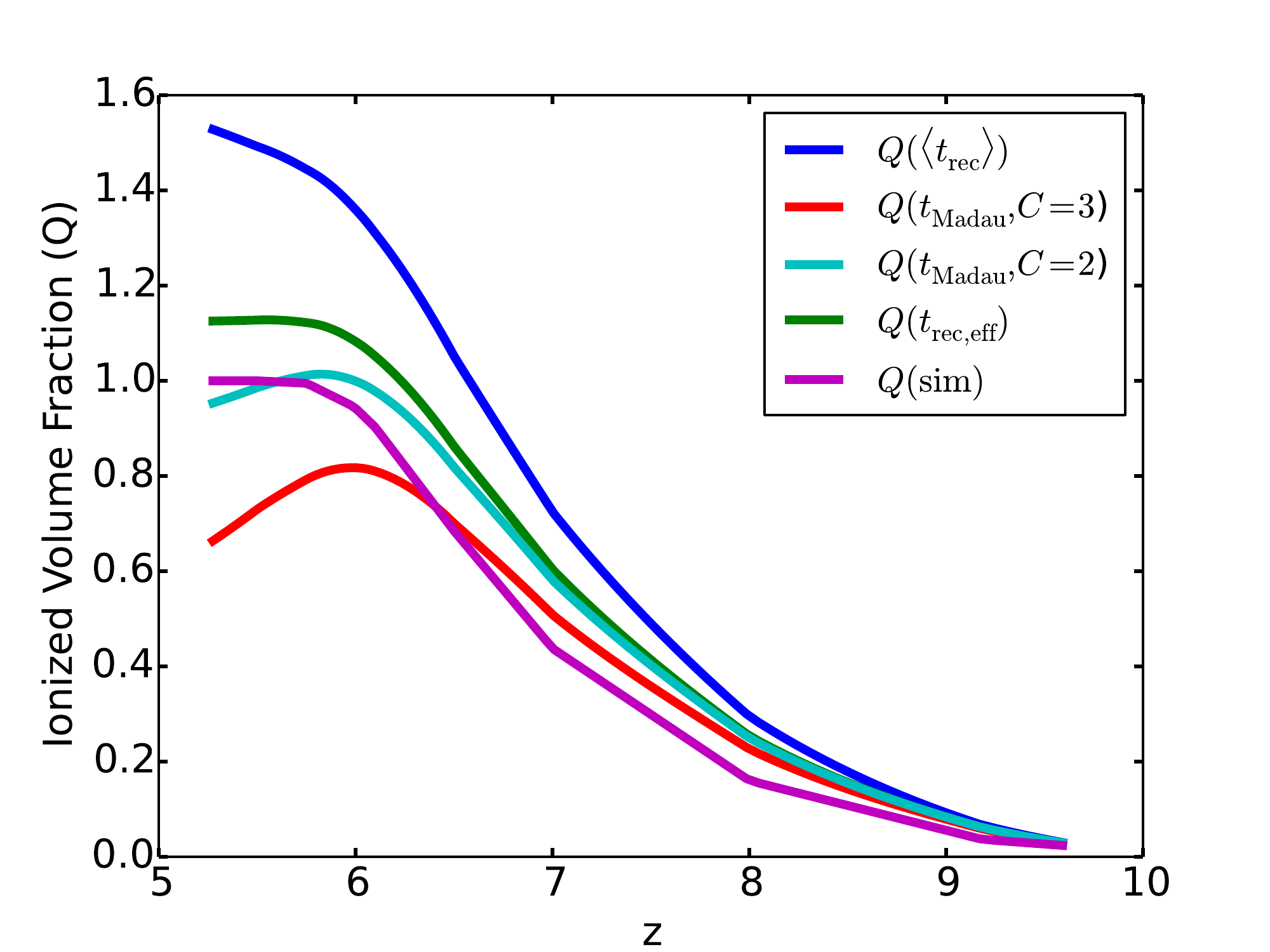}
    \includegraphics[width=0.5\textwidth]{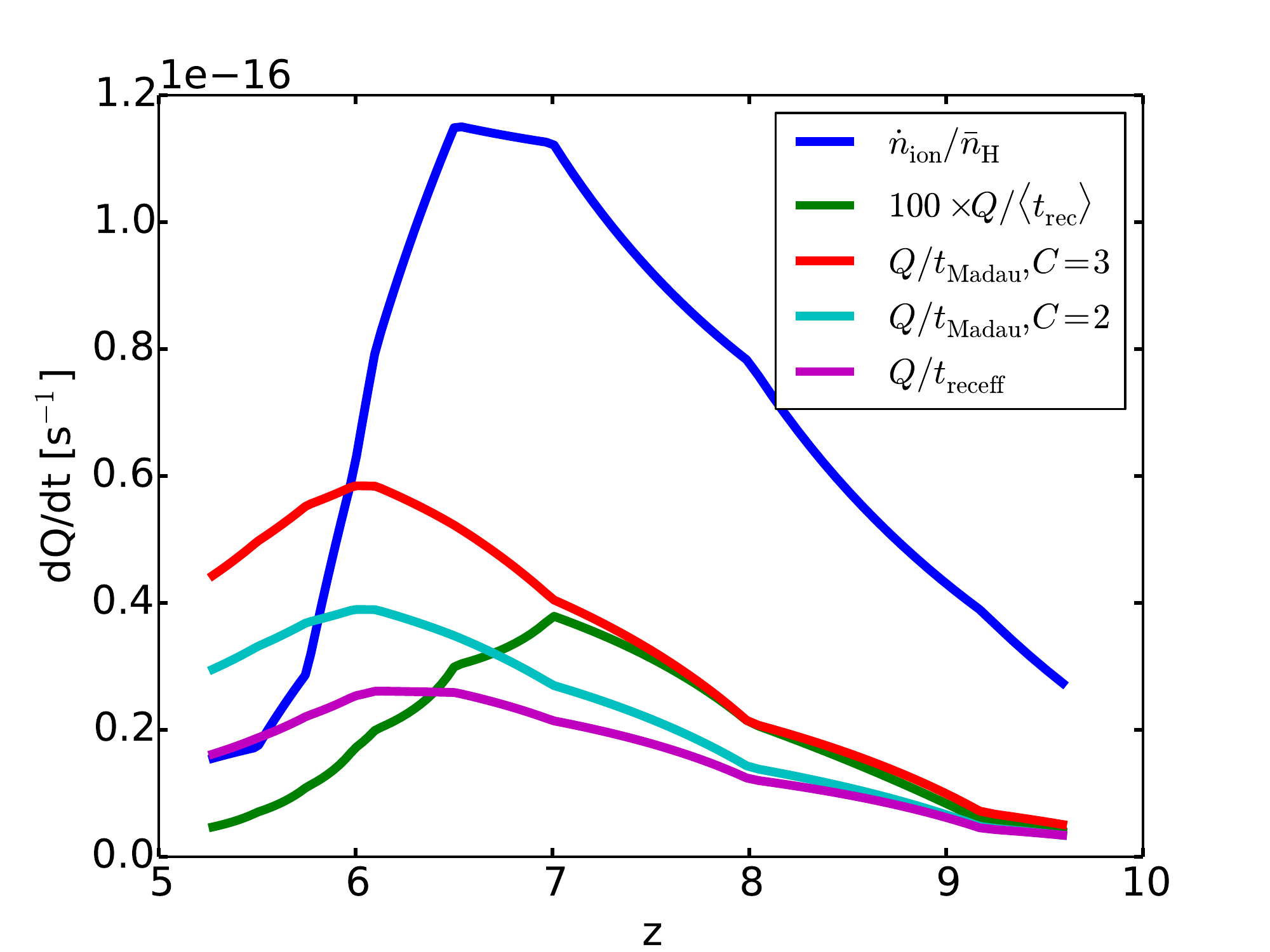}
    \caption{{\em Top}: Comparison of the evolution of the ionized volume fraction Q from our simulation with the analytic model introduced by \cite{MadauEtAl1999}.  Q(sim) is calculated directly from counting the cells satisfying the Well Ionized threshold of $f_i>0.999$. The other curves are calculated from integrating Equation \eqref{eq:integration} with the different expressions for $\bar{t}_{rec}$ in Term$_2$, as described in the text.  {\em Bottom}:  Plot of Term$_1$ and Term$_2$ individually using the different expressions for $\bar{t}_{rec}$.}
    \label{Qeffv1}
\end{figure}

To examine these issues we plot in Figure \ref{Qeffv1} $Q(z)$ from our simulation, as well as theoretical curves obtained by integrating Equation \eqref{eq:dQdt} under various assumptions. The curve labelled $Q(sim)$ is the ionized volume fraction from our simulation that is at least 99.9\% ionized (Well Ionized). The other four curves are obtained by integrating Equation \eqref{eq:dQdt} setting $\dot{n}_{ion}=\dot{N}_t$ for various choices for $\bar{t}_{rec}$ (we investigate the $\dot{n}_{ion}=\dot{N}_{IGM}$ case at the end of this section.) The integral is approximated by summing a piecewise linear interpolation of the two terms on the RHS of Equation  \eqref{eq:dQdt} using the trapezoidal rule:
\begin{align}
Q(t) &= \int_{t*}^{t} \frac{dQ}{dt}dt \approx \sum \frac{dQ}{dt} \Delta t \notag\\
&= \sum_{i}(\mathrm{Term_1} - \mathrm{Term_2})_i \Delta t_i
\label{eq:integration}
\end{align}
where $t*$ is the time when the first star forms in the simulation.

The curve labeled $Q(\langle t_{rec}\rangle)$ uses the volume averaged recombination time (volume average of Equation \ref{recombtime}).  The two curves labeled $Q(t_\mathrm{Madau})$ use Equation \eqref{eq:tmadau} to evaluate $\bar{t}_{rec}$ for $C=2$ and $3$, substituting $\bar{n}_\mathrm{H}$ for $\langle n_\mathrm{H\,II} \rangle$ and assuming a constant T=10$^4$K for the IGM. 
%is using the definition of $\bar{t_{rec}}$ in \cite{MadauEtAl1999}, which is using the volume averaged number density of H {\footnotesize II} along with the clumping factor as $[(1+2\chi)\langle n_\mathrm{H\,II}\rangle \alpha_\mathrm{B} C]^{-1}$, assuming a constant T=10$^4$K which gives $\alpha_B\sim2.59\times10^{13}$cm$^{-3}$s$^{-1}$ in the IGM.  
The curve labeled $Q(t_{rec,eff})$ uses the effective recombination time definition 
\begin{equation}
	\label{eq:treceff}
	\bar{t}_{rec}=t_{rec,eff}\equiv \frac{\langle n_\mathrm{H\,II}\rangle}{\langle n_\mathrm{H\,II} n_e \alpha_B(T)\rangle}
\end{equation}
This particular definition makes the last line of Equation \eqref{eq:trecpart2} true trivially, with no assumption about the IGM temperature or ionization state of the hydrogen.  It involves no {\em ad hoc} clumping factors, and represents the actual appropriately averaged recombination time in the simulation. All the above volume averaged quantities have the threshold of $\Delta_b<100$ applied, and thus exclude dense gas bound to halos.   
%The curve $Q(sim)$ is the actual ionized volume fraction to Well Ionized level vs. z.  
Several of the curves derived from integrating $\frac{dQ}{dt}$ reach values above unity at the end of the overlapping phase. While it is physically impossible to have $Q>1$ it is not mathematically forbidden, and so we show the complete curves because they give us some insight about the relative contribution of the recombination term (Term$_2$) as compared to the ionization term (Term$_1$).

The $Q(\langle t_{rec}\rangle)$ curve ionizes the quickest, reaching  $Q=1$ at $z\sim 6.5$, which is substantially before the simulation which achieves it at $z\approx 5.8$. The reason for this, as we will analyze shortly, is that recombinations play essentially no role in this model. The $Q(t_{rec,eff})$ curve has the same shape as the $Q(sim)$, but is everywhere higher, and crosses $Q=1$ at $z\sim 6.1$. Given that this integration uses the actual ionization rate density and effective recombination time in the simulation, this discrepancy demands an explanation. We address this below. Finally the $Q(t_\mathrm{Madau})$ curves do not match the shape of the $Q(sim)$ curve, ionizing more quickly at early times, and exhibiting a maximum value for $Q$ at $z \sim 6$. 

To understand this behavior more fully we plot in Figure \ref{Qeffv1} {\em bottom} the values for Term$_1$ and Term$_2$ in Equation \eqref{eq:dQdt}. 
%We are curious as to why the integrated curves reach values above unity, so we look at the relative contribution of each of the two terms in Equation \eqref{eq:dQdt}, and we plot the result in Figure \ref{term1vsterm2}.  
The blue curve is Term$_1$ of Equation \eqref{eq:dQdt}. The other four curves plot Term$_2$ with their respective values for $\bar{t}_{rec}$.  The ionization curve dominates all the recombination curves at high redshifts, and reaches a maximum at $z\sim 6.5$. 
This is a partial reflection of the plateauing and subsequent decline of the SFRD shown in Figure \ref{SFR}. More fundamentally, it is a reflection of the rapid drop in the neutral fraction of the IGM as overlap is approached.  
%The physical reason for this is the suppression of star formation in low mass halos post-reionization, which is the topic of a future paper in this series.  
The curve using the volume averaged recombination time $\langle t_{rec}\rangle$ yields such low values compared to the others that we multiply it by 100 to make it more visible.  Although this is not the relevant recombination time to use, since it weights low density regions, it is effectively the limiting case $\bar{t}_{rec}\rightarrow \infty$.  We can therefore interpret the blue curve in Figure \ref{Qeffv1}a as an integration of the ionization term only. It is significantly higher than the $Q(sim)$ curve, suggesting that recombinations are important in the simulation at some level. The ionization term dominates the recombination term by factors of $6-10$ in the $t_{rec,eff}$ curve until just before overlap, and the two terms come into balance after overlap. The two $t_\mathrm{Madau}$ recombination curves are subdominant to the ionization term until $z \sim 6$, and at lower redshifts they become dominant. This explains the turnaround in the corresponding $Q$ curves in Figure \ref{Qeffv1}a. 

\begin{figure}
    \includegraphics[width=0.5\textwidth]{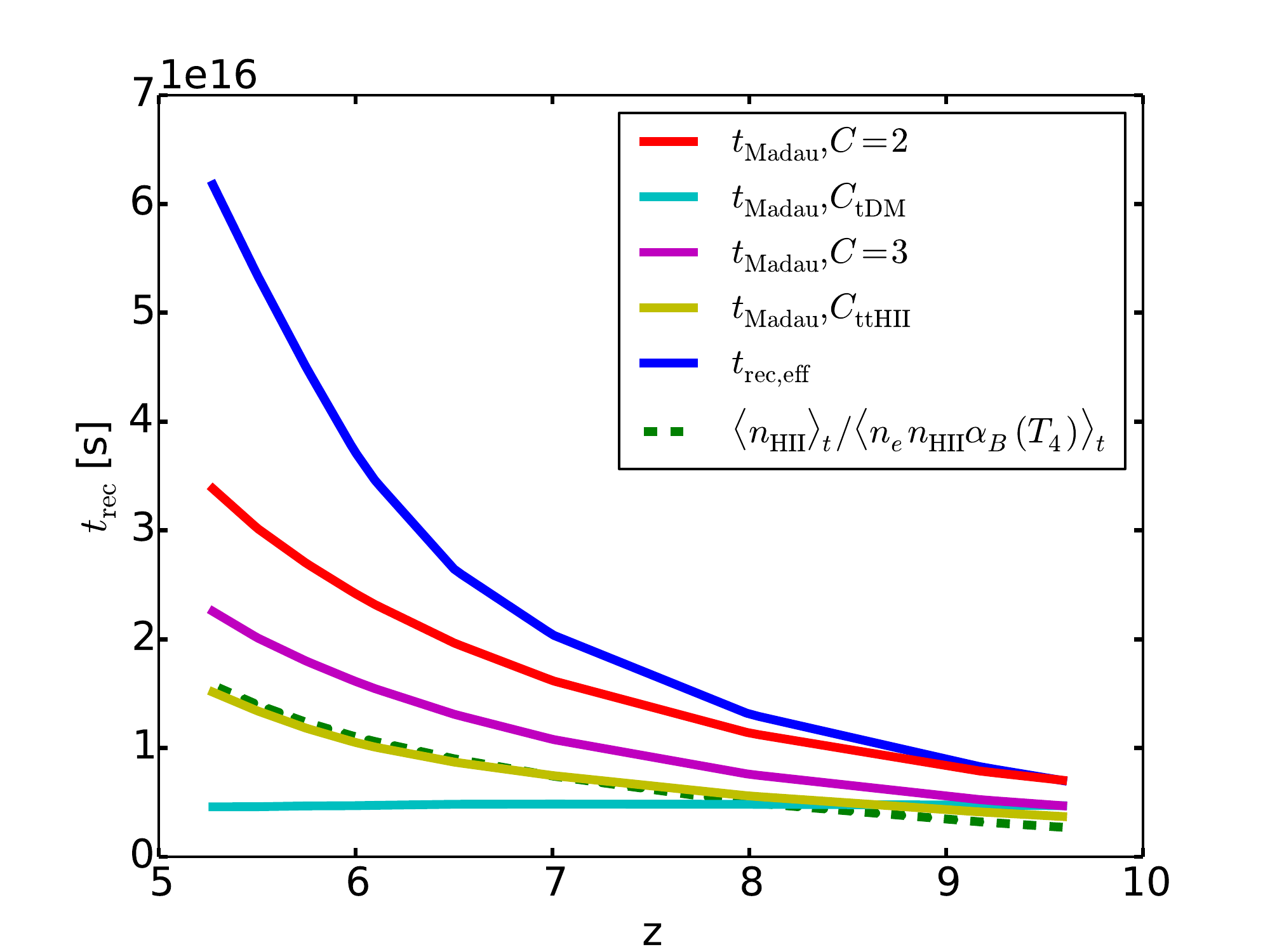}
    \includegraphics[width=0.5\textwidth]{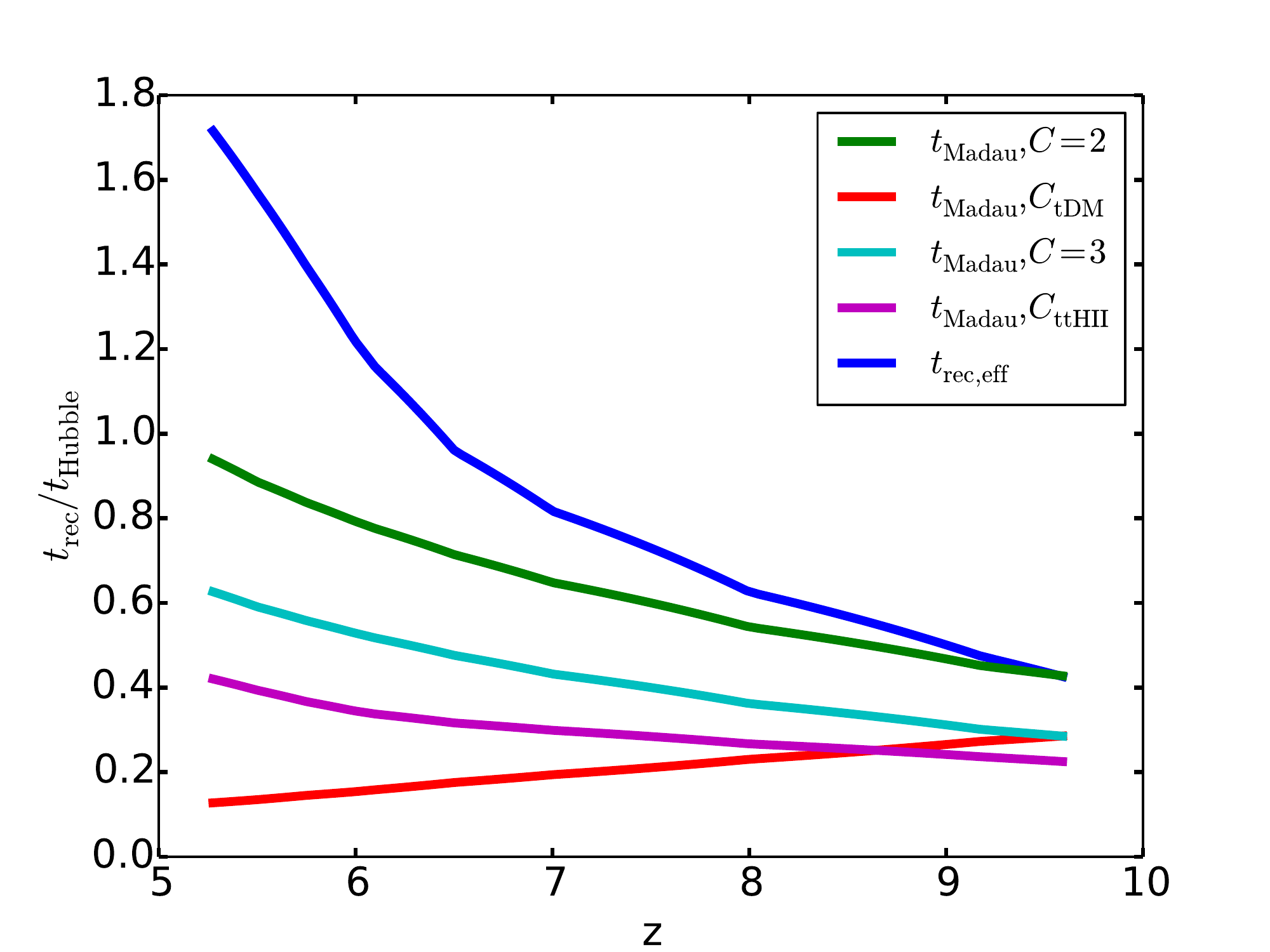}
    \caption{{\em Top}: Recombination time versus redshift, for various expressions for $\bar{t}_{rec}$ as described in the text. Curve labeled $t_{rec,eff}$ is the characteristic recombination time measured directly in the simulation. Curves labeled $t_{Madau}$ evaluate Eq. \eqref{eq:tmadau} for various choices for the clumping factor C. {\em Bottom}: Recombination time versus redshift normalized by the Hubble time, for various expressions for $\bar{t}_{rec}$.}
    \label{treceffhubble}
\end{figure}

The differences in the magnitude of the recombination curves in Figure \ref{Qeffv1}b, especially at higher redshifts, is directly attributable to the magnitude of $\bar{t}_{rec}$. For completeness we plot $\bar{t}_{rec}$ versus redshift in Figure \ref{treceffhubble}, both unnormalized and normalized by $t_\mathrm{Hubble}$. In addition to the three curves for $t_{rec,eff}$ and $t_\mathrm{Madau}$ for $C=2, 3$, we also plot $t_\mathrm{Madau}$ for $C=C_\mathrm{ttH\,II}$ and $C=C_\mathrm{tdm}$. We see that all the curves with the exception of the Madau formula curve using the thresholded dark matter clumping factor exhibit an increasing recombination time with decreasing redshift, in line with our expections. The latter curve shows the opposite trend, which is due to the fact that the dark matter clumping factor increases with decreasing redshift, even if it is thresholded to exclude halos (see Figure \ref{thresholded} bottom). Among the remaining curves the $t_{rec,eff}$ has the highest values, and increases more sharply than the $t_\mathrm{Madau}$ curves due to the temperature of the IGM. To demonstrate that, we plot one additional curve (dashed curve) for $t_{rec,eff}$ evaluated assuming a constant $T=10^4$K in the recombination rate coefficient.

We now comment on the often-made assumption in reionization models that $\bar{t}_{rec} \ll t$. \cite{MadauEtAl1999} make this assumption in order to derive Equation \eqref{eq:ndot}. It is this assumption that allows for an instantaneous analysis of the photon budget to maintain the universe in an ionized state while ignoring history dependent effects.
Referring to  Figure \ref{treceffhubble}b we see this is never true for $t_{rec,eff}$ and it is not true for $t_\mathrm{Madau}$ at redshifts approaching overlap for any sensible value of $C$. We therefore conclude that history-dependent effects cannot be ignored, and that this is the reason Equations \eqref{eq:ndot}, \eqref{eq:updatedNdot} and \eqref{eq:ShullNdot} mis-predict the epoch of reionization completion. For the same reason applying these formulae at lower redshifts is highly suspect.  

%Clearly, the first term is dominant over different versions of the second term.  However, the second term with $Q/t_{rec,eff}$ is not negligible, and contributes the most by having recombination curb the growth of $Q$ from the first term.

%\begin{figure}
%\includegraphics[width=0.5\textwidth]{term1vsterm2.eps}
%	\caption{test}
%\label{term1vsterm2}
%\end{figure}

%The best case scenario is represented by using $t_{rec,eff}$, and to focus on the difference, we plot each $t_{rec}$ with respect to the hubble time $t_\mathrm{Hubble}$ in Figure \ref{treceffhubble}.  This time, we multiply $t_{rec,eff}/t_\mathrm{Hubble}$ by a factor of 100 to maintain visibility along with the other two curves.  The figure shows $t_{rec,eff}$ to be the lowest, which tells us that recombinations are happening on a quicker time scale using this definition as oppose to other definitions.  Also notice that the other two curves with $\rangle t_\mathrm{rec}\langle$ and $t_\mathrm{Madau}$ divided by $t_\mathrm{Hubble}$ both have values much greater than unity, therefore the assumptions made in \cite{MadauEtAl1999} that says only the instantaneous ionization rate is important and not its history is not valid here.  Even though the $t_\mathrm{rec,eff}$ curve started below 1.0, it is not an order of magnitude lower, and crosses above 1.0 around $z\sim 6.5$, so the claim $t_{rec} \ll t_\mathrm{Hubble}$ is never satisfied.

Returning to the discrepancy between the $Q(sim)$ and $Q(t_{rec,eff})$ curves in Figure \ref{Qeffv1}a, since the most sensible choice for $t_{rec}$ did not give us satisfactory agreement, we wondered what the origin of the discrepancy could be.  Since we have shown that recombinations are relatively unimportant at high redshifts, but that the discrepancy is already present at high redshifts, the only possibility is that there is something wrong with the first term of Equation \eqref{eq:integration}. When looking at the derivation for Equation \eqref{eq:dQdt} in \cite{MadauEtAl1999}, it is stated that it ``approximately holds for every isolated source of ionizing photon in the IGM.''  That got us to think that our calculation of $\bar{n}_\mathrm{H}$ may be off from what is originally intended if it is a global average over the entire simulation box.  Since the original $\frac{dQ}{dt}$ is derived from the analytical Str\"{o}mgren sphere model, it assumed a single ionizing source at the center of the volume, and the the average density of the box is just the uniform density everywhere, we thought that might be the discrepancy.  In an Inside-out model, I-fronts are not initially propagating in a gas with an average density given by $\bar{n}_H$, but somewhat higher density. Would agreement improve if instead of using $\bar{n}_H$ in the first term of Equation \eqref{eq:dQdt}, we used the local average density?

%We attempt to use a $\delta_b\equiv \langle \rho_b\rangle_{tt}/\langle \rho_b\rangle_{t}$ as a correction factor to the right hand side term 1 in Equation \eqref{eq:dQdt}.  The volume average $\langle\rangle$ with subscript $t$ is the usual $\Delta_b<100$ threshold, the double subscript $tt$ indicates the additional threshold of $x>0.1$, therefore the correction factor $\delta_b$ tells us how much denser are the ionized gas that the radiation front has passed through, compare to the rest of the box.  The denser average number density of hydrogen will be enhanced by a factor of $\delta_b$ as follows:

We therefore modify Equation \eqref{eq:dQdt} as follows:
\begin{equation}
	\frac{dQ}{dt} = \frac{\dot{n}_{ion}}{\delta_b\bar{n}_\mathrm{H}}-\frac{Q}{\bar{t}_{rec}}
	\label{eq:dQdtdb}
\end{equation}
where we have introduced in the denominator of the first term a factor $\delta_b \geq 1$ which corrects for the higher mean density within ionized bubbles. We measure $\delta_b$ from each redshift output as follows: $\delta_b = \langle \rho_b\rangle_{tt}/\langle \rho_b\rangle_{t}$. The volume average $\langle\rangle$ with subscript $t$ is the usual $\Delta_b<100$ threshold, the double subscript $tt$ indicates the additional threshold of $x_e>0.1$. Thus $\delta_b$ is the average baryon overdensity within Ionized regions excluding gas inside halos. Figure \ref{deltabvsQfit5} shows a plot of $\delta_b$ versus $Q$ together with a simple fitting formula which fits the data extremely well over the domain $0.01 \leq Q \leq 1$. 

\begin{figure}
	\includegraphics[width=0.5\textwidth]{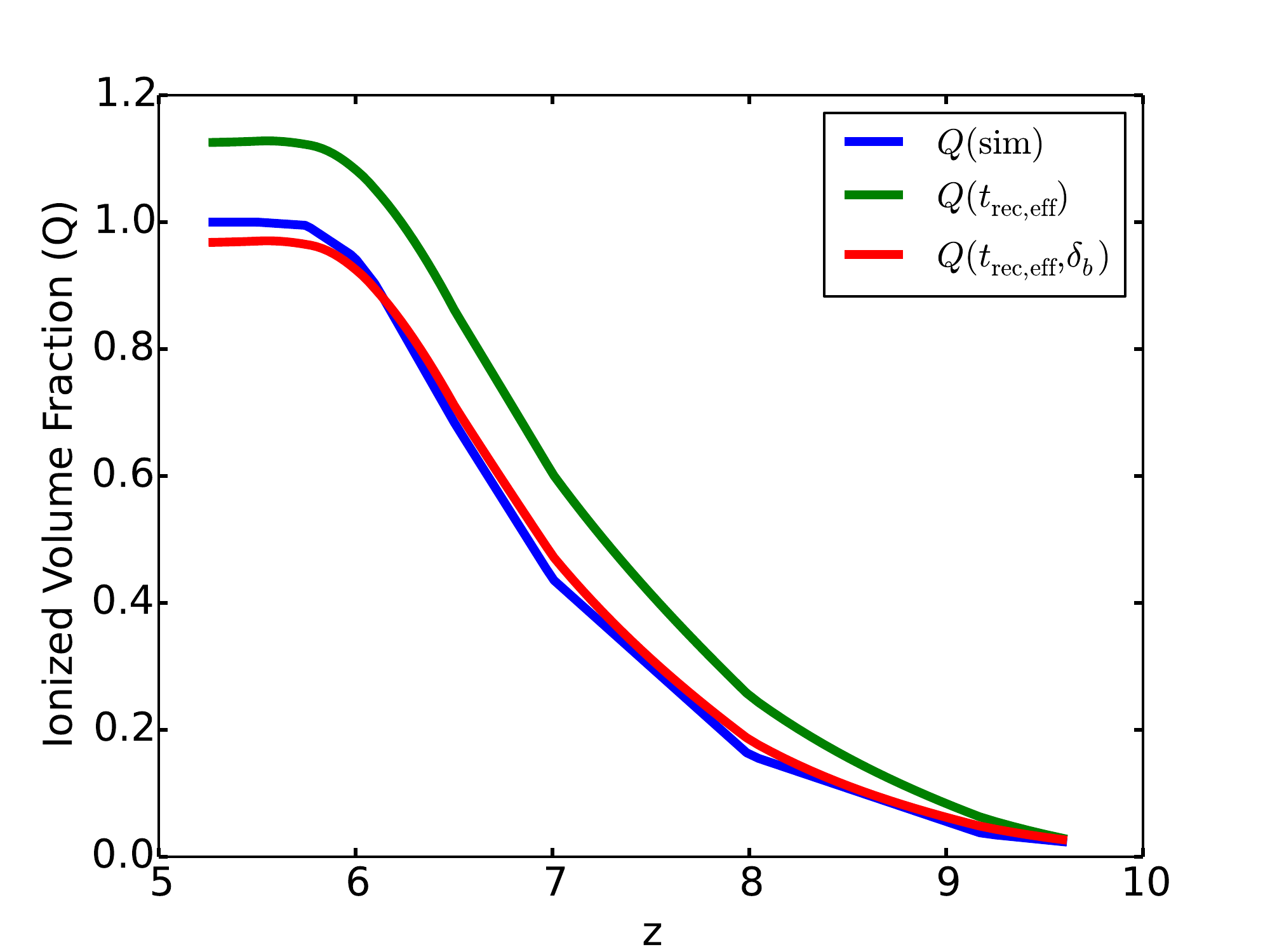}
	\caption{Improved agreement between theory and simulation. Green and blue curves are as in Fig. \ref{Qeffv1}. Red curve is obtained by integrating modified evolution equation for Q taking into account the overdensity effect of Inside-out reionization (Equation \eqref{eq:dQdtdb}).}
	\label{Qeffv2}
\end{figure}

\begin{figure}
	\includegraphics[width=0.5\textwidth]{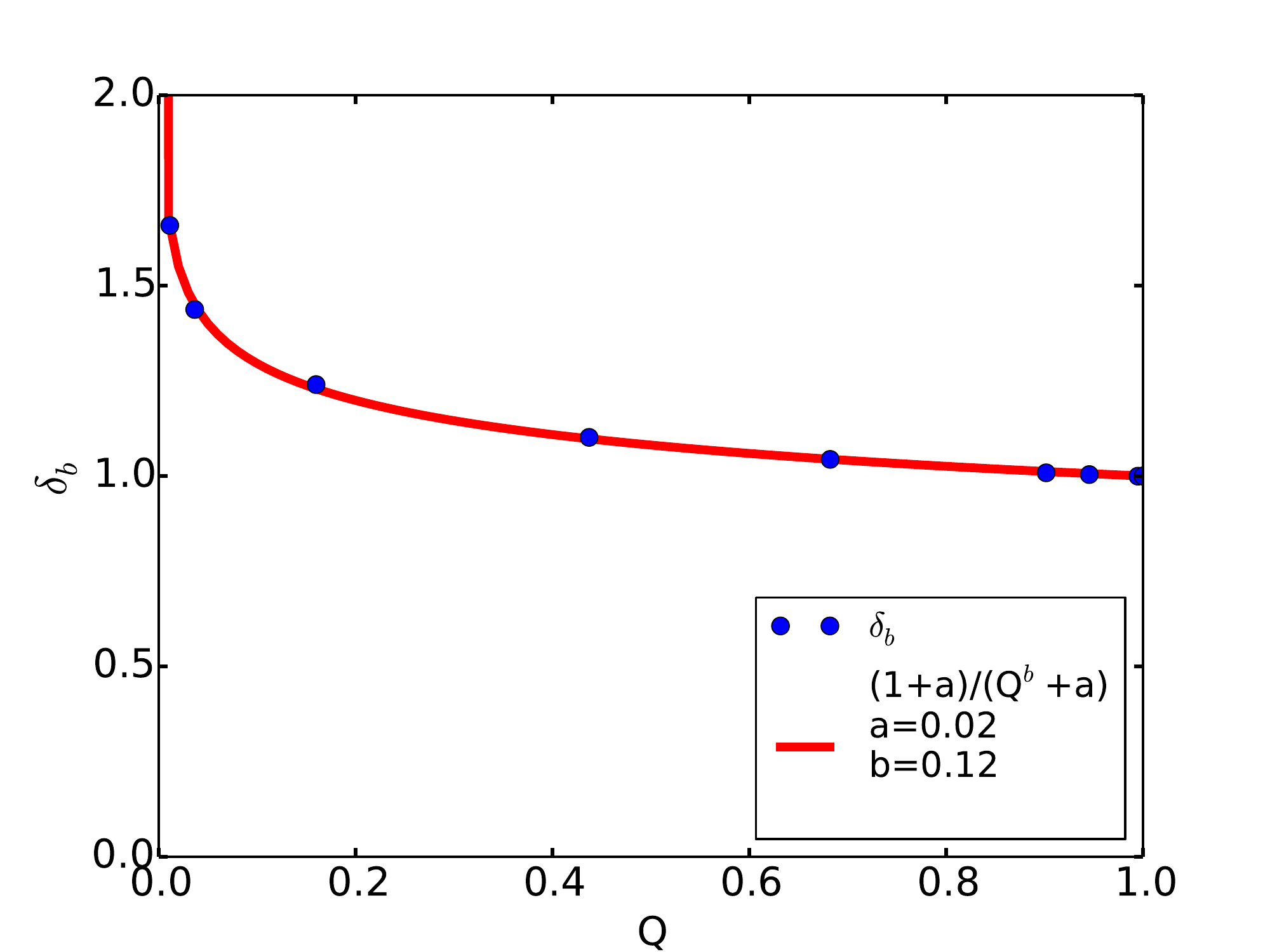}
	\caption{Mean baryon overdensity of ionized gas as a function of the ionized volume filling fraction Q. Blue points are measured in the simulation by averaging over the doubly thresholded cells obeying $\Delta_b<100$ and $x_e > 0.1$. Red curve is a fit to the data.}
	\label{deltabvsQfit5}
\end{figure}

To see if this formulation improves agreement with our simulated data, in Figure \ref{Qeffv2} we integrate Equation \eqref{eq:dQdtdb} again setting $\dot{n}_{ion}=\dot{N}_t$ and using $t_{rec,eff}$ to evaluate the second term. For comparison we show the curve obtained setting $\delta_b=1$, which repeats a curve already presented in Figure \ref{Qeffv1}.  Although the simulated and integrated analytic model curves do not agree exactly, the $Q(\delta_b,t_{rec,eff})$ curve shows much better agreement with the simulation, with error on the order of 1\% instead of 10\%.

\begin{figure}
	\includegraphics[width=0.5\textwidth]{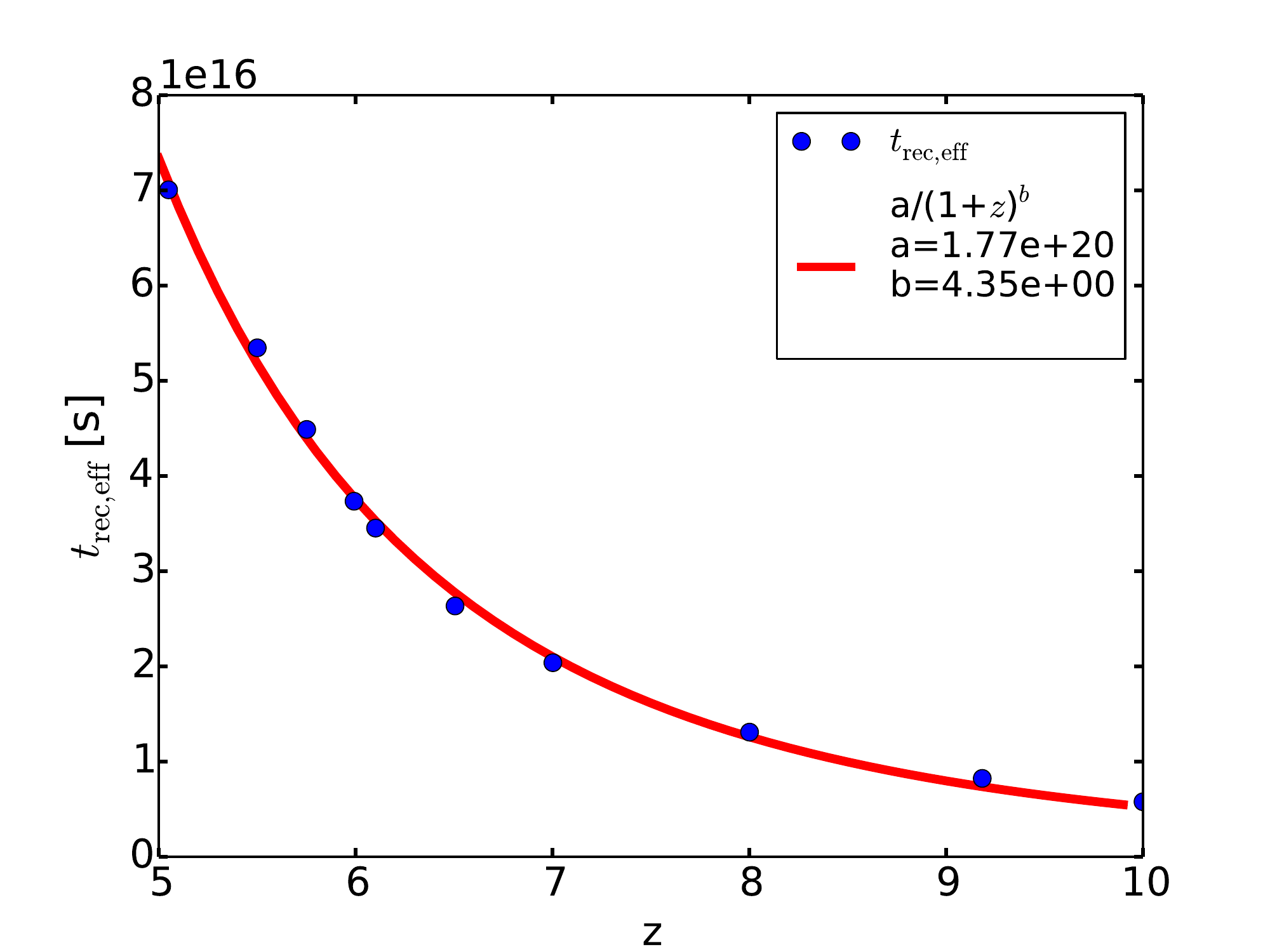}
	\caption{Analytic fit to $t_{rec,eff}$ (red line) , evaluated using simulation data (blue points) via Equation \eqref{eq:treceff}.}
	\label{treceffvszfit}
\end{figure}

By not assuming a constant emissivity and using the modified differential form in determining the volume filling fraction of Equation \eqref{eq:dQdtdb}, we are able to more accurately model the evolution of the simulated volume filling fraction of H {\footnotesize II} to the Well Ionized level. For completeness we plot in Figure \ref{treceffvszfit} the evolution of $t_{rec,eff}$ used in the above integration, including a reasonably good fit to the data. 

%For the volume filling fraction of H {\footnotesize II} $Q$, we are able to get a good agreement ($\sim O(1\%)$) with simulation using slight modification to the differential equation resulting in Equation \eqref{eq:dQdtdb}.  Even though the volume fraction is originally derived for a single radiating source in a uniform medium, we were able to modify it for the cosmological case.  We believe by using $\delta_b$ we are considering a more representative gas density that is actually affected by the radiation, and using $t_{rec,eff}$ is a more accurate representation of the recombination time for the hydrogen being ionized in the IGM.  Here we show the simple fitting formula for $\delta_b$ and $t_{rec,eff}$ respectively as Figure \ref{deltabvsQfit5} and \ref{treceffvszfit}.

\begin{figure}
	\includegraphics[width=0.5\textwidth]{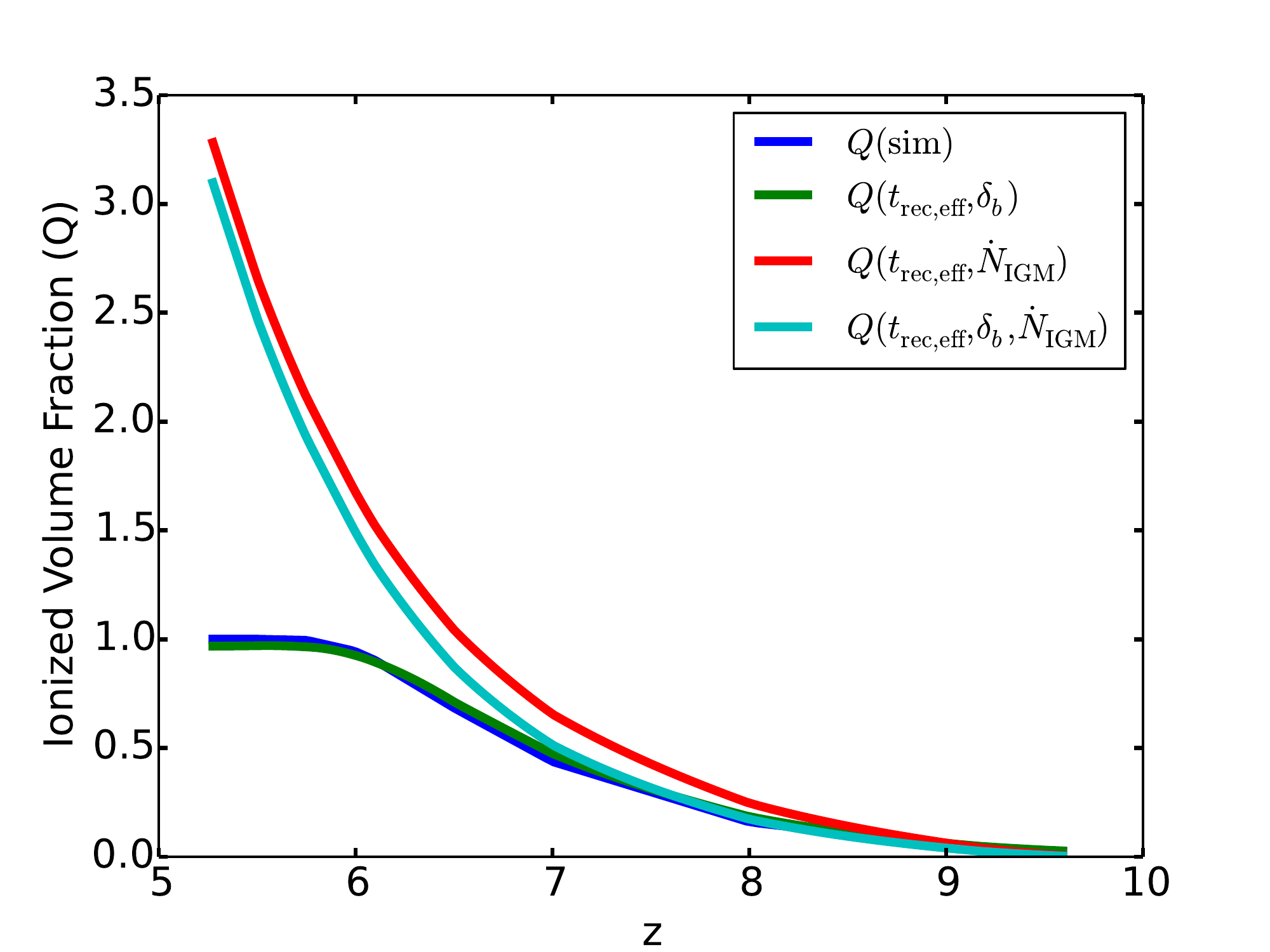}
	\caption{Dependence of analytic models on the choice for $\dot{n}_{ion}$. Red and teal curves assume $\dot{n}_{ion}=\dot{N}_{IGM}$; i.e., the photon injection rate into the IGM. Green curve assumes $\dot{n}_{ion}=\dot{N}_{t}$; i.e., the measured photoionization rate in the IGM. Blue curve is $Q(sim)$--the measured ionized volume filling fraction in the simulation. The green and teal curves take into account the overdensity effect of inside-out reionization (Equation \eqref{eq:dQdtdbg}), while the red curve assumes $\delta_b=1$. All models assume $\bar{t}_{rec}=t_{rec,eff}$ as measured in the simulation (Fig. \ref{treceffvszfit}.}
	\label{Qeffv3}
\end{figure}

Finally, we return to the question of what is the appropriate choice for $\dot{n}_{ion}$ in Equation \eqref{eq:dQdtdb}. This is commonly taken to be the rate at which ionizing photons are injected into the IGM (e.g., Haardt \& Madau 2012, \S9.3), because this can be connected to the observed UV luminosity density $\rho_{UV}$ by the formula $\dot{n}_{ion}=f_{esc}\xi_{ion}\rho_{UV}$, where $f_{esc}$ is the escape fraction for ionizing radiation, and $\xi_{ion}$ is the rate of ionizing photons per unit UV (1500 \AA{}) luminosity for the stellar population \citep{RobertsonEtAl2013}. However we have obtained excellent agreement between simulation and Equation \eqref{eq:dQdtdb} using the mean ionization rate density in the IGM  $\dot{N}_t$, which differs from the ionizing photon injection rate density $\dot{N}_{IGM}$ as $Q \rightarrow 1$. In Figure \ref{Qeffv3} we show the result of integrating Equations \eqref{eq:dQdt} and \eqref{eq:dQdtdb} with the choice $\dot{n}_{ion}=\dot{N}_{IGM}$, as originally proposed by \citep{MadauEtAl1999}. Also plotted in Figure \ref{Qeffv3} is $Q(sim)$ (blue line) and our best agreeing model (green line). The red line ignores the $\delta_b$ correction, and deviates to the high side of $Q(sim)$ almost immediately, for reasons we discussed earlier. It crosses $Q=1$ at $z\approx 6.6$, which is too early by $\Delta z =0.8$. The teal line includes the $\delta_b$ correction, and tracks the $Q(sim)$ closely to $z \approx 7$, and thereafter deviates on the high side. It crosses $Q=1$ at $z\approx 6.4$, which is too early by $\Delta z =0.6$. Both curves show an accelerated change in $Q$  as z decreases, which is characteristic of standard analytic ionization models (e.g., Haardt \& Madau 2012, Fig.14). By contrast, the simulation and our best fit model using $\dot{n}_{ion}=\dot{N}_t$ show a decelerated change in $Q(z)$ as $Q \rightarrow 1$. This is clearly due to the fact that the ration of ionizations to emitted photons decreases as $Q \rightarrow 1$, as illustrated in Figure 22. The consequence of this flattening in the $Q(z)$ curve is a delay in redshift of overlap of $\Delta z=0.6-0.8$, relative to the predictions of Equations \eqref{eq:dQdtdb} and \eqref{eq:dQdt}, respectively, using the photon injection rate as the source term. 

\begin{figure}
	\includegraphics[width=0.5\textwidth]{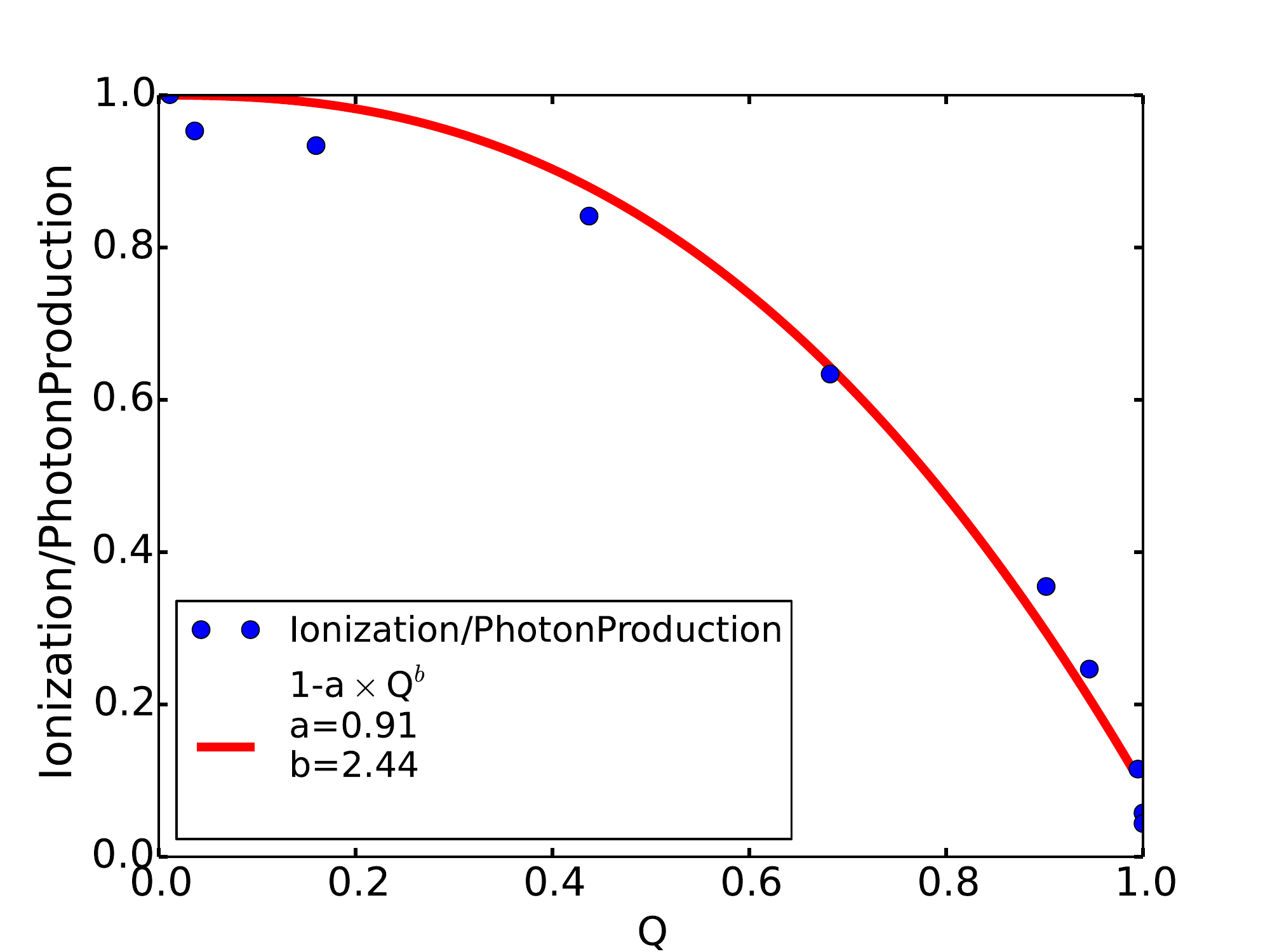}
	\caption{Ratio of the volume averaged \hi photoionization rate to photon injection rate in the IGM as a function of $Q$. Data points are measured from the simulation; line is a simple powerlaw fit. }
	\label{RatiovsQfit}
\end{figure}

We have seen above that the ionization rate density is the appropriate quantity to use to source the $dQ/dt$ equation, independent of $\delta_b$ corrections. Because the ionization rate density is not directly observable, but since $\dot{n}_{ion}$ can be derived from observables, we introduce a correction factor to convert from one to the other. Defining 
\begin{equation}
\gamma \equiv \frac{\langle n_{HI}\Gamma_{HI}^{ph}\rangle}{\dot{n}_{ion}} = \frac{\dot{N}_t}{\dot{n}_{ion}}
\label{gamma}
\end{equation}
\\where the angle brackets denote an average over the singly thresholded volume (IGM), then we can recast Equation \eqref{eq:dQdtdb} into a form useful for observers:
\begin{equation}
	\frac{dQ}{dt} = \frac{\gamma\dot{n}_{ion}}{\delta_b\bar{n}_\mathrm{H}}-\frac{Q}{\bar{t}_{rec}}, 
	\label{eq:dQdtdbg}
\end{equation}
\\where $\gamma$ and $\delta_b$ are functions of $Q$. 
In Fig. \ref{RatiovsQfit} we plot data values for $\gamma(Q)$ taken from our simulation, as well as a simple powerlaw fit. The fit is not meant to be definitive, but merely illustrative. More simulations need to be performed under various circumstances, and better fits made, to see whether our $\gamma(Q)$ is approximately universal, or merely anecdotal. 

\section{Discussion}
\label{Discussion}

%\begin{itemize}
%\item what we have done
\subsection{Significance of our Main Results}
We have carried out a fully-coupled radiation hydrodynamic cosmological simulation of hydrogen reionization by stellar sources using an efficient flux-limited diffusion radiation transport solver coupled to the Enzo code (Paper I). This method has the virtue of a high degree of scalability with respect to the number of sources, which allows us to simulate reionization in large cosmological volumes including hydrodynamic and radiative feedback effects self-consistently. In this paper we have presented first results from a simulation in a cosmological volume of modest size--20 Mpc comoving--to investigate the detailed radiative transfer, nonequilibrium photoionization, photoheating and recombination processes that operate during reionization and dictate its progress. In a future paper we apply our method to larger volumes to examine the large scale structure of reionization, evolution of the bubble size distribution, etc. 

The simulation presented here is carried out on a uniform mesh of $800^3$ cells and with an equivalent number of dark matter particles. As such, the mass resolution is sufficiently high to evolve a dark matter halo population which is complete down to ($M_{halo} \approx 10^8 M_{\odot}$) which cools via H and He atomic lines. However, a spatial resolution of 25 kpc comoving poorly resolves internal processes within early galaxies, but does an excellent job of resolving the Jeans length in the photoionized IGM \citep{BryanEtAl1999}. Our simulation is most appropriately thought of as a high redshift IGM simulation which evolves an inhomogeneous ionizing radiation field sourced by star-forming early galaxies. Star formation is modeled using a modified version of the Cen \& Ostriker (1992) recipe that can be tuned to reproduce the observed star formation rate density (SFRD) \citep{SmithEtAl2011}. We have tuned our simulation to roughly match the observed SFRD \citep{BouwensEtAl2011,RobertsonEtAl2013} for $z\geq 7$, but due to the small boxsize, it somewhat underpredicts the SFRD for $z < 7$. Our simulation also matches the observed $z=6$ galaxy luminosity function well, which gives us some confidence that our ionizing souce population is representative of the real universe. However a substantial fraction of our ionizing flux comes from sources that are too faint to be observed; we defer a discussion of this topic to Paper III in this series (So et al., {\em in prep.})

Our goal was not to predict the precise redshift of ionization completion, as this would depend on details such as escape fraction of ionizing radiation from galaxies and their stellar populations that we do not model directly. Rather our goal was to examine the mechanics of reionization in its early, intermediate, and late phases within a model which is calibrated to the observed source population. Nonetheless, we present a model in which reionization completes at $z\approx 6$, consistent with observations. 

At early and intermediate times we find that reionization proceeds ``inside-out", confirming the results of many previous investigations \citep{Gnedin2000,RazoumovEtAl2002,SokasianEtAl2003,FurlanettoEtAl2004,IlievEtAl2006,TracCen2007,TracEtAl2008}. However, at late times isolated islands of neutral gas are ionized from the outside-in as they have no internal sources of ionization. Even this characterization is somewhat oversimplified when {\em degree of ionization} is considered, as we discussed in Sec. \ref{IOOI}. It accurately depicts how reionization proceeds for a low degree of ionization (> 0). However for high degrees of ionization, ``inside-out-middle" is more appropriate, as filaments lag behind low and high density regions, as discussed by \cite{FinlatorEtAl2009}. 

Our most interesting findings concerns the widely used analytic model of reionization introduced by \cite{MadauEtAl1999}. Both the instantaneous (Equation \ref{eq:ndot}) and time-dependent (Equation \ref{eq:dQdt}) versions of this model underpredict the time (overpredict the redshift) when reionization completes, when applied to our simulation. There are two reasons for this having to do with the detailed mechanics of reionization at early and late times respectively. At early times, I-fronts are propagating in regions of higher density than the cosmic mean since the first sources are highly biased. Higher densities translate into slower bubble expansion rates, retarding $Q_{\hii}(z)$ relative to a solution which assumes the cosmic mean density (Figure \ref{Qeffv2}). At late times, which we loosely define as $Q_{\hii} > 0.5$, conversion of ionizing photons into new ionized hydrogen atoms becomes inefficient. This can be seen by forming this ratio directly from the simulation data (Figure \ref{Ndot_Ratio}), or by defining a global \hi  ionization parameter (Equation \eqref{eq:IP} and Figure \ref{IP}). The consequence of this dropping ionization efficiency, which is as low as 0.05 at overlap in our simulation, is to further retard $Q_{\hii}(z)$ relative to a solution which assumes an ionization efficiency of unity (Figure \ref{Qeffv3}).

We have introduced a modified version of \cite{MadauEtAl1999}'s time-dependent analytic reionization model in Equation \eqref{eq:dQdtdbg}. Modifications which correct for the above-mentioned effects apply to the source term only, {\em not to the recombination term}. These corrections are therefore totally independent of issues like clumping factors and the temperature of the IGM, which enter into the characteristic recombination time of the IGM. The modifications are introduced as correction factors to the mean density of baryons in the vicinity of ionizing sources at early times ($\delta_b$), and the conversion efficiency of ionizing photons emitted to \hi photoionization rate at late times ($\gamma$). Fits of these two correction factors versus $Q_{\hii}$ are presented in Figures \ref{deltabvsQfit5} and \ref{RatiovsQfit} for consumption by other researchers. At this point we do not know how general these results are. However we have indications based on another simulation we have analyzed with a softer source SED that the functional forms are representative of this class of reionization model. 

The significance of these results to high redshift galaxy observers is the following. Setting $Q_{\hii} = 1$ and $\delta_b = 1$ in Equation \eqref{eq:dQdtdbg}, we derive
\begin{equation}
\dot{n}_{ion} = \frac{1}{\gamma} \frac{\bar{n}_H}{\bar{t}_{rec}}.
\label{eq:ndotgamma}
\end{equation}
This differs from the usual expression used to assess whether a given ionizing photon injection rate can maintain an ionized IGM by the factor $1/\gamma$, which is a factor of $\sim 20$ at overlap in our simulation. If this result is correct, then it means that the required UV luminosity density to maintain an ionized IGM has been underestimated by a factor of approximately 20. However, a more precise statement would be that the UV luminosity density required to maintain the IGM in a {\em highly ionized state; $f_n =10^{-5}$} is 20 times higher than what has been previously estimated. Lower levels of UV luminosity density than that specified in Equation \eqref{eq:ndotgamma} could still maintain the IGM in an ionized state, but one with a higher neutral fraction. 

As we showed in Figure \ref{treceffhubble}, the effective recombination time at and after overlap in our model is comparable to the Hubble time, whether we use the Madau formula to evaluate it for reasonable values for the clumping factor, or we evaluate it directly from our simulation data. This fact casts in doubt the entire instantaneous photon counting argument which is the basis of Equation \ref{eq:ndot}, and the equation becomes less useful for the purposes to which it has been applied (e.g., Robertson et al. 2013). It means that the ionization state of the IGM has a memory on the timescale of $\bar{t}_{rec}$ which is always a significant fraction of $t_{Hubble}$ before overlap, and of order the Hubble time after overlap. We therefore recommend observers use the time-dependent version Equation \eqref{eq:dQdtdbg} in future assessments of high redshift galaxy populations and their role in reionization. 

\subsection{Limitations of the Simulation}

We conclude this section with a brief discussion of the known limitations of our simulation and a comparison of our results with others in the published literature. First the limitations. The principal limitation is the use of a uniform grid, which prevents us from resolving processes occuring inside galaxy halos. The main defect this introduces is an inability to calculate the ionizing escape fraction directly, as is done in some high resolution simulations; e.g., \cite{WiseCen2009,FernandezShull2011}. In our simulation, we calibrate our star formation recipe to match the observed SFRD, and then use that that to calculate UV feedback cell-by-cell via Equation \eqref{eq:emissivity}. We use a value for $\epsilon_{UV}$ taken from \cite{RicottiEtAl2002} for an unattenuated low metallicity stellar population. We underestimate the amount of internal attenuation of ionizing flux due to our limited resolution within halos, and we do not incorporate an explicit escape fraction parameter in Equation \eqref{eq:emissivity}. Effectively, we assume $f_{esc}(ISM)=1$. Using a lower value for $f_{esc}$ would result in a lower overlap redshift \citep{PetkovaSpringel2011a}. Clearly, it would be desirable to vary this parameter in future studies. 

A second limitation of our simulation is that we have presented only one realization in a relatively small box. Previous studies have shown that \hii bubbles reach a characteristic size of $\sim 10$ Mpc comoving in the lates stages of reionization \citep{FurlanettoEtAl2004,ZahnEtAl2007,ShinEtAl2008}. At 20 Mpc on a side, our box is scarcely larger than this. Therefore one can ask how robust our results are to boxsize. We have addressed this by carrying out a simulation of identical physics, spatial, and mass resolution in a volume 64 times as large as the one described in this paper. The simulation is carried out in a box 80 Mpc on a side on a uniform mesh of $3200^3$ cells, and with an equivalent number of dark matter particles. Results of this simulation will be presented in a forthcoming paper (So et al., in preparation). For the present we merely state that the $Q_{\hii}(z)$ curve for the $800^3$ simulation falls within the $\pm 1 \sigma$ band for the larger simulation, where this band is obtained by subdividing the large simulation into 64 cubes of size 20 Mpc on a side, and calculating the mean and standard deviation. While the larger box begins to ionize at a slightly earlier redshift, due to the presence of higher sigma peaks forming galaxies, both simulations complete reionization at the same redshift, $z_{reion} = 5.8$. The $Q_{\hii}(z)$ curve for the $800^3$ simulation is near the lower edge of the band, which means that at intermediate redshifts ($7 \leq z \leq 8$), where the difference is largest, the small box simulation underestimates the fraction of the volume that is ionized by about 20\%, with differences smoothly decreasing to lower and higher redshift. 

A third limitation is that our SFRD systematically deviates from observations below $z \sim 7$, flattening and then decreasing slightly, rather than continuing to rise (Figure \ref{SFR}). The large box simulation does not show this effect, but rather tracks the observed SFRD over the entire range of redshifts. The difference in the mean SFRD between the large and small box simulations increases smoothly from 0.1 dex at $z=9$ to 0.3 dex at $z=6$. The higher levels of star formation in the large box simulation accounts for the higher ionized volume fraction at intermediate redshifts. Nonetheless, the two simulations complete reionization at virtually the same redshift, which is a curious result which we address in a subsequent paper.

Another limitation of our method is the use of flux-limited diffusion (FLD) to transport radiation. It is well known that FLD does not cast shadows behind opaque blobs. This could potentially overestimate how rapidly the IGM ionizes, and hence overestimate $z_{reion}$. In Paper I we showed through a direct comparison between FLD and an adaptive ray tracing method incorporated in the {\em Enzo} code on a standard test problem that the differences in the volume- and mass-weighted ionized volume fraction are small. This was for a rather small volume with a small number of ionizing sources. The differences will likely be even smaller as larger volumes containing larger numbers of sources are considered. At the present time, no fully-coupled radiation hydrodynamic simulations of reionization using ray tracing in large volumes are available to compare our method against, to confirm or deny this conjecture. 

\subsection{Comparison with Other Self-Consistent Simulations}

Finally, we compare our results to the results of several recent fully-coupled simulations of reionization including hydrodynamics, star formation, and radiative transfer. \cite{PetkovaSpringel2011a} simulated a (10 Mpc/h)$^3$ volume with the {\tt Gadget-2} code coupled to a variable tensor Eddington factor moment method for the ionizing radiation field sourced by star forming galaxies. They carried out a suite of simulations with $2 \times 128^3$ gas and dark matter particles, varying the ionizing escape fraction and the mean energy per photon from hot, young stars. The also performed one simulation at $2 \times 256^3$ resolution to check for convergence. Our simulation has 80/10 times superior mass resolution as their $128^3/256^3$ simulations. Because {\tt Gadget} is a Lagrangian code, our Eulerian simulation has 8/16 times lower resolution in the highest density regions, but 4.46/2.23 times higher resolution at mean density, and even higher resolution compared to the {\tt Gadget} simulations in low density voids. Our method also has a more accurate adaptive subcycling timestepping scheme for the coupled radiation-ionization-energy equations, obviating the need to model nonequilibrium effects by means of a gas heating parameter $\epsilon$. 

Morphologically, our results are qualitatively similar, as are the neutral hydrogen fraction versus overdensity phase diagrams. As might be expected from the two methods, the phase diagrams show some differences at the highest and lowest overdensities which is likely a resolution effect. The SFRD in the \cite{PetkovaSpringel2011a} simulation is about an order of magnitude higher than observed, making a direct comparison on $Q_{\hii}(z)$ somewhat problematic. However, since they vary the ionizing escape fraction, we can roughly compare their $f_{esc}=0.1$ case with our results. Their model completes reionization at $z \approx 5$ compared to our own which completes at $z \approx 5.8$. They plot the quantity $log[1-Q_{\hii}(z)]$, which makes the end of reionization look abrupt. We plot $Q_{\hii}(z)$, which makes the end of reionization look slow. When we plot $log[1-Q_{\hii}(z)]$ using our data, it looks very similar to their curves, and shows a rapid plunge in the average neutral fraction at late times.  \cite{PetkovaSpringel2011a} do not compare with the predictions of the \cite{MadauEtAl1999} model, nor do they investigate the evolution of clumping factors, recombination times, or the number of photons per H atom to achieve overlap as we do. We do not investigate the properties of the $z=3$ IGM via Lyman $\alpha$ forest statistics, as they do. Therefore further comparisons are not possible at this time. 

\cite{FinlatorEtAl2012} examined some of the same issues we have, hence a comparison with their results is informative. They carried out a suite of {\tt Gadget-2} simulations in small volumes (3, 6)Mpc/h coupled to a variable tensor Eddington factor moment method. Unlike \cite{PetkovaSpringel2011a}, the radiation transport is solved on a uniform Cartesian grid, rather than evaluated using the SPH formalism. The results presented in \cite{FinlatorEtAl2012} use $2 \times 256^3$ dark matter and gas particles, which given their small volumes, yields a similar mass resolution to our simulation, superior spatial resolution in high density regions, and slightly coarser spatial resolution at mean density and below. However, their radiation transport is done on coarse $16^3$ mesh, which in their fiducial run is $536$ comoving kpc $\approx 20 \times$ as coarse as ours. Their simulation thus coarse-grains the radiation field relative to the density field, which necessitates the introduction of a sub (radiation) grid model for unresolved self-shielded gas (i.e., Lyman limit systems). The effect of their subgrid model is to remove some gas in the overdensity regime $1 \leq \Delta_b \leq 50$ in the calculation of the \hii clumping factor, thereby lowering it. Since our radiation field is evolved on the same grid as the density field, we have not included an explicit subgrid model for unresolved self-shielded gas. Lyman limit systems, with neutral column densities of $\sim 10^{17}$ cm$^{-2}$, have a characteristic size of 10 physical kpc \citep{Schaye2001,McQuinnEtAl2011}. At $z=6$ this is 70 comoving kpc, which is resolved by 3 grid cells in our simulation. While this is lower than one would ideally like (5-10 cells), we believe we can make an apples-to-apples comparison between our resolution-matched simulation results and Finlator et al.'s results. 

Our results are in broad agreement with those of \cite{FinlatorEtAl2012}, with some minor quantitative differences.  We both find that the unthresholded baryon clumping factor $C_b$ significantly overestimates the clumping in ionized gas at redshifts approaching overlap, and therefore that it should not be used to estimate the mean recombination rate in the IGM. We confirm their findings that properly accounting for the ionization state and temperature of gas of moderate overdensities lowers the clumping factor to less than $\approx 6$ (in our case less than 5; see Figure \ref{ClumpingFactors}).  Finlator et al. quote a value for $C_{\hii}$ of 4.9 at $z=6$ taking self-shielding into account, which is in good agreement with our value of $C_{tt\hii} \approx 4.8$. However, they favor a lower value for $C$ of 2.7-3.3 taking temperature corrections into account. This can be compared with our value for $C_{RR} \approx 2.3$, which includes temperature corrections but also excludes gas with $\Delta_b<1$. Including this low density gas, as Finlator et al. do, would raise this value somewhat since a larger range of densities enter into the average. We conclude therefore that clumping factors derived from our simulation are in good agreement with those reported by \cite{FinlatorEtAl2012}. 

We find that approximately 2 photons per hydrogen atom ($\gamma/H\approx 2$) are required to reionize gas satisfying $\Delta_b<100$--our proxy for the fluctuating IGM. \cite{FinlatorEtAl2012} quote a model-dependent value for $\gamma/H$ which depends on the redshift at which the IGM becomes photoheated and thereby Jeans smoothed  (their Fig. 7). For $z=6$,  $\gamma/H \approx 5$, significantly higher than our number evaluated directly from the simulation. However, for $z=8$, when our box is already significantly ionized,  $\gamma/H \approx 3$. Because there are many model-dependent assumptions that go into the Finlator et al. estimate, we consider this reasonably good agreement. However we point out that our estimate is the first to be derived from a self-consistent simulation of reionization with no subgrid models aside from the star formation/radiative feedback recipe. 

Finally, \cite{FinlatorEtAl2012} compare $Q_{\hi}(z)=1-Q_{\hii}(z)$ for their fiducial model with the time-dependent model of \cite{MadauEtAl1999}. They point out the sensitivity of the redshift of overlap on the choice of clumping factor, which enters into the recombination time, and showed that $C_{\hii}$ provides better agreement with theory at early times than $C_b$, consistent with our findings. Since small discrepancies in $Q_{\hii}(z)$ at early times are masked by plotting $Q_{\hi}(z)$, Finlator et al. did not discover the need for our overdensity correction $\delta_b$. Similar to us, they found that even with the best clumping factor estimate the analytic model predicts that reionization completes earlier than the simulation by $\Delta z \approx 1$. They ascribe this delay to finite speed-of-light effects (which can only account for $\Delta z =0.1$), while we ascribe it to nonequilibrium ionization effects. \cite{FinlatorEtAl2012} did not propose modifications to the \cite{MadauEtAl1999} model to improve agreement with simulation, as we do in Equation \eqref{eq:dQdtdbg}.

%\item simulation assumptions
%\item what we have found
%\item comparison with previous work (agreement/disagreement)
%\item implications for connecting high-z galaxies to reionization
%\item areas for improvement

%\end{itemize}
\section{Summary and Conclusions}
\label{Conclusions}

We now summarize our main results.
\begin{enumerate}
\item
We use a fully self-consistent simulation including self-gravity, dark
matter dynamics, cosmological hydrodynamics, chemical ionization and
flux limited diffusion radiation transport, to look at the epoch of hydrogen
reionization in detail. By tuning our star formation recipe to approximately match the observed high redshift star formation rate density and galaxy luminosity function, we have created a fully coupled radiation hydrodynamical realization of hydrogen reionization which begins to ionize at $z \approx 10$ and completes at $z \approx 5.8$ without further tuning. While our goal is not the detailed prediction of the redshift of ionization completion, the simulation is realistic enough to analyze in detail the role of recombinations in the clumpy IGM on the progress of reionization. 
\item
We find that roughly 2 ionizing photons per H atom are required to convert the neutral IGM to a well ionized state ($f_i>0.999$), which supports the ``photon starved'' reionization scenario discussed by \cite{BoltonHaehnelt2007}.
\item
Reionization proceeds initially ``inside-out", meaning that regions of higher mean density  ionize first, consistent with previous studies. However the late stages of reionization are better characterized as ``outside-in" as isolated neutral islands are swept over by externally driven I-fronts. Intermediate stages of reionization exhibit both characteristics as I-fronts propagate from dense regions to voids to filaments of moderate overdensity. In general, the appropriateness of a given descriptor depends on the level of ionization of the gas, and the reionization process is rather more complicated that these simple descriptions imply. 
\item
The evolution of the ionized volume fraction with time $Q_{\hii}(z)$ depends on the level of ionization chosen to define a parcel of gas as ionized. The curves for ionization fractions $f_i = 0.1$ and $f_i =0.999$ are very similar, but the curve for $f_i =0.99999$ is significantly lower at a given redshift, amounting to a delay of $\Delta z \approx 1$ relative to the other curves for $Q_{\hii} \ll 1$, smoothly decreasing to 0 as the redshift of overlap is approached.

\item
Before overlap, 30-40\% of the total recombinations occur outside halos in our simulation, where this refers to gas with $\Delta_b < 100$. After overlap, this fraction decreases to 20\% and continues to decrease to lower redshifts. 
\item
Before and after overlap, 3-4\% of the total recombinations occur in voids (defined as $\Delta_b < 1$.) While this is a small fraction of all recombinations, it is about 10\% of the recombinations in the IGM before overlap, increasing to about 20\% by $z=5$. The contribution of voids to the ionization balance of the IGM is therefore not negligible. 
\item
The formula for the ionizing photon production rate needed to maintain the IGM in an ionized state derived by \cite{MadauEtAl1999} (Eq. \ref{eq:ndot}) should not be used to predict the epoch of reionization completion because it ignores history-dependent terms in the global ionization balance which are not ignorable. While not originally intended for this purpose, it is being used by observers to assess whether increasingly higher redshift populations of star forming galaxies can account for the ionized state of the IGM. A direct application of the formula to our simulation predicts an overlap redshift of $z=7.4$ compared to the actual value of $z=5.8$. 
\item
Estimating the recombination rate density in the IGM before overlap through the use of clumping factors based on density alone is unreliable because it ignores large variations in local ionization state and temperature which increase the effective recombination time compared to density-based estimates. For a currently popular value of the clumping factor $C=3$ \citep{ShullEtAl2012}, the formula for $\bar{t}_{rec}$ from \cite{MadauEtAl1999}(Eq. \ref{eq:tmadau}) understimates by $2\times$ at all redshifts the effective recombination time measured directly from the simulation. If we adjust $C$ downward so that Eq. \ref{eq:tmadau} matches $t_{rec,eff}$ from the simulation, then it is too low by 60\% at $z=6$ due to the aforementioned effects. 
\item
The assumption that $\bar{t}_{rec}/t \ll 1$ which underlies the derivation of Eq. \ref{eq:ndot} is never valid over the range of reionization redshifts explored by our simulation (Fig. \ref{treceffhubble}). Depending on how $\bar{t}_{rec}$ is evaluated, $\bar{t}_{rec}/t$ increases from $0.3-0.4$ at $z=9.7$ to $\geq 1$ at overlap. This means that an instantaneous analysis of the ionization balance in the IGM post overlap is invalid because recombination times are so long. 
\item
Retaining time-dependent effects is important for the creation of analytic models of global reionization. The analytic model for the evolution of $Q_{\hii}$ introduced by \cite{MadauEtAl1999}(Eq. \ref{eq:dQdt}) retains important time-dependent effects, and predicts well the shape of our simulated curve, but overpredicts $Q_{\hii}$ at all redshifts because it does not take into account that reionization begins in overdense regions consistent with the inside-out paradigm. It also assumes every emitted ionizing photon results in a prompt photoionization, which is not true in our simulation at late times $Q_{\hii}>0.5$. The Madau model, which ignores these effects, predicts a universe which reionizes too soon by $\Delta z \approx 1$. When we introduce correction factors for these effects into Eq. \ref{eq:dQdtdbg} the simulation and model curves agree to approximately 1\% accuracy. We recommend researchers use Eq. \ref{eq:dQdtdbg} for future analytic studies of reionization. 
\item
Finally, we present in Figs. \ref{deltabvsQfit5}, \ref{treceffvszfit}, and \ref{RatiovsQfit} fitting functions for the overdensity correction $\delta_b(Q)$, the effective recombination time derived from our simulation, and the ionization efficiency parameter $\gamma(Q)$ which may be useful for other researchers in the field. 
\end{enumerate}

This research was partially supported by National Science Foundation grants AST-0808184 and AST-1109243
and Department of Energy INCITE award AST025 to MLN and DRR. Simulations were performed on the {\em Kraken}
supercomputer operated for the Extreme Science and Engineering Discovery Environment (XSEDE)
by the National Institute for Computational Science (NICS), ORNL with support from XRAC allocation MCA-TG98N020 to MLN.
MLN, DRR and GS would like to especially acknowledge the tireless devotion to this project by our co-author Robert Harkness
who passed away shortly before this manuscript was completed.

%%Succinct format, expanded version of the abstract.  What we did and significance is.

\bibliography{ref}
\bibliographystyle{apj}
\end{document}